\documentclass[12pt]{article}

\parskip=\medskipamount

\usepackage{tensor}
\usepackage{physics}
\usepackage{slashed}
\usepackage{amsmath,amssymb,calc, amsthm,bbm, epsfig,psfrag, mathtools}
\usepackage[normalem]{ulem} 
\newtheorem{theorem}{Theorem}

\theoremstyle{definition}
\newtheorem{corollary}{Corollary}

\def\cO{{\mathcal{O}}}

\usepackage{latexsym,bm,amsfonts}
\usepackage{graphicx, enumerate}
 \usepackage[all,cmtip]{xy}
\usepackage[numbers,sort&compress]{natbib}
\usepackage[dvipsnames]{xcolor}
\usepackage{mathrsfs}
\usepackage{tikz,tikz-cd}

\allowdisplaybreaks

\newcommand{\Comment}[1]{{}}
\definecolor{darkblue}{rgb}{0.15,0.35,0.55}
\definecolor{reddish}{rgb}{0.65, 0.2, 0.2}
\usepackage[linktocpage=true]{hyperref}
\hypersetup{
colorlinks=true,
citecolor=darkblue,
linkcolor=reddish,
urlcolor=darkblue,
pdfauthor={},
pdftitle={},
pdfsubject={}
}

\def\be{\begin{equation}}
\def\ee{\end{equation}}
\def\bea{\begin{eqnarray}}
\def\eea{\end{eqnarray}}

\makeatletter
\renewcommand\section{\@startsection {section}{1}{\z@}%
 {-3.5ex \@plus -1ex \@minus -.2ex}
                                   {2.3ex \@plus.2ex}%
                                   {\normalfont\large\bfseries}}
\renewcommand\subsection{\@startsection{subsection}{2}{\z@}%
                                     {-3.25ex\@plus -1ex \@minus -.2ex}%
                                     {1.5ex \@plus .2ex}%
                                     {\normalfont\bfseries}}
\makeatother

\newfont{\goth}{ygoth.tfm scaled 1200}                   

 \numberwithin{equation}{section}

\def\1{{(1)}}
\def\2{{(2)}}
\def\3{{(3)}}

\newcommand{\overbar}[1]{\mkern 1.5mu\overline{\mkern-1.5mu#1\mkern-1.5mu}\mkern 1.5mu}

\def\TT{{T\overbar{T}}}

\newcommand {\cD}{{\cal D}}
\newcommand {\cE}{{\cal E}}

\newcommand {\cH}{{\cal H}}

\newcommand {\cM}{{\cal M}}
\newcommand {\cN}{{\cal N}}

\newcommand {\cR}{{\cal R}}

\newcommand {\cT}{{\cal T}}

\def\a{\alpha}
\def\b{\beta}

\def\d{\delta}
\def\e{\epsilon}

\def\g{\lambda}

\def\l{\lambda}
\def\m{\mu}
\def\n{\nu}
\def\o{\omega}

\def\q{\theta}
\def\r{\rho}
\def\s{\sigma}

\def\D{\Delta}

\def\L{\Lambda}

\def\ri{{\rm i}}
\def\re{{\rm e}}

\newcommand{\ad}{{\dot{\alpha}}}
\newcommand{\bd}{{\dot{\beta}}}

\newcommand{\sU}{\mathsf{U}}

\newcommand{\ve}{\varepsilon}
\newcommand{\cDB}{{\bar\cD}}

\newcommand{\pa}{\partial}
\newcommand{\hf}{\frac12}

\newcommand{\vf}{\varphi}

\newcommand{\non}{\nonumber}
\newcommand{\ba}{\begin{array}}
\newcommand{\ea}{\end{array}}

\def\double #1{#1{\hbox{\kern-2pt $#1$}}}

\newcommand{\bsubeq}{\begin{subequations}}
\newcommand{\esubeq}{\end{subequations}}

\newcommand{\ul}{\underline}

\newcommand{\rd}{\mathrm d}
\def\tr{{\rm tr}}

\begin{document}
\begin{titlepage}
\begin{flushright}
\today
\end{flushright}
\vspace{5mm}

\begin{center}
{\Large \bf
Interacting Chiral Form Field Theories and $\TT$-like Flows in Six and Higher Dimensions}
\end{center}

\begin{center}

{\bf
Christian Ferko,${}^{a}$
Sergei M. Kuzenko,${}^{b}$
Kurt Lechner,${}^{c, d}$
Dmitri P. Sorokin,${}^{d, c}$
Gabriele Tartaglino-Mazzucchelli${}^{e}$
} \\
\vspace{5mm}

\footnotesize{
${}^{a}$
{\it
Center for Quantum Mathematics and Physics (QMAP),
\\ Department of Physics \& Astronomy,  University of California, Davis, CA 95616, USA
}
 \\~\\
 ${}^{b}$
{\it
Department of Physics M013, The University of Western Australia \\
35 Stirling Highway, Perth W.A. 6009, Australia}
\\~\\
${}^{c}$
{\it
Dipartimento di Fisica ed Astronomia “Galileo Galilei” \\
Universit\`a degli Studi di Padova, Via Marzolo 8, 35131 Padova, Italy
}
 \\~\\
${}^{d}$
{\it
INFN, Sezione di Padova, Via Marzolo 8, 35131 Padova, Italy
}
  \\~\\
${}^{e}$
{\it
School of Mathematics and Physics, University of Queensland,
\\
 St Lucia, Brisbane, Queensland 4072, Australia}
}
\vspace{2mm}
~\\
\texttt{caferko@ucdavis.edu,
sergei.kuzenko@uwa.edu.au,
kurt.lechner@pd.infn.it,
dmitri.sorokin@pd.infn.it, g.tartaglino-mazzucchelli@uq.edu.au}\\
\vspace{2mm}

\end{center}

\begin{abstract}
\baselineskip=14pt
\noindent
In this paper we initiate the study of six-dimensional non-linear chiral two-form gauge theories as deformations of free chiral two-form gauge theories driven by stress-tensor $\TT$-like flows. To lay the background for this study, we elaborate on the relationship between different Lagrangian formulations of duality-invariant $p$-form theories and corresponding $\TT$-like flows in various dimensions.
To this end we propose a new formulation which (i) is a generalization of the four-dimensional construction by Ivanov, Nurmagambetov and Zupnik (INZ) and (ii) turns into the PST formulation upon integrating out an auxiliary self-dual field. We elucidate space-time covariant properties of the PST formulation by clarifying and making use of its relation to the INZ-type formulation and to a so-called ``clone" construction.

\end{abstract}
\vspace{5mm}

\vfill
\end{titlepage}

\newpage
\renewcommand{\thefootnote}{\arabic{footnote}}
\setcounter{footnote}{0}

\tableofcontents{}
\vspace{1cm}
\bigskip\hrule

\allowdisplaybreaks

\section{Introduction}

\subsection{Stress tensor deformations of field theories}

Recently, the study of deformations of quantum field theories involving operators constructed from the energy-momentum tensor $T_{\mu \nu}$ has received much attention. Several interesting new research directions have emerged, shedding new light on integrable quantum field theories, conformal invariance, supersymmetry, holography, and string theory.

The most famous deformation driven by composite operators built out of the energy-momentum tensor is the $\TT$ deformation of two-dimensional ($2d$) quantum field theories \cite{Zamolodchikov:2004ce, Smirnov:2016lqw,Cavaglia:2016oda}.
A remarkable property of $\TT$ is that, despite being an irrelevant operator, in the sense of renormalisation group flows, it is quantum mechanically well-defined. As such, $\TT$ is \emph{universal}, in the sense that it can be used to deform  any translation-invariant $2d$ QFT, and, in particular, it is part of the spectrum of any $2d$ quantum field theory. Moreover, the $\TT$ deformation has been shown to be \emph{solvable}, meaning that quantities in the deformed theory can often be computed in terms of corresponding quantities in the undeformed theory. This was shown for the finite volume spectrum \cite{Smirnov:2016lqw,Cavaglia:2016oda}, the $S$-matrix \cite{Dubovsky:2017cnj}, the torus partition function \cite{Cardy:2018sdv,Datta:2018thy,Aharony:2018bad}, and correlation functions \cite{Cardy:2019qao,Kraus:2018xrn}. It has also been shown that the $2d$ $\TT$ deformation preserves many \emph{symmetries} and other desirable properties of a seed theory. This includes integrability \cite{Smirnov:2016lqw} and supersymmetry \cite{Baggio:2018rpv,Chang:2018dge,Jiang:2019hux,Chang:2019kiu, Coleman:2019dvf,Ferko:2019oyv,Ferko:2021loo,Ebert:2022xfh}. The $\TT$ deformation of $2d$ CFTs leads to special classes of quantum field theories that, though not scale invariant due to the presence of a dimensionful coupling, prove to be invariant under a deformation of the Virasoro algebra \cite{Kruthoff:2020hsi,Guica:2020uhm,Georgescu:2022iyx,Guica:2022gts}, hence retaining an infinite dimensional set of symmetries.
The literature on the subject has been steadily growing since 2016, and we refer to \cite{Jiang:2019epa,monica_notes,Ferko:2021loo} and references therein for an introduction to the subject.

Considering the remarkable properties of $\TT$ in two space-time dimensions, it is natural to ask whether similar deformations exist in $d > 2$. To date, it remains unclear whether irrelevant local composite operators constructed out of the energy-momentum tensor can have the same quantum mechanical properties as $\TT$ in $d>2$ (a hope is that supersymmetry might help to find an example). Despite that, various generalisations of $\TT$ flows have been introduced in the literature \cite{Taylor:2018xcy,Bonelli:2018kik}. If one focuses even just on classical theories (or effective field theories), one inspiring property of $\TT$ is that its Lagrangian flow equation  (with $T_{\mu\nu}$ being the energy-momentum of a theory deformed by the parameter $\lambda$),
\begin{align}\label{TTbar_flow-0}
    \frac{\partial \mathcal{L}^{(\lambda)}}{\partial \lambda} = \frac{1}{4} \left( T^{\mu \nu} T_{\mu \nu} - \left( \tensor{T}{^\mu_\mu} \right)^2 \right) \, ,
\end{align}
takes free scalar fields in $d = 2$ into the theory of a gauge-fixed Nambu-Goto string \cite{Cavaglia:2016oda}.
A similar classical flow equation in general space-time dimension,
\begin{align}\label{TTbar_flow}
    \frac{\partial \mathcal{L}^{(\lambda)}}{\partial \lambda}
    = \cO_{\TT}
    :=\frac{1}{2d} \left( T^{\mu \nu} T_{\mu \nu} - \frac{1}{d} \left( \tensor{T}{^\mu_\mu} \right)^2 \right) \, ,
\end{align}
once specialised to the $d=4$ case, was proven to deform the free Maxwell Lagrangian, $\mathcal{L}_0 = - \frac{1}{4} F_{\mu \nu} F^{\mu \nu}$, into the Born-Infeld theory describing the effective gauge dynamics on a brane \cite{Conti:2018jho}.
It was successively shown that Born-Infeld theory in $d = 3$ satisfies a classical $\TT$-like flow connected to the free Maxwell theory  \cite{Ferko:2023sps}
\begin{align}\label{BI_TT_3d}
\frac{\partial \mathcal{L}^{(\lambda)}}{\partial \lambda} = \frac{1}{6} T^{\mu \nu} T_{\mu \nu}
- \frac{1}{9} \left( \tensor{T}{^\mu_\mu} \right)^2
    + \frac{1}{9} \left( \tensor{T}{^\mu_\mu} \right)\cR
\, ,
\end{align}
where $\cR$ is given by the root-$\TT$ operator
\begin{align}\label{general_rTT}
    \mathcal{R} = \sqrt{ \frac{1}{d} T^{\mu \nu} T_{\mu \nu} - \frac{1}{d^2} \left( \tensor{T}{^\mu_\mu} \right)^2 } \, ,
\end{align}
evaluated at $d=3$.
Other $\TT$-like flows have then been identified. For instance, it has been shown that a $\TT$-like flow generates the Nambu-Goto action from a free scalar field in any space-time dimension \cite{Ferko:2023sps}.
All these results have given hints that $\TT$-like deformations appear to be related to theories of strings and branes.\footnote{Relations between $\TT$ and strings have been noticed in several contexts, see for example the single trace $\TT$ deformation of \cite{Giveon:2017nie,Asrat:2017tzd,Giveon:2017myj,Apolo:2019zai,Chang:2023kkq}.}

The root-$\TT$ operator $\cR$ mentioned above has been proven to have several remarkable properties in various space-time dimensions.
A key property is that $\cR$ can be associated, at least classically, with new classes of non-analytic marginal deformations of $d$-dimensional conformal field theories. In fact, coming back to $d=4$, it was shown \cite{Babaei-Aghbolagh:2022uij,Ferko:2022iru} that the flow equation
\begin{align}\label{root_TTbar_flow}
    \frac{\partial \mathcal{L}^{({\gamma)}}}{\partial \gamma} = \frac{1}{\sqrt d} \sqrt{ T^{\mu \nu} T_{\mu \nu} - \frac{1}{d} \left( \tensor{T}{^\mu_\mu} \right)^2 } \, ,\qquad d=4\,,
\end{align}
with the free Maxwell Lagrangian density as the initial value at $\gamma=0$, gives as the solution the following one-parameter family of Lagrangians
\begin{align}\label{modmax_lagrangian}
    \mathcal{L}_{\text{ModMax}} =
    - \frac{1}{4} \cosh ( \gamma ) F^{\mu \nu} F_{\mu \nu}
    + \frac{1}{4} \sinh ( \gamma )
    \sqrt{
    \left( F^{\mu \nu} F_{\mu \nu} \right)^2 + \left( F^{\mu \nu}{F}^*_{\mu \nu} \right)^2
    } \, ,
\end{align}
which is the Modified Maxwell (or ModMax) theory introduced in \cite{Bandos:2020jsw}. ModMax  has attracted a wealth of attention in the last few years since it is the unique conformally invariant and electromagnetic duality-invariant extension of Maxwell electrodynamics (in the assumption that the Lagrangian does not contain derivatives of the field strength $F_{\mu\nu}$).\footnote{See \cite{Sorokin:2021tge} for an introduction to theories of non-linear electrodynamics, including ModMax.}
By letting the ModMax Lagrangian also flow under the deformation in equation \eqref{TTbar_flow}, one obtains the Lagrangian density of the so-called $\gamma$-Born-Infeld (or ModMax-Born-Infeld) theory \cite{Bandos:2020jsw,Bandos:2020hgy},
\bea\label{gBI}
    \mathcal{L}_{\gamma {\rm BI}} = \frac{1}{\lambda}\Bigg\{
    ~1
    -\sqrt{1
    +\frac{\lambda}{2}\left[
    \cosh(\gamma)F^2
    -\sinh(\gamma)
    \sqrt{(F^2)^2
    +(F F^*)^2}\right]
    -\frac{\lambda^2}{16}(F F^*)^2}
    ~\Bigg\}
    ~.~~~~~~
    \label{gBI-0}
\eea
Interestingly, the above Lagrangian is an example which satisfies two commuting $\TT$-like flows with respect to the parameters\footnote{In the remainder of the paper, we will always use the symbol $\gamma$ for dimensionless couplings associated with marginal deformations. We use the symbol $\lambda$ either for the dimensionful coupling in an irrelevant $\TT$-like deformation, or for a generic flow parameter in an arbitrary deformation.} $\lambda$ and $\gamma$ \cite{Babaei-Aghbolagh:2022uij,Ferko:2022iru}. Analogues of ModMax theories and their flows have been studied in several works, including theories with supersymmetry, sigma-models, and models in dimensions less than four \cite{Babaei-Aghbolagh:2020kjg,Babaei-Aghbolagh:2022uij,Babaei-Aghbolagh:2022itg,Ferko:2022iru,Ferko:2023ruw,Bandos:2021rqy,Kuzenko:2021cvx,Bandos:2020hgy,Garcia:2022wad,Ferko:2022cix,Ferko:2023ozb, Ferko:2023iha,Kuzenko:2023ysh}.

Commenting more about the case of four-dimensional non-linear electrodynamics, it is worth stressing one surprising result obtained in \cite{Ferko:2023wyi}. There, three of us have proven that all (non-higher-derivative) duality invariant theories of non-linear electrodynamics are equivalent to $\TT$-like flows. More precisely, every such theory is a solution of a flow of the following type
\begin{align}\label{TTbar_flow_general}
    \frac{\partial \mathcal{L}^{(\lambda)}}{\partial \lambda} = f\left( T_{\mu \nu},\lambda\right) \, ,
\end{align}
for some Lorentz scalar function of the energy-momentum tensor $f\left( T_{\mu \nu},\lambda\right)$.\footnote{Flows of the form \eqref{TTbar_flow_general} are precisely what we refer to in general as $\TT$-like flows in $d$ space-time dimensions.}
This result, on the one hand, explains why various $\TT$-like flows for non-linear electrodynamics have been discovered in the last five years. On the other hand, the results of \cite{Ferko:2023wyi} provide a new framework to construct duality-invariant theories of non-linear electrodynamics, for which only a few explicit examples of Lagrangians are known in closed form. One of the main aims of our work is to extend the boundaries of this framework and show that the majority of these results can be extended to $d=6$. In fact, to the best of our knowledge, our paper is the first to provide examples of all-orders solutions of $\TT$-like flows in $d>4$.

Before moving to our main six-dimensional playground, it is worth mentioning some further properties of the non-analytic field theories that arise by studying root-$\TT$ flows in $d=2$. Considering the importance played by $2d$ conformal field theories, it is natural to ask what kind of quantum field theories are generated by flows driven by the root-$\TT$ operator (\ref{general_rTT}) and passing through a CFT$_2$ seed theory. This question was asked in \cite{Ferko:2022cix} where it was shown that classical conformal symmetry is preserved along the flow, indicating that root-$\TT$ is a classically marginal operator. Moreover, when evaluated on a CFT$_2$, root-$\TT$ precisely coincides with the square root of the determinant of the energy-momentum tensor in $d=2$, directly justifying its name. Independently, \cite{Conti:2022egv} showed that the dimensional reduction of ModMax leads to a modified theory of scalars in $2d$ that is exactly a root-$\TT$ deformation of a system of free real scalar fields. The same operator was also considered as an ingredient to obtain ultra-relativistic limits of $2d$ CFTs, hence relating to the construction of BMS$_3$ field theories \cite{Rodriguez:2021tcz,Bagchi:2022nvj,Tempo:2022ndz}. It has also been established that $2d$ root-$\TT$ deformations preserve classical integrability for certain $2d$ models \cite{Borsato:2022tmu}. Though it remains an open question whether the $2d$ root-$\TT$ operator can be defined at the quantum level and whether it is exactly marginal, interesting results have been obtained by employing a holographic approach \cite{Ebert:2023tih} that generalise those for $\TT$  \cite{Guica:2019nzm,Ebert:2022ehb}.

Returning to $\TT$-like deformations in $d>2$, one might wonder which other flows could be identified with properties analogous to the ones described above. We have stressed how the marginal $\gamma$-flow \eqref{root_TTbar_flow} and irrelevant $\lambda$-flow \eqref{TTbar_flow} play a central role in describing ModMax-Born-Infeld theory in $d=4$. As was found in \cite{Bandos:2020hgy}, in six space-time dimensions there is a two parameter family of theories of interacting chiral $2$-form fields that share several similarities with their $4d$ counterparts and reduce to the latter upon dimensional reduction.
This is a consequence of the fact that both in $4d$ and $6d$, the (on-shell) functionals describing these theories only depend upon two Lorentz invariant algebraic structures (details will be given later).
It is then natural to ask whether these $6d$ models obey the same type of $\TT$-like flows as their $4d$ cousins in eqs. \eqref{TTbar_flow} and \eqref{root_TTbar_flow}. As we will prove in Section \ref{TTbarf} of this paper, the answer is yes.
Moreover, we will show that all the properties and theorems regarding stress-energy flows of general duality invariant non-linear electrodynamics in $d=4$ found in \cite{Ferko:2023wyi} straightforwardly translate to generic self-interacting chiral 2-form theories in $d=6$. As we will see, the most natural formulation for the construction of the stress-tensor flows of the $6d$ chiral two-form fields is the Hamiltonian one, or its space-time-covariant extension \`a la PST. We will also address problems of $\TT$-like flows in duality-invariant $p$-form theories in $d>6$.

To this end we will revisit and elaborate on Hamiltonian and Lagrangian formulations of $d=2(p+1)$ chiral $p$-form gauge theories. These have a long history of developments and applications on their own.

\subsection{Chiral $p$-form gauge theories}

Chiral $2n$-form gauge fields appear as part of various theories, for instance, chiral scalars in $2d$ theories and string theory, chiral 2-form gauge fields in $6d$ supergravities and on the worldvolume of the M-theory 5-brane, and the chiral 4-form gauge field in type IIB $d=10$ supergravity.

A distinguishing feature of the chiral $2n$-form gauge fields is that their $(2n+1)$-form field strengths satisfy, on the mass shell, certain (in general non-linear) self-duality conditions which are first-order in derivatives of the gauge fields. The problem of obtaining these self-duality conditions with the use of an action principle turned out to be highly non-trivial (see e.g. \cite{Marcus:1982yu}). For the chiral $2n$-forms obeying linear Hodge self-duality conditions in $d=2(2n+1)$ this problem was solved by Henneaux and Teitelboim \cite{Henneaux:1987hz,Henneaux:1988gg} in a Hamiltonian formulation in which the space-time invariance of the theory is not manifest. The Henneaux-Teitelboim formulation is the generalization to higher dimensions of the duality-symmetric Hamiltonian formulation of Maxwell's electrodynamics by Deser and Teitelboim  \cite{Deser:1976iy} and the Floreanini-Jackiw description of $2d$ chiral scalars \cite{Floreanini:1987as}. In the context of a duality-symmetric formulation of string worldsheet dynamics, chiral scalars were exploited in \cite{Tseytlin:1990nb,Tseytlin:1990va}. See also \cite{Schwarz:1993vs} for the Hamiltonian-like description of fields related by Hodge duality in $d=4$ similar to \cite{Henneaux:1987hz,Henneaux:1988gg}, and \cite{Perry:1996mk} for a generalization of the Henneaux-Teitelboim-like formulation to the description of the non-linear chiral two-form on the worldvolume of the M5-brane.

Since it is desirable to have a manifestly space-time covariant formulation of relativistic field theories, in particular for its consistent coupling to gravity and other relativistic fields, several approaches have been proposed for the construction of covariant duality-symmetric actions for gauge fields in various space-time dimensions with the use of auxiliary fields \cite{Siegel:1983es,Kavalov:1986ki,McClain:1990sx,Pasti:1995tn,Pasti:1996vs,Bengtsson:1996fm,Berkovits:1996tn,Belov:2006jd,Ivanov:2014nya,Sen:2015nph,Mkrtchyan:2019opf,Townsend:2019koy}. Comparisons and relations between some of these approaches were discussed e.g. in \cite{Bandos:2020hgy,Arvanitakis:2022bnr,Evnin:2022kqn,Bansal:2023pnr}. Among these approaches, the PST formulation \cite{Pasti:1995tn,Pasti:1996vs,Pasti:1997gx} is the direct and most economical covariantization of the constructions of \cite{Deser:1976iy,Floreanini:1987as,Henneaux:1988,Henneaux:1988gg,Schwarz:1993vs,Perry:1996mk} with the use of a single Stueckelberg-like auxiliary scalar field. Fully-fledged actions for the M-theory 5-brane \cite{Bandos:1997ui,Aganagic:1997zq} and the type IIA NS5-brane \cite{Bandos:2000az}, $d=10$ type IIB supergravity \cite{Dall'Agata:1997ju,Dall'Agata:1998va}, $d=6$ supergravities \cite{Dall'Agata:1997db,Riccioni:1998pj,DePol:2000re}, and other theories involving chiral $p$-forms were first constructed with the use of this formulation.
In Section \ref{PSTf}, we will review several aspects of the PST formulation of interacting theories of chiral two-form gauge theories in $d=6$, and their relation to their Hamiltonian formulation, that are necessary for the rest of our analyses.

We will see that the PST formulation is also the most economical one for the study of $\TT$-like flows in $6d$ chiral 2-form gauge theories and directly produces the corresponding Hamiltonian flows in these theories.

To lay the background for this study, we elaborate on the relationship between different Lagrangian formulations of duality-invariant $p$-form theories and corresponding $\TT$-like flows in various dimensions.
To this end we propose a novel formulation which (i) is a generalization of a four-dimensional formulation by Ivanov, Nurmagambetov and Zupnik (INZ) \cite{Ivanov:2014nya} and (ii) turns into the PST formulation upon integrating out an auxiliary self-dual $3$-form field. We elucidate space-time covariant properties of the PST formulation by clarifying and making use of its relation to the INZ-type formulation and to a so-called ``clone" construction \cite{Evnin:2022kqn} of \cite{Mkrtchyan:2019opf,Bansal:2021bis,Avetisyan:2021heg,Avetisyan:2022zza}.

Regarding \cite{Ivanov:2014nya}, we would like to note that in that paper, for $d=4$ duality-symmetric non-linear electrodynamics, there was proposed a way of adding, to the free-field PST Lagrangian, duality-invariant non-linear interaction terms that automatically respect local PST symmetries. It is based on the earlier approach by Ivanov and Zupnik
 \cite{Ivanov:2001ec, Ivanov:2002ab, Ivanov:2003uj}
to formulate $\sU(1) $ duality-invariant, $4d$ non-linear electrodynamics with the use of an auxiliary antisymmetric $2$-form tensor field. An important feature of this formulation is that the most general duality-invariant interaction term in the action is a completely arbitrary function of a single quartic invariant of components of the auxiliary antisymmetric tensor field.\footnote{The Ivanov-Zupnik approach is a reformulation of the Gaillard-Zumino-Gibbons-Rasheed  (GZGR) formalism
\cite{Gaillard:1981rj, Gibbons:1995cv, Gibbons:1995ap, Gaillard:1997zr, Gaillard:1997rt}. It has been extended to $\sU(1)$ duality-invariant theories of gauge $(2n-1)$-forms in $d=4n$ dimensions in \cite{Kuzenko:2019nlm}. In the $d=4$ case, it has been extended to $\cN=1$ and $\cN=2$
supersymmetric non-linear electrodynamics \cite{Kuzenko:2013gr, Ivanov:2013ppa}, to higher-spin conformal gauge fields on conformally flat backgrounds and some of their $\cN$-extended  superconformal cousins \cite{Kuzenko:2021qcx, Kuzenko:2023ebe}.
}
This is in contrast to the conventional formulation \`a la Gaillard and Zumino \cite{Gaillard:1981rj} in which the Lagrangian must satisfy a duality-invariance condition \cite{BialynickiBirula:1984tx,Gibbons:1995cv,Gaillard:1997zr}.

A more recent approach, which is conceptually similar and on-shell equivalent to that of \cite{Ivanov:2003uj,Ivanov:2014nya}, was developed  for $d=4$ non-linear duality-symmetric electrodynamics in \cite{Avetisyan:2021heg} and for $p$-form gauge fields in higher dimensions in \cite{Avetisyan:2022zza}.

In this paper we will generalize the INZ formulation \cite{Ivanov:2014nya} to  six- (Section \ref{sec:INZ}) and higher-dimensional (Section \ref{hdINZ}) chiral form fields and elaborate on its relation to the PST formulation and that of \cite{Avetisyan:2022zza} described in Section \ref{clone}. We will derive and compare the energy-momentum tensors of $6d$ chiral two-forms in these formulations and use them for the analysis of $\TT$-like flows in Section \ref{TTbarf}. We will also show that the Lagrangian formulations  of duality-symmetric theories considered in \cite{Ivanov:2003uj,Ivanov:2014nya,Avetisyan:2021heg,Avetisyan:2022zza} do not include the Lagrangian description of the $4d$ conformal Bialynicki-Birula electrodynamics \cite{Bialynicki-Birula:1992rcm} and its six-dimensional counterpart \cite{Gibbons:2000ck,Townsend:2019ils,Bandos:2020hgy}.

Yet another approach to the Lagrangian formulation of chiral $p$-form fields was proposed by Sen \cite{Sen:2015nph,Sen:2019qit}. It was used for the description and study of theories with chiral $p$-forms e.g. in \cite{Lambert:2019diy,Chakrabarti:2020dhv,Vanichchapongjaroen:2020wza,Andriolo:2021gen,Andrianopoli:2022bzr,Phonchantuek:2023iao,Hull:2023dgp,Chakrabarti:2022lnn,Chakrabarti:2023czz}. A peculiar feature of this formulation is the presence of an additional dynamical ghost-like self-dual field which however decouples from the physical chiral form field and from other physical fields in the theory including gravity. The covariant generalization of Sen's approach with the use of a second metric was recently proposed in \cite{Hull:2023dgp}. Because of this peculiarity, which affects the structure of the energy-momentum tensor, we will not study $\TT$-like flows in this formulation, but only mention that in two-dimensional theories such an analysis has been carried out in \cite{Chakrabarti:2020dhv,Chakrabarti:2022lnn,Chakrabarti:2023czz}.

We have already discussed the content of Sections  \ref{PSTf}--\ref{hdINZ}. The rest of the paper is organised as follows. In Section \ref{Conclusion} we give concluding comments. We also provide two Appendices that contain results related to the analysis of the main body of our paper. In particular, in Appendix \ref{AppendixA} we discuss $\TT$-like flows for self-interacting gauge $(2n-1)$-forms in $d=4n$ dimensions by using three approaches: the Gaillard-Zumino-Gibbons-Rasheed-type formalism (Section \ref{AppendixA1}); the Ivanov-Zupnik-type approach (Section \ref{AppendixA2}); and a PST-type Lagrangian description (Section \ref{AppendixA3}).
Appendix \ref{AppendixB} contains a discussion of
deformations of $\sU(1)$ duality-invariant supersymmetric theories in $d=4$ based on the general formalism of \cite{Kuzenko:2000tg, Kuzenko:2000uh,Kuzenko:2002vk}
and extension of $\TT$-like flow results and ideas of \cite{Baggio:2018rpv,Chang:2018dge,Jiang:2019hux,Chang:2019kiu,Ferko:2019oyv,Ferko:2022iru,Ferko:2023ruw,Ferko:2023wyi,Jiang:2019trm}.

{\bf Notation and conventions.} We use lower case Greek letters as $d$-dimensional space-time indices $\mu,\nu,\ldots=(0,1,\ldots, d-1)$, and lower case Latin letters as $(d-1)$-dimensional spatial indices  $i,j,\ldots=(1,\ldots,d-1)$. The signature of the Minkowski metric is chosen to be ``mostly plus'' i.e. $(- , + , + , \cdots , +)$. The (anti)-symmetrization of indices is performed with weight one, so that
\begin{align}\begin{split}
    X_{[\mu_1\mu_2\ldots\mu_p]} &= \frac 1{p!}(X_{\mu_1\mu_2\ldots \mu_p}-X_{\mu_2\mu_1\ldots \mu_p}+\cdots) \, , \\
    X_{(\mu_1\mu_2\ldots\mu_p)} &= \frac 1{p!}(X_{\mu_1\mu_2\ldots \mu_p}+X_{\mu_2\mu_1\ldots \mu_p}+\cdots) \, .
\end{split}\end{align}

\section{PST formulation} \label{PSTf}
As has been mentioned in the Introduction, the PST approach to the construction of space-time covariant Lagrangians for chiral $p$-form gauge fields \cite{Pasti:1996vs,Pasti:1997gx} is the most economical prescription for the covariantization of the Hamiltonian formulation of these theories \cite{Henneaux:1987hz,Henneaux:1988gg} and/or of a similar formulation by Perry and Schwarz \cite{Perry:1996mk} (see \cite{Bandos:2020hgy} for the details on the relation between these formulations).

Let us overview the main features of the PST formulation for a chiral two-form gauge field $A_{\mu\nu}(x)$ in six-dimensional space-time.

\subsection{Action, symmetries and equations of motion} \label{PSTac}
On the mass shell the three-form field-strength $F_{\mu\nu\rho}=3\partial_{[\mu}A_{\nu\rho]}$ of $A_{\mu\nu}$ obeys a non-linear self-duality condition which is obtained as the general solution of the equation of motion of $A_{\mu\nu}$ by varying the following action defined on a $6d$ manifold with metric $g_{\mu\nu}$
\be\label{PST}
\mathcal S_{\text{PST}} = \int d^6x \mathcal L_{\text{PST}} =\int d^6x\sqrt{-g}\left(\frac 14 E^{\mu\nu}B_{\mu\nu}-\mathcal H(s,p)\right)\,,
\ee
where
\bea\label{Bp}
E^{\mu\nu}=F^{\mu\nu\rho}v_\rho\,,\qquad
B^{\mu\nu}=\frac 1{2\sqrt{-g}}\varepsilon^{\mu\nu\rho\lambda\sigma\kappa}\partial_{\lambda}A_{\sigma\kappa}v_\rho
=F^{*\mu\nu\rho}v_\rho,
\eea
are analogues of the electric and magnetic field of $4d$ electrodynamics,
and the Levi-Civita symbol $\varepsilon^{\mu\nu\rho\lambda\sigma\kappa}$
is defined such that
$$
\varepsilon^{012345}=-\varepsilon_{012345}=-1 \, .
$$
Notice that by construction
$$
E_{\mu\nu}v^\nu=B_{\mu\nu}v^\nu=0\,.
$$

The vector $v_\mu$ is the normalized derivative $\partial_\mu a$ of an auxiliary scalar field $a(x)$, such that $v_\mu$ is a unit time-like vector which in the mostly plus metric convention is
\bea\label{cool}
 v_\mu=\frac{\partial_\mu a}{\sqrt{-g^{\nu\lambda}\partial_\nu a\partial_\lambda a}}, \qquad v_{\mu}v^\mu=-1\,.
\eea
Finally, $\mathcal H(s,p)$ is a function of two independent Lorentz invariants which one may construct with the components of the anti-symmetric tensor $B_{\mu\nu}$,
\be\label{sp}
 s=\frac 14 B_{\mu\nu}B^{\mu\nu},\qquad  p=\sqrt{p_\mu p^\mu}, \qquad p^{\mu} =-\frac 1{8\sqrt{-g}}\varepsilon^{\mu\nu\rho\lambda\sigma\kappa}B_{\rho\lambda}B_{\sigma\kappa}v_{\nu}\,.
\ee
As we will explain in more detail in Section \ref{BLMreview}, the presence of the nowhere vanishing time-like vector $v^\mu$ defined by \eqref{cool}  implies non-trivial conditions on the global causal structure of space-time in which the theory is constructed. See also \cite{Bandos:2014bva} for the study of the PST formulation in topologically non-trivial backgrounds.

\subsubsection*{\it Examples of chiral 2-form theories}

\begin{description}
\item{i)}
The free chiral two-form theory, whose field strength is Hodge self-dual on the mass-shell
\be\label{lsd}
F_{\mu\nu\rho}=F^{*}_{\mu\nu\rho},
\ee
is described by the Langrangian density \eqref{PST} with
\be\label{freePST}
\mathcal H=s=\frac 14 B_{\mu\nu}B^{\mu\nu}\,.
\ee
\item{ii)}
The chiral form field on the worldvolume of the M5-brane has an interaction function $\mathcal H(s,p)$ of a Born-Infeld-like form \cite{Perry:1996mk,Pasti:1997gx},
\be\label{BI}
\mathcal H_{\text{BI}}= \sqrt{T^2 + 2Ts + p^2} -T\,,
\ee
where $T$ has the dimension of a $6d$ energy density and is associated with the ${\rm M}5$-brane tension.
\item{iii)}
The tensionless limit ($T\to 0$) of this theory is the $6d$ counterpart \cite{Gibbons:2000ck,Townsend:2019ils} of the $4d$ conformal electrodynamics of Bialynicki-Birula \cite{Bialynicki-Birula:1992rcm}, which has
\be\label{BB}
\mathcal H_{\text{BB}} = p\,.
\ee
The conformal invariance of the theory requires that
\be\label{CC}
s\mathcal H_s+p\mathcal H_p=\mathcal H,
\ee
which is satisfied by \eqref{freePST} and \eqref{BB}.

\item{iv)} The only other example of a conformal $6d$ chiral two-form field theory is
the $6d$ counterpart of the $4d$ ModMax theory \cite{Bandos:2020jsw}. It has \cite{Bandos:2020hgy}  %
\be\label{MM}
\mathcal H_{\text{MM}} = \cosh(\gamma)\,s-\sinh(\gamma)\,\sqrt{s^2-p^2}\,,
\ee
where $\gamma$ is a dimensionless parameter.  Like the $d=4$ case \cite{Bandos:2020jsw}, this
one-parameter family of theories is the unique non-linear conformal extension of the free chiral 2-form theory \eqref{freePST}, to which it reduces at $\gamma=0$.\footnote{This so far has only been proven for any action that is a function of $F_{\mu\nu\rho}$ through $s$ and $p$, but no higher-derivatives of the $3$-form field strength.} Note that $s^2-p^2$ is always non-negative, as can be easily shown by setting $v_\mu=\delta^0_\mu$ and choosing an $SO(1,5)$ basis in which $B_{\mu\nu}$ has only two independent non-zero components, e.g. $B_{12}$ and $B_{34}$. Then $s=\frac 12 (B_{12}^2+B^2_{34})$, $p=|B_{12}B_{34}|$ and $s^2-p^2=\frac 14(B_{12}^2-B_{34}^2)^2$.

The following comment is in order. In $d=4$ the ModMax Hamiltonian density has a form similar to \eqref{MM}, but with $s=\frac 12(\mathbf D^2+\mathbf B^2)$ and $p=|\mathbf D\times \mathbf  B|$, where $\mathbf D$ is the electric displacement vector and $\mathbf B$ is the magnetic field. As was explained in \cite{Bandos:2020jsw}, for the theory described by the Lagrangian density \eqref{modmax_lagrangian} to be causal, the parameter $\gamma$ must be non-negative. The Lagrangian density \eqref{modmax_lagrangian} is obtained by the involutive Legendre transform of the Hamiltonian density if $\mathcal H$ is a convex function of $\mathbf D$. This requires $\mathcal H_s\geq 0$, which implies  $s\geq p\cosh\gamma$ and hence the lower bound  on the Hamiltonian density $\mathcal H\geq p$. Since the $4d$ ModMax Hamiltonian density is obtained by a straightforward dimensional reduction of its $6d$ counterpart, in what follows we will assume that the $6d$ ModMax Hamiltonian density is subject to the same conditions, i.e. $\gamma \geq 0$ and $\mathcal H_s\geq 0$. For the later $T\overbar{T}$ analysis of Section \ref{TTbarf} let us also point out the relation
\be\label{Hg=-Hs}
\partial_\gamma \mathcal H=-\sqrt{s^2-p^2}\,\mathcal H_s \, ,
\ee
and hence $\partial_\gamma \mathcal H\leq 0.$

\item{v)}
A chiral form theory which combines together the $6d$ counterparts of BI and ModMax \cite{Bandos:2020hgy} has
\be\label{MMBI}
\mathcal H_{\text{MMBI}} = \sqrt{T^2 + 2T\mathcal H_{MM} + p^2} -T\,.
\ee
\end{description}
In the general case, in addition to the conventional gauge invariance $\delta A_{\mu\nu}=2\partial_{[\mu}\lambda_{\nu]}$, the local symmetries of the PST action \eqref{PST} are
\be\label{PST1}
\delta A_{\mu\nu}=2v_{[\mu}\Phi_{\nu]}\,, \qquad \delta a=0,
\ee
and
\be\label{PST2}
\delta a =\varphi(x), \qquad \delta {A_{\mu\nu}}=-\frac{\varphi}{\sqrt{-\partial_\mu a\partial^\mu a}}(E_{\mu\nu}- H_{\mu\nu}) \, ,
\ee
where $\Phi_\nu(x)$ and $\varphi(x)$ are local symmetry parameters and
\be\label{Hmn}
H_{\mu\nu}=2\frac{\partial{\mathcal H}}{\partial B^{\mu\nu}}=({\cal H}_s + 2s p^{-1} {\cal H}_p) B_{\mu\nu} + p^{-1}{\cal H}_p (B^3)_{\mu\nu}\, ,
\ee
with $\mathcal H_s:=\partial_s\mathcal H$, $\mathcal H_p:=\partial_p\mathcal H$,
and
$(B^3)_{\mu\nu}:=B_{\mu\rho}B^{\rho\tau}B_{\tau\nu}$.

The action \eqref{PST} is invariant under the local transformation \eqref{PST2} provided the following relation holds
\be\label{HH=BB}
\varepsilon^{\mu\nu\rho\sigma\kappa\delta}v_\nu H_{\rho\sigma}H_{\kappa\delta}
=\varepsilon^{\mu\nu\rho\sigma\kappa\delta}v_\nu B_{\rho\sigma}B_{\kappa\delta} \, .
\ee
A similar condition for duality-invariant $p$-form theories in higher dimensions was obtained in \cite{Buratti:2019cbm,Buratti:2019guq}.

Substituting the relation \eqref{Hmn} into \eqref{HH=BB} one finds that the latter holds if and only if
\be\label{Li}
\mathcal{H}_s^2 + \frac{2s}{p} \mathcal{H}_s \mathcal{H}_p + \mathcal{H}_p^2=1\,.
\ee
The local shifts \eqref{PST2} of the scalar $a(x)$ can be used to impose
the gauge $\partial_\mu a = \tensor{\delta}{^0_\mu}$. Then, if for simplicity we consider the flat space-time case $g_{\mu\nu}=\eta_{\mu\nu}$, $B^{\mu\nu}$ defined in \eqref{Bp} reduces to the antisymmetric tensor $-B_{ij}$ with spatial indices ($i,j=1,\ldots,5$). In this case the PST Lagrangian density reduces to one which is first-order in time derivatives, and which is equivalent to the Hamiltonian formulation of a chiral two-form theory with the Hamiltonian density $\mathcal H(s,p)$,
\be\label{LH}
\mathcal L_{\text{H}} =\frac 14 B_{ij}\partial_0 A^{ij}- \mathcal H(s,p)\,.
\ee
Therefore, the Hamiltonian formulation of the chiral $p$-form fields is a gauge-fixed counterpart of the PST formulation, and all the results obtained in the PST formulation straightforwardly hold in the Hamiltonian formulation and vice versa. In the Hamiltonian formulation the condition \eqref{Li} ensures the relativistic invariance of the theory (see e.g. \cite{Bandos:2020hgy} for a review).

In a generic non-linear chiral 2-form theory
the PST equations of motion of $A_{\mu\nu}$ produce, upon fixing the symmetry \eqref{PST1} (see \cite{Pasti:1996vs,Pasti:1997gx} for details), the non-linear self-duality condition
\be\label{PSTsd}
E_{\mu\nu}-H_{\mu\nu}(B)=0\,.
\ee
In the gauge $v_{\mu} = \tensor{\delta}{^{0}_\mu}$ it reduces to the self-duality condition of the Hamiltonian formulation,
\be\label{Hsd}
E_{ij}=\dot{A_{ij}}-\partial_{i}A_{0j}+\partial_{j}A_{0i}=H_{ij}\,.
\ee
It is important to note that the equation of motion of the auxiliary field $a(x)$ is not independent, but is a square of the left-hand-side of \eqref{PSTsd}. Namely,
\be\label{aeom}
\frac{\delta\mathcal L}{\delta a(x)}\sim\varepsilon^{\mu\nu\rho\kappa\lambda\sigma}(E_{\mu\nu}-H_{\mu\nu})\partial_\rho \Big((E_{\kappa\lambda}-H_{\kappa\lambda})v_{\sigma}\Big)=0\,.
\ee
So the $a(x)$-field equation is identically satisfied when \eqref{PSTsd} holds. It can actually be seen that this equation is a consequence of the $A_{\mu\nu}$-field equation of motion even if the symmetry \eqref{PST1} is not fixed \cite{Buratti:2019guq}.
  This reflects the fact that $a(x)$ is a non-dynamical pure gauge field.

Note that, in view of the identity
\be\label{HBPST}
H_{\mu\nu}B^{\mu\nu}=4s\mathcal H_{s}+4p\mathcal H_p\,,
\ee
 on the mass-shell \eqref{PSTsd} the PST Lagrangian density \eqref{PST} is
\be\label{PSTonshell}
\mathcal L_{\text{PST}}^{\text{on-shell}} = (s\mathcal H_s+p\mathcal H_p)-\mathcal H(s,p).
\ee
The quantity (\ref{PSTonshell}) vanishes in the Maxwell, ModMax and  Bialynicki-Birula cases due to the conformal invariance of these theories.

\begin{corollary}\label{C1}
    An important consequence of the property that the variation \eqref{aeom} of the PST Lagrangian density with respect to the auxiliary field $a(x)$ vanishes when the non-linear self-duality condition is satisfied is that the on-shell Lagrangian density \eqref{PSTonshell} is manifestly $6d$ relativistic invariant and {\it independent} of the auxiliary vector field $v^\mu$. Further evidence for this Corollary will be given by the comparison of the on-shell PST Lagrangian density with those in other formulations of the chiral-form theories.
\end{corollary}

\subsection{PST energy-momentum tensor}\label{PSTEMon}
A generic chiral 2-form energy-momentum tensor obtained by varying \eqref{PST} with respect to the metric has the following form \cite{Bandos:1997gm}
\be\label{EMPST}
T_{\mu\nu}=-\frac 2{\sqrt{-g}}\frac{\delta L}{\delta g^{\mu\nu}}=-g_{\mu\nu}\left(\mathcal H-\frac 12\,H_{\rho\lambda}B^{\rho\lambda}\right )+\frac 12 v_\mu v_\nu (H_{\rho\lambda}B^{\rho\lambda} )-H_{\mu}{}^{\rho}B_{\nu\rho}- 2 v_{(\mu}p_{\nu)}.
\ee
Note that the last term in \eqref{EMPST} is obtained from the metric variation of the first term in \eqref{PST} with the use of the identity
\cite{Bandos:1997gm}
\be\label{id}
F^{\mu\nu\rho}=-3v^{[\mu}E^{\nu\rho]}-\frac 1{2\sqrt{-g}}\varepsilon^{\mu\nu\rho\lambda\sigma\delta}v_\lambda B_{\sigma\delta}\,.
\ee
Namely,
$$
-\frac 12\frac {\delta(\sqrt{-g} E_{\rho\lambda}B^{\rho\lambda})}{\sqrt{-g}\,\delta g^{\mu\nu}}=-\frac 12 v_{(\mu}F_{\nu)\lambda\rho}B^{\lambda\rho}-\frac 12v_{\mu}v_{\nu}\,E_{\lambda\rho}B^{\lambda\rho}=- 2 v_{(\mu}p_{\nu)}\,.
$$

 We will now show that on the mass shell, i.e. when the non-linear self-duality condition \eqref{PSTsd} is satisfied, the energy-momentum tensor does not depend on the auxiliary field $v_\mu$. Indeed, with \eqref{PSTsd} taken into account, eq. \eqref{EMPST} takes the form
\be\label{EMPSTOS1}
T_{\mu\nu}=-g_{\mu\nu}\mathcal H+\frac 12(g_{\mu\nu}+v_\mu v_\nu)\,E_{\rho\lambda}B^{\rho\lambda}-E_{\mu}{}^\rho B_{\nu\rho}- 2 v_{(\mu}p_{\nu)} \, .
\ee
Note that $E_{\mu}{}^\rho B_{\nu\rho}$ is symmetric on the mass shell because $H_{\mu}{}^\rho B_{\nu\rho}$ is symmetric.

Now, using the identity  \eqref{id}
and its dual, one finds that
\bea\label{FF*}
\frac 14F_{\mu\rho\lambda}F_\nu^{*\,\,\rho\lambda}&=&\frac 14(g_{\mu\nu}+2v_\mu v_\nu)\,E_{\rho\lambda}B^{\rho\lambda}-E_{(\mu}{}^\rho B_{\nu)\rho}\nonumber\\
&&+\frac 18v_{\mu}\varepsilon_{\nu\rho\lambda\delta\sigma\kappa}E^{\rho\lambda}E^{\delta\sigma}v^\kappa+\frac 18v_{\nu}\varepsilon_{\mu\rho\lambda\delta\sigma\kappa}B^{\rho\lambda}B^{\delta\sigma}v^\kappa\,.
\eea
So, on the mass shell, i.e. taking into account the dynamical self-duality equation \eqref{PSTsd}, the ``PST-invariance" condition \eqref{HH=BB}  and eq. \eqref{PSTonshell}, one gets\footnote{For the M5-brane the on-shell energy momentum tensor in this form was given in \cite{Sorokin:1998xx}.}
\be\label{EMPSTOS}
T_{\mu\nu}|^{\text{on-shell}} = g_{\mu\nu}\left(\frac 14\,H_{\lambda\rho}B^{\lambda\rho}-\mathcal H\right) +\frac 14F_{\rho\lambda(\mu}F_{\nu)}^{*\,\,\rho\lambda}=g_{\mu\nu}(\sqrt{-g})^{-1}\mathcal L|^{\text{on-shell}} +\frac 14 F_{\rho\lambda(\mu}F_{\nu)}^{*\,\,\rho\lambda}\,.
\ee
We see that the on-shell energy momentum tensor is $6d$ space-time covariant and $v^\mu$ independent, since we have already proved  that the Lagrangian density is covariant and independent of $v^\mu$ on the mass shell (see Corollary \ref{C1} below eq. \eqref{PSTonshell}).

Three explicit (trivial) examples of this general statement about $v^\mu$-independence and $6d$ space-time covariance of the chiral 2-form on-shell Lagrangian density are the conformally invariant cases of (1) the free theory, (2) the ModMax chiral $2$-form theory, and (3) the $6d$ counterpart of the Bialynicki-Birula theory. In all three of these cases, the on-shell Lagrangian density vanishes and is thus obviously covariant and independent of $v^\mu$.

With the help of the superembedding formulation of the M5-brane \cite{Howe:1996yn,Howe:1997fb}, the independence of the covariant on-shell Lagrangian from $v^\mu$ was also explicitly shown in \cite{Ko:2013dka} for different versions of the M5-brane action. Namely, (modulo normalization factors), the on-shell Lagrangian density for the chiral two-form in this case is
\be\label{OSBI}
\mathcal L^{\text{on-shell}}_{\text{BI}} = \sqrt{-g}\,T\left(1-\left(1-\frac 2{3T^2}\,I_h\right)^{-1}\right) \, ,
\ee
where
$$
I_h=k_{\mu\nu}k^{\mu\nu}\,,\qquad k_{\mu\nu}=h_{\mu\lambda\rho}h_\nu{}^{\lambda\rho}\,,\qquad h_{\mu\nu\rho}=h^*_{\mu\nu\rho}\,,
$$
with $h_3$ being a self-dual three form to which the physical 3-form field strength is related as follows
\bea\label{F=h}
F_{\mu\nu\rho}&=&\left(1-\frac 2{3T^2}I_h\right)^{-1}\left(\delta_\mu^{\kappa}+\frac 2{T}k_{\mu}{}^{\kappa}\right)h_{\kappa\nu\rho}\,,\nonumber\\
F^*_{\mu\nu\rho}&=&\left(1-\frac 2{3T^2}I_h\right)^{-1}\left(\delta_\mu^{\kappa}-\frac 2{T}k_{\mu}{}^{\kappa}\right)h_{\kappa\nu\rho}\,.
\eea
See \cite{Howe:1997fb,Howe:1997vn,Howe:1997vn,Gibbons:2000ck} for more details on the relations between $F_{\mu\nu\rho}$ and $h_{\mu\nu\rho}$, and on the form of the non-linear self-duality conditions on $F_{\mu\nu\rho}$ derived from the superembedding formulation. As we will show, the auxiliary self-dual three-form $h_3$ is related in a non-linear way to an auxiliary self-dual three form $\Lambda_3$ of the INZ-type formulation of the BI-like theory which we will consider in the next Section.

\section{INZ-type formulation of chiral two-form theories}\label{sec:INZ}

In the previous Section we have seen that the PST formulation requires the non-linear function $\mathcal H(s,p)$ in the chiral two-form field action \eqref{PST} to satisfy the equation \eqref{Li}, which ensures the relativistic invariance of the theory, and, in particular, the (non-manifest) relativistic invariance of the non-linear self-duality condition \eqref{PSTsd} independently of the choice of the auxiliary vector $v^\mu$. The argument regarding the $v^\mu$-independence of the on-shell PST action given in Corollary \ref{C1} implies that it should be always possible to rewrite the non-linear self-duality condition \eqref{PSTsd} in a manifestly covariant form which does not contain $v^\mu$ (see also \cite{Pasti:2012wv} where the same argument was used for the case of duality-symmetric non-linear electrodynamics in $d=4$).

As we will now show, the conversion of the non-linear PST self-duality condition \eqref{PSTsd} into a covariant one containing only the physical gauge field can be achieved with the use of the approach of Ivanov, Nurmagambetov and Zupnik \cite{Ivanov:2014nya} originally proposed in $d=4$, which we will generalize to higher-dimensional $p$-form theories. An important feature of this formulation is that the most general duality-invariant interaction term in the action is a completely arbitrary function of a single quartic invariant of components of an extra auxiliary tensor field. This is in contrast to the Hamiltonian and PST formulations in which the non-linear Lagrangian must satisfy a relativistic-invariance condition similar to \eqref{Li} (see \cite{Bandos:2020hgy} for a review).

A generalization of the INZ formulation \cite{Ivanov:2014nya} to self-interacting chiral $2n$-form fields in $d=4n+2>6$ dimensions will be  described in Section \ref{hdINZ}. In the $d=6$ case the construction of this novel formulation goes as follows.

\subsection{Action, symmetries and equations of motion}
Take the $6d$ PST Lagrangian density for the free chiral 2-form field, eq. \eqref{PST} with $\mathcal H = \frac{1}{4} B_{\mu \nu} B^{\mu \nu}$ as in \eqref{freePST}, and add to this Lagrangian density the following term containing a self-dual auxiliary three-form field $\Lambda_{\mu\nu\rho}=\Lambda^*_{\mu\nu\rho}$:
\bea\label{INZ}
\mathcal L^{\text{free}}_{\text{INZ}} &=&\mathcal L^{\text{free}}_{\text{PST}} + \frac {\sqrt{-g}}2 (B_{\mu\nu}+\Lambda_{\mu\nu})(B^{\mu\nu}+\Lambda^{\mu\nu})\nonumber\\
&=&\frac {\sqrt{-g}}4(E_{\mu\nu}-B_{\mu\nu})B^{\mu\nu}+\frac {\sqrt{-g}}2 (B_{\mu\nu}+\Lambda_{\mu\nu})^2,
\eea
where
\be\label{Lmn}
\Lambda_{\mu\nu}=\Lambda_{\mu\nu\rho}v^\rho=\Lambda^*_{\mu\nu\rho}v^\rho\,.
\ee
Modulo a total derivative, the INZ-type Lagrangian density is invariant under the local symmetry \eqref{PST1} and a counterpart of the second PST symmetry \eqref{PST2} which is modified as follows
\be\label{PST2INZ}
\delta a =\varphi(x), \qquad \delta {A_{\mu\nu}}=-\frac{\varphi}{\sqrt{-\partial_\mu a\partial^\mu a}}(F_{\mu\nu\rho}v^\rho+ B_{\mu\nu}+2\Lambda_{\mu\nu})\,, \qquad  \delta \Lambda_{\mu\nu\rho}=0.
\ee
Note that $\Lambda_{\mu\nu\rho}$ remains intact under the action of \eqref{PST1} and \eqref{PST2INZ}. This fact -- that, off the mass shell, the auxiliary tensor field is invariant under the PST local symmetries -- was used in \cite{Ivanov:2014nya} to construct the most general manifestly duality-invariant non-linear interaction term in $4d$ electrodynamics. Such a general interaction can be parameterized in terms of an arbitrary function of a single independent fourth-order Lorentz-invariant and duality-invariant quantity built from a $4d$ anti-symmetric tensor $\hat\Lambda_{\mu\nu}$. A similar picture holds in $6d$ in which, as was shown in \cite{Avetisyan:2022zza}, there is a single independent invariant which one can construct from a self-dual three-form tensor $\Lambda_{\mu\nu\rho}$. It has the following form
\be\label{I}
I=M_{\mu}{}^{\nu}M_{\nu}{}^{\mu}\,,
\ee
where
\be\label{M}
M_{\mu}{}^{\nu}=\Lambda_{\mu\rho\lambda}\Lambda^{\nu\rho\lambda}.
\ee
Because of the self-duality of $\Lambda_{\mu\nu\rho}$, the following useful identities hold (see e.g. \cite{Howe:1997vn}, \cite{Gibbons:2000ck} or \cite{Avetisyan:2022zza})
\be\label{3L}
\Lambda^{\mu\nu\sigma} \Lambda_{\sigma\lambda\kappa}=M_{[\lambda}{}^{[\mu}\delta^{\nu]}_{\kappa]}\,,
\ee
\be\label{MI1}
M_{\mu}{}^{\nu}M_{\nu}{}^{\rho}=\frac 16 \delta^\mu_\nu\,I\,.
\ee
As in the $4d$ analysis, to construct a generic non-linear chiral 2-form Lagrangian density in the INZ formulation, we add to
\eqref{INZ} an arbitrary function $\mathcal V(I)$ of $I$:
\be\label{INZNL}
\mathcal L_{\text{INZ}} = \mathcal L^{\text{free}}_{\text{PST}} + \sqrt{-g}\left(\frac 12 (B_{\mu\nu}+\Lambda_{\mu\nu})^2-\mathcal V(I)\right)\,.
\ee
The variation of the Lagrangian density \eqref{INZNL} with respect to $\Lambda_{\mu\nu\rho}$ gives its (algebraic) equation of motion, which in view of the self-duality of $\Lambda_{\mu\nu\rho}$ has the following form
\be\label{LambdaeomNL}
3v^{[\rho}\Lambda^{\mu\nu]}-\frac 12 \varepsilon^{\rho\mu\nu\lambda\sigma\kappa}v_\lambda \Lambda_{\sigma\kappa}-\mathcal V^{\mu\nu\rho}=-3v^{[\rho}B^{\mu\nu]}+\frac 12 \varepsilon^{\rho\mu\nu\lambda\sigma\kappa}v_\lambda B_{\sigma\kappa}\,\,,
\ee
where
\be\label{dV}
\mathcal V^{\mu\nu\rho}=6\frac{\partial \mathcal V}{\partial \Lambda_{\mu\nu\rho}}=24\,\mathcal V_I\Lambda^{\rho\mu}{}_{\sigma}M^{\sigma\nu}\mathcal\,, \qquad \mathcal V_I=d\mathcal V/dI \,.
\ee
Note that the term $\Lambda^{\rho\mu}{}_{\sigma}M^{\nu\sigma}$ in $\mathcal V^{\mu\nu\rho}$ is automatically anti-symmetric and anti-self-dual with respect to the indices $\mu,\nu,\rho$\, because of the identity \eqref{3L}.

Contracting \eqref{LambdaeomNL} with $v_\rho$ we find that
\be\label{L=B}
\Lambda_{\mu\nu}-24\mathcal V_I\Lambda_{\mu\sigma}M^{\sigma}{}_\nu=-B_{\mu\nu}\,.
\ee
Note that in the free theory case $\mathcal V=0$, the above equation reduces to $\Lambda_{\mu\nu}=-B_{\mu\nu}$, the second term in \eqref{INZ} vanishes and the free INZ-type action reduces to that of PST.

In the general case, using the identity
\be\label{Lambdaid}
\Lambda^{\mu\nu\rho}=-3v^{[\rho}\Lambda^{\mu\nu]}-\frac 12 \varepsilon^{\rho\mu\nu\lambda\sigma\kappa}v_\lambda \Lambda_{\sigma\kappa},
\ee
we may write
\be\label{MLL}
M^{\sigma}{}_{\nu}=\Lambda^{\rho\mu\sigma}\L_{\rho\mu\nu}
=(\delta_\nu^\sigma +2v^\sigma v_\nu)\Lambda_{\mu\rho}\Lambda^{\mu\rho}-4\Lambda^{\sigma\mu}\Lambda_{\rho\mu}-4(v^\sigma \rho_\nu+v_\nu \rho^\sigma)\,,
\ee
with
\be\label{rho}
\rho^{\mu}=-\frac 1{8\sqrt{-g}}\varepsilon^{\mu\nu\rho\lambda\sigma\kappa}\Lambda_{\rho\lambda}\Lambda_{\sigma\kappa}v_{\nu}\,, \qquad \Lambda_{\mu\nu}\rho^\nu\equiv 0,
\ee
and
\be\label{I=L4}
I=M_\mu{}^{\nu}M_\nu{}^\mu=6\left((\Lambda_{\mu\nu}\Lambda^{\mu\nu})^2-16\rho^2\right)
=6\left(4\,\tr\Lambda^4-(\tr\Lambda^2)^2\right)\,,
\ee
where under the traces we use the standard multiplication law of the matrices $\Lambda_{\mu}{}^{\nu}$. Note that $I$ is non-negative, as its form is similar to non-negative $s^2-p^2$ in \eqref{MM}.

Then equation \eqref{L=B} takes the following form
\bea\label{B=L}
B_{\mu\nu}&=&-\Lambda_{\mu\nu}-24 \mathcal V_I\,\left(4\Lambda_\mu{}^{\sigma}\Lambda_{\sigma\lambda}\Lambda^{\lambda}{}_\nu+\Lambda_{\mu\nu}\,(\Lambda_{\rho\sigma}\Lambda^{\rho\sigma})\right)\nonumber\\
&=&-\Lambda_{\mu\nu}+\frac{\partial\mathcal V}{\partial\Lambda^{\mu\nu}}=-\Lambda_{\mu\nu}+\mathcal V_I\frac{d I}{d\Lambda^{\mu\nu}}.
\eea
From this equation one gets the following useful relations
\begin{align}
    s &= - 2 I \mathcal{V}_I - \frac{1}{4} \tr \left( \Lambda^2 \right) \left( 1 + 96 I \mathcal{V}_I^2 \right) \, , \label{s=L^2} \\
    p^2 &= - \frac{1}{96} \left( I - 6 \left( \tr \left( \Lambda^2 \right) \right)^2 \right) \left( 1 - 96 I \mathcal{V}_I^2 \right)^2 \, . \label{p^2=L^4}
\end{align}
Excluding the singular case of the Bialynicki-Birula theory, equation \eqref{B=L} can be inverted (at least formally and/or order by order) to write
\be\label{L=Binv}
\Lambda_{\mu\nu}(B)=B_{\mu\nu}\,f_1(s,p)
+
(B^3)_{\mu\nu}\,f_2(s,p),
\ee
where $f_1(s,p)$ and $f_2(s,p)$ are functions of the two invariants \eqref{sp}, whose form is related to the form of $\mathcal V(I)$.

Substituting the expression \eqref{L=Binv} for $\L_{\mu\nu}$ into the INZ Lagrangian density \eqref{INZNL}, we get a non-linear PST Lagrangian density \eqref{PST}, depending only on $E_{\mu\nu}$ and $B_{\mu\nu}$. The resulting interaction function $\mathcal H(s,p)$ is written in terms of $B_{\mu\nu}$ and $\Lambda_{\mu \nu} ( B )$, where the latter is itself a function of $B_{\mu \nu}$ due to \eqref{L=Binv}. Explicitly, one finds
\bea\label{HPST}
\mathcal H(s,p)&=&\frac 14 B_{\mu\nu}B^{\mu\nu}-\frac 12 (\Lambda_{\mu\nu}(B)+B_{\mu\nu})(\Lambda^{\mu\nu}(B)+B^{\mu\nu})+\mathcal V\left(I(\Lambda_{\mu\nu}(B))\right)\\
&=&\frac 14 B_{\mu\nu}B^{\mu\nu}-48\,(\Lambda_{\mu\nu}\Lambda^{\mu\nu})\,I\,\mathcal V_I^2\,+\mathcal V \, .
\non
\eea
Here we used \eqref{B=L} to compute $(\Lambda_{\mu\nu}+B_{\mu\nu})(\Lambda^{\mu\nu}+B^{\mu\nu})=96\,(\Lambda_{\mu\nu}\Lambda^{\mu\nu})\,I\,\mathcal V_I^2$ and again we stress that in all previous expressions $\Lambda_{\mu\nu}=\Lambda_{\mu\nu}(B)$ and $I(\Lambda_{\mu\nu}(B))$, are functions of $B_{\mu\nu}$.

We will give further details on the relation between the PST and INZ formulation in Section \ref{PST-INZ}.

Let us now consider the dynamical equation of motion of the 2-form potential $A_{\mu\nu}$ in the INZ formulation. It is obtained by varying \eqref{INZNL} with respect to $A_{\mu\nu}$ and results (with the use of the local symmetry \eqref{PST1}) in the relation
\be\label{INZsd}
E^{\mu\nu}+B^{\mu\nu}+2\Lambda^{\mu\nu}=0\, \quad \to \quad E^{\mu\nu}=-B^{\mu\nu}-2\Lambda^{\mu\nu}.
\ee
Substituting the expression \eqref{L=Binv} for $\Lambda$ in terms of $B$ into \eqref{INZsd}, one gets the PST non-linear self-duality condition \eqref{PSTsd} associated with the ``Hamiltonian" density \eqref{HPST}.

Regarding the equation of motion of the scalar field $a(x)$ which follows from \eqref{INZNL},
$$
\varepsilon^{\mu\nu\rho\kappa\lambda\sigma}(E_{\mu\nu}+B_{\mu\nu}+2\Lambda_{\mu\nu})\,\partial_\rho \Big((E_{\kappa\lambda}+B_{\kappa\lambda}+2\Lambda_{\kappa\lambda})v_{\sigma}\Big)=0,
$$
it is identically satisfied  when eq. \eqref{INZsd} holds. Thus $a(x)$ is a pure gauge field, just as in the PST formulation.

One can easily see that, because of the self-duality of $\Lambda_3$ and $F_3+F^*_3$, eq. \eqref{INZsd} is equivalent to
\be\label{L=F*}
\Lambda_{\mu\nu\rho}=-\frac 12 (F_{\mu\nu\rho}+F^*_{\mu\nu\rho})=-F^+_{\mu\nu\rho}\,.
\ee
Then, using \eqref{LambdaeomNL}, one finds that
\be\label{VI}
24{\mathcal V_I}\Lambda^{\mu\nu}{}_{\sigma}M^{\sigma\rho}=\frac 12(F^{\mu\nu\rho}-F^{*\mu\nu\rho})=F^{-\mu\nu\rho}\,.
\ee
Replacing $\Lambda$ on the left side of this equation with its on-shell value \eqref{L=F*}, we get the covariant non-linear self-duality condition on the physical field strength,
\be\label{NLsdM}
F^{-\mu\nu\rho}=-6\frac {\partial\mathcal V}{\partial F^+_{\mu\nu\rho}}\,,
\ee
which is the same as the one obtained in the formulation of \cite{Avetisyan:2022zza} modulo a conventional coefficient (see Section \ref{clone} for further details).

\subsection{Energy-momentum tensor}
The energy-momentum tensor of the INZ-type formulation obtained by varying \eqref{INZNL} with respect to the metric is
\bea\label{EMTINZ}
T_{\mu\nu}^{\text{INZ}}&=&T^{\text{free\,PST}}_{\mu\nu}+2(B_{\mu\rho}+\Lambda_{\mu\rho})(B_{\nu}{}^{\rho}+\Lambda_{\nu}{}^{\rho})-\frac 12 (\eta_{\mu\nu}+2v_\mu v_\nu) (B_{\rho\sigma}+\Lambda_{\rho\sigma})^2 \nonumber\\
&&+\frac 12\eta_{\mu\nu}(B+\Lambda)_{\rho\sigma}\Lambda^{\rho\sigma}-2\Lambda_{(\mu}{}^\rho(B+\Lambda)_{\nu)\rho}-v_{(\mu}\Lambda_{\nu)\rho\lambda}(B+\Lambda)^{\rho\lambda}\,\nonumber\\
&&
+\eta_{\mu\nu}(2I\,\mathcal V_I-\mathcal V)\,.
\eea
Upon performing some algebraic manipulations it can be written in the following form
\bea\label{EMTINZn2}
T_{\mu\nu}^{\text{INZ}}
&=&-\frac 14 (\eta_{\mu\nu}+2v_\mu v_\nu) (B_{\rho\sigma}+2\Lambda_{\rho\sigma})B^{\rho\sigma} +B_{(\nu}{}^{\rho}(B_{\mu)\rho}+2\Lambda_{\mu)\rho})\nonumber\\
&&+\frac 18v_{(\mu}\varepsilon_{\nu)\rho\lambda\delta\lambda\kappa}(B+2\Lambda)^{\rho\lambda}\mathcal (B+2\Lambda)^{\delta\lambda}v^\kappa+\frac 18v_{(\mu}\varepsilon_{\nu)\rho\lambda\delta\lambda\kappa}B^{\rho\lambda} B^{\delta\lambda}v^\kappa
\nonumber\\
&&
+\eta_{\mu\nu}(2I\,\mathcal V_I-\mathcal V)\,.
\eea
We can now use the on-shell relation \eqref{INZsd} and the identity \eqref{FF*}
to get the on-shell expression for the energy-momentum tensor,
\bea\label{EMTINZnl3}
T_{\mu\nu}^{\text{INZ}}|_{\text{on-shell}}&=&\frac 14 F_{\rho\lambda(\mu}F^*_{\nu)}{}^{\rho\lambda}
+g_{\mu\nu}\mathcal (\sqrt{-g})^{-1} \mathcal{L}_{\text{INZ}}^{\text{on-shell}} \,,
\eea
where the on-shell value of the INZ Lagrangian density is
\be\label{osINZL}
(\sqrt{-g})^{-1}\mathcal L^{\text{on-shell}}_{\text{INZ}} = 2I\,\mathcal V_I-\mathcal V(I)\,.
\ee
Comparing \eqref{EMTINZnl3} with the on-shell PST energy-momentum tensor \eqref{EMPSTOS},
we can thus assert that the on-shell Lagrangian densities of the PST and INZ formulation are equal and related as follows
\be\label{PST=INZ}
(\sqrt{-g})^{-1}\mathcal L^{\text{on-shell}} = 2I\,\mathcal V_I-\mathcal V(I)=s\mathcal H_s+p\mathcal H_p-\mathcal H(s,p)\,,
\ee
which provides further evidence that the on-shell PST Lagrangian density is $v^\mu$-independent and Lorentz invariant. In fact, the relation
\begin{align}\label{INZ=PST1}
    2I\,\mathcal V_I-\mathcal V(I)=s\mathcal H_s+p\mathcal H_p-\mathcal H(s,p)
\end{align}
between $\mathcal V$ and $\mathcal H$ is valid {\it off-shell},\footnote{When we say that a relation holds ``off-shell,'' we mean that this relation is valid assuming the {\it algebraic} equation of motion \eqref{LambdaeomNL} (or \eqref{L=B}) which relates the auxiliary field $\Lambda_{\mu\nu\rho}$ to the physical field $B_{\mu\nu}$, but {\it not} with the use of the dynamical field equation \eqref{INZsd}.} as we will prove in the next Section, but of course it does not describe the off-shell Lagrangian.

\subsection{Off-shell relation between the INZ-type and PST formulation}\label{PST-INZ}
As we have already pointed out, solving the algebraic equation \eqref{B=L} for $\Lambda_{\mu\nu}$ as a function of $B_{\mu\nu}$ (see eq. \eqref{L=Binv}) and substituting the solution back into the Lagrangian density \eqref{INZNL}, one gets a PST Lagrangian density with $\mathcal H(s,p)$ defined in \eqref{HPST}.

Taking the derivative of \eqref{HPST} with respect to $B^{\mu\nu}$ and using eq. \eqref{B=L} we get the following {\it off-shell} relation
\be\label{H=f}
H_{\mu\nu}=2\frac {\partial\mathcal H}{\partial B^{\mu\nu}}=-B_{\mu\nu}-2\Lambda_{\mu\nu}(B).
\ee
The equation \eqref{H=f} provides an explicit form of the formal expression \eqref{L=Binv} of $\Lambda$ as a function of $B$. In particular, using \eqref{H=f} one gets the following expression for the invariant $I$ of equation \eqref{I=L4},
\begin{align}\label{I-2}
   I &=6  \left( s^2 - p^2 \right) \left( \mathcal{H}_p^2 - \left( 1 + \mathcal{H}_s \right)^2 \right)^2 \, .
\end{align}

In Section \ref{PSTf} we saw that for the PST theory to be invariant under the local symmetry \eqref{PST2} the tensor $H_{\mu\nu}$ must satisfy the condition \eqref{HH=BB} which is equivalent to \eqref{Li}. It can be easily checked that $H_{\mu\nu}$ defined in \eqref{H=f}, with $B_{\mu\nu}$ and $\Lambda_{\mu\nu}$ related as in \eqref{B=L}, identically satisfies \eqref{HH=BB}.\footnote{A more involved proof of this fact was given in \cite{Avetisyan:2022zza} with the use of a dimensional reduction of the $6d$ theory to $d=5$. } Indeed, for the quantities related by \eqref{H=f} and \eqref{B=L}, the condition \eqref{HH=BB} reduces to
\be\label{LIL}
\varepsilon^{\mu\nu\rho\sigma\kappa\delta}v_{\nu}\Lambda_{\rho\sigma}\frac{dI}{d\Lambda^{\kappa\delta}}=0,
\ee
where (see \eqref{B=L})
\be\label{dI}
\frac{dI}{d\Lambda^{\mu\nu}}=-24 \left(4\Lambda_\mu{}^{\sigma}\Lambda_{\sigma\lambda}\Lambda^{\lambda}{}_\nu+\Lambda_{\mu\nu}\,(\Lambda_{\rho\sigma}\Lambda^{\rho\sigma})\right)\,.
\ee
The equation \eqref{LIL} is identically satisfied by the self-dual tensor $\Lambda_{\mu\nu\rho}$. This can be checked using  the identities \eqref{Lambdaid} and \eqref{3L}.

Therefore, the INZ-type formulation is directly related to the PST formulation and ensures
that the PST-invariance constraint  \eqref{Li} on the non-linear term \eqref{HPST} of the PST Lagrangian density is identically satisfied by any choice of $\mathcal V(I)$.
Vice versa, for a given PST function $\mathcal H(s,p)$, one finds the corresponding INZ interaction function
\be\label{V=H1}
\mathcal V(I)=\mathcal H(s,p)-(s\mathcal H_s+p\mathcal H_p) -\frac{s^2-p^2}p\mathcal H_s\mathcal H_p\,.
\ee
On the right hand side of this relation it is assumed that $B_{\mu\nu}$ is a function of $\Lambda_{\mu\nu}$ obtained by inverting the equation \eqref{H=f} for $\Lambda$ as a function of $B$.

The  {\it off-shell} relation \eqref{V=H1} is obtained from \eqref{HPST} in which one uses the equality \eqref{H=f}, namely,
\be\label{L+B=H-B}
\frac 12 (\Lambda_{\mu\nu}+B_{\mu\nu})(\Lambda^{\mu\nu}+B^{\mu\nu})=\frac 18 (H_{\mu\nu}-B_{\mu\nu})(H^{\mu\nu}-B^{\mu\nu})
\ee
and then the identity \eqref{HBPST} and
\be\label{HHPST}
H_{\mu\nu}H^{\mu\nu}=4(s\mathcal H_s^2+2p\mathcal H_s\mathcal H_p+s\mathcal H_p^2)=4s-\frac{8(s^2-p^2)}p \mathcal H_s\mathcal H_p\,.
\ee
Note also that in view of the relation\footnote{The relation \eqref{IVI} is obtained by the comparison of the results of the multiplication of \eqref{B=L} with $\Lambda^{\mu\nu}$, and \eqref{H=f} with $B^{\mu\nu}$ and the consequent use of \eqref{Li}.}
\be\label{IVI}
\frac{s^2-p^2}p\mathcal H_s\mathcal H_p=-2I\,\mathcal V_I\,,
\ee
the equation \eqref{V=H1} is equivalent to \eqref{INZ=PST1}.

Finally, substituting the relations \eqref{H=f}, \eqref{V=H1} and \eqref{IVI} into the INZ energy momentum tensor \eqref{EMTINZn2} and using \eqref{HBPST}, we find that the resulting expression coincides with that of the PST {\it off-shell} energy-momentum tensor \eqref{EMPST}.

We have thus shown that the PST and INZ-type formulation are directly related to each other via the algebraic relations \eqref{L=B} and \eqref{H=f} between the auxiliary field $\Lambda_{\mu\nu\rho}$ and the components $B_{\mu\nu}$ of the field strength  of the physical chiral field $A_{\mu\nu}$.

\subsection{Examples} \label{examples}

\subsubsection*{\it $6d$ conformal chiral 2-form (ModMax-like) theory}

In this case the function $\mathcal H(s,p)$, associated with the PST (Hamiltonian) density of the theory, has the form given in \eqref{MM}. Using \eqref{MM} one can directly compute the right-hand side of \eqref{V=H1}
and get
\be\label{V=Hmm}
\mathcal V(I)= -\frac{s^2-p^2}p\mathcal H_s\mathcal H_p=\sinh ( \gamma ) \,\partial_\gamma\mathcal H_{\text{MM}}= (\cosh ( \gamma ) \,\mathcal H_{\text{MM}}-s) \,.
\ee
On the other hand, from \eqref{I} we find that in the case under consideration
\be\label{corrected_I}
     I=96\cosh^4 \left( {\frac\gamma 2} \right)  (\mathcal\partial_\gamma \mathcal H_{\text{MM}})^2 .
\ee
Comparing \eqref{corrected_I} with \eqref{V=Hmm} and taking into account that $\partial_\gamma\mathcal H_{\text{MM}}\leq 0$ (see the comment in item (iv) of Section \ref{PSTac}) we finally get
\be\label{VMM}
\mathcal V(I(\Lambda))=- \frac 1{2} \left(\tanh{\frac \gamma 2}\right)\sqrt{\frac I6}\,.
\ee
Vice versa, if one starts from the conformal INZ theory with $\mathcal V$ given in \eqref{VMM}, and integrates out the auxiliary field $\Lambda$, one gets $\mathcal H$ of the PST formulation of $6d$ ModMax \eqref{MM}. This can be straightforwardly checked by substituting  \eqref{VMM} into \eqref{HPST} and using \eqref{s=L^2} and \eqref{p^2=L^4}.

Without giving details, the function $\mathcal V(I)=\delta \sqrt{I}$ (with a constant $\delta$)  for the conformal $6d$ counterpart of the ModMax theory, which is similar to \eqref{VMM}, was suggested in \cite{Avetisyan:2022zza} in a formulation put forward in \cite{Mkrtchyan:2019opf}. We will briefly discuss this formulation in \hbox{Section \ref{clone}}.
Earlier, for the $d=4$ ModMax theory, the corresponding form of $\mathcal V(I)$ in the Ivanov-Zupnik formulation \cite{Ivanov:2003uj} (which will be given in equation (\ref{IZ_4d_modmax})) was first obtained in \cite{Kuzenko:2021cvx}, and then in \cite{Avetisyan:2021heg} in the approach of \cite{Mkrtchyan:2019opf}. The reason that, for the $4d$ and $6d$ conformal duality-invariant theories, the function $\mathcal V(I)$ must be proportional to the square root of the forth-order invariant $I$ is very simple. It is the only possible non-linear duality-invariant term (constructed with the self-dual field) in the Lagrangian density which respects conformal invariance in these formulations, thus ensuring the uniqueness of these theories.

For completeness, let us present an explicit form of the $6d$ conformal chiral two-form field equations in the INZ formulation. This follows from \eqref{NLsdM}, with $\mathcal V(I)$ given in  \eqref{VMM} upon some algebraic manipulations; one finds
\be\label{eomMk'''}
F^{*\,\mu\nu\rho}=F^{\nu\rho\lambda}\left(\cosh ( \gamma )  \, \tensor{\delta}{_\lambda^\mu} -\frac {24\,{\sinh^2 ( \gamma ) }\,}{F^2}\,T_\lambda{}^{\mu}\right)\,,
\ee
where
\be\label{TMM}
T_{\mu\nu}=\frac 14F_{\rho\lambda(\mu}F_{\nu)}^{*\,\,\rho\lambda}= \frac 1{4\cosh ( \gamma ) }\left(F_{\mu\rho\lambda}F_{\nu}{}^{\rho\lambda}-\frac 16\eta_{\mu\nu}F^2\right).
\ee

\subsubsection*{\it Bialynicki-Birula theory \cite{Bialynicki-Birula:1992rcm,Gibbons:2000ck,Townsend:2019ils,Bandos:2020hgy}}

For this theory, substituting \eqref{BB} into  \eqref{I} we find that $I=0$. This is an {\it off-shell} constraint on the form of the Bialynicki-Birula field strength $F_{\mu\nu\rho}$.  Therefore, in this  case we cannot get $\mathcal V(I)$ as a result of the canonical `Legendre' transform of the PST (Hamiltonian) function \eqref{BB}.

On the other hand, in \cite{Avetisyan:2021heg,Avetisyan:2022zza} it was claimed that the Bialynicki-Birula theory and its $6d$ counterpart are described (modulo a conventional factor) by \eqref{VMM} in which \linebreak $\mathcal V(I)=\pm \frac 1{2}\sqrt{\frac I6}$ , i.e. in the limit $\gamma\to \infty$. However, using \eqref{s=L^2} and \eqref{p^2=L^4} for computing the form of $\mathcal H$ in \eqref{HPST} for this choice of $\mathcal V(I)$, we find that
\be\label{pH=0}
p=\mathcal H_{\text{BB}} = 0\,.
\ee
Note that these conditions are valid {\it off the mass shell}.
This is not the case for the BB theory described by \eqref{BB} even on-shell.

Instead, this is just the $\gamma \to \infty$ limit of the ModMax theory \eqref{MM}. Indeed, let us rewrite \eqref{MM} as
$$
\mathcal H_{\text{MM}}=\cosh ( \gamma ) \,(s-\tanh ( \gamma ) \sqrt{s^2-p^2}) \, ,
$$
and require that $\mathcal H_{\text{MM}}$ remain finite in the $\gamma\to \infty$ limit. This is only possible if
$$
\frac 1{\cosh ( \gamma ) }\mathcal H_{MM} \big\vert_{\gamma\to \infty}=s-\sqrt{s^2-p^2}=0\,\qquad \to \qquad p=0,
$$
which agrees with \eqref{pH=0}. We conclude that the INZ-type Lagrangian formulation and the ``clone" formulation of duality-invariant field theories considered in the next Section do not include the Lagrangian description of the Bialynicki-Birula theory.

\subsubsection*{\it BI-like (M5-brane) theory}

In this case, as we will discuss in Section \ref{Lflows}, the interaction function $\mathcal V(I)$ takes the form of a certain hypergeometric function given in equation (\ref{hypergeometric_solution}). This function can be alternatively described by giving its ``Legendre transform'' from the variable $I$ to the quartic invariant $I_h=h_{\mu\rho\sigma} h^{\rho\sigma\lambda}h_{\lambda\kappa\delta}h^{\kappa\delta\mu}$ of the auxiliary self-dual field $h_3$ of the superembedding formulation; the resulting function is very simple and is given in \eqref{OSBI}. Moreover, comparing the on-shell expression for $\Lambda_3$ in \eqref{L=F*} with \eqref{F=h} we see that
\be\label{L=h}
\Lambda_{\mu\nu\rho}=-\left(1-\frac 2{3T^2}I_h\right)^{-1}h_{\mu\nu\rho} \, ,
\ee
and hence
\be\label{IL=Ig}
I(\Lambda)=\left(1-\frac 2{3T^2}I_h\right)^{-4}\,I_h\,.
\ee
Inverting the above relation, one can (at least perturbatively) find an expression for $I_h$ as a function of $I(\Lambda)$. Upon substitution into \eqref{L=h}, this gives an expression for $h_3$ as a function of $\Lambda_3$, while upon substitution into \eqref{OSBI}, it produces the relation between the expressions of the interaction term  as functions of $I(\Lambda)$ and $I_h$
\begin{equation}\label{ILIh}
    2I(\Lambda)\mathcal V_I-\mathcal V(I(\Lambda)) = \frac 2{3T}\frac {I_h}{\frac 2{3T^2}\,I_h-1} \, \equiv \mathcal{W} ( I_h ) \, .
\end{equation}
From (\ref{ILIh}), one can reconstruct $\mathcal V_{\text{BI}}(I(\Lambda))$ order-by-order and verify that it matches the closed-form result (\ref{hypergeometric_solution}) which will be obtained by solving a $\TT$-like flow equation.

\subsubsection*{\it Relationship with $4d$ Ivanov-Zupnik Formalism}

It turns out that equation \eqref{ILIh} expresses exactly the same relationship as the one between the interaction functions of $4d$ Born-Infeld electrodynamics formulated in two related auxiliary field presentations of \cite{Ivanov:2003uj}, which we referred to as the $\nu$- and $\mu$-frames. Although we will not give a comprehensive discussion of these formalisms in the current work, let us briefly state the main ingredients in order to make the connection with our $6d$ analysis clear. We refer the reader to the original work \cite{Ivanov:2003uj}, a later survey \cite{Ivanov:2013jba}, or to Sections 5 and 6 of \cite{Ferko:2023wyi} for a more detailed treatment.

Within this formalism, in addition to the field strength $F_{\mu \nu}$ of an Abelian gauge field in four spacetime dimensions, one introduces an auxiliary $2$-form field $V_{\mu \nu} = - V_{\nu \mu}$, and then converts both $F_{\mu \nu}$ and $V_{\mu \nu}$ to spinor indices as
\begin{align}
    \tensor{F}{_\alpha^\beta}
    &= -\frac{1}{4}(\sigma^{\mu})_{\alpha\dot\beta}(\tilde{\sigma}^{\nu})^{\dot\beta\beta}F_{\mu\nu}
    ~,\quad \tensor{\overbar{F}}{_{\dot\alpha}^{\dot\beta}}
    = \,\frac{1}{4}(\tilde{\sigma}^{\mu})^{\dot\beta\beta}(\sigma^{\nu})_{\beta\dot\alpha}F_{\mu\nu}~, \\
    \tensor{V}{_\alpha ^\beta} &= -\frac{1}{4}(\sigma^{\mu})_{\alpha\dot\beta}(\tilde{\sigma}^{\nu})^{\dot\beta\beta} V_{\mu\nu} ~,\quad \tensor{\overbar{V}}{_{\dot{\alpha}}^{\dot{\beta}}} = \frac{1}{4}(\tilde{\sigma}^{\mu})^{\dot\beta\beta}(\sigma^{\nu})_{\beta\dot\alpha} V_{\mu\nu} \, ,
\end{align}
where the $\sigma^\mu$, $\tilde{\sigma}^\mu$ are Pauli matrices. We also define the scalars
\begin{align}
    \varphi &= F^{\alpha \beta} F_{\alpha \beta} \, , & \overbar{\varphi} &= \overbar{F}_{\dot{\alpha} \dot{\beta}} \overbar{F}^{\dot{\alpha} \dot{\beta}} \, , \nonumber \\
    \nu &= V^{\alpha \beta} V_{\alpha \beta} \, , & \overbar{\nu} &= \overbar{V}_{\dot{\alpha} \dot{\beta}} \overbar{V}^{\dot{\alpha} \dot{\beta}} \, , \nonumber \\
    V \cdot F &= V^{\alpha \beta} F_{\alpha \beta} \, , & \overbar{V} \cdot \overbar{F} &= \overbar{V}_{\dot{\alpha} \dot{\beta}} \overbar{F}^{\dot{\alpha} \dot{\beta}} \, .
\end{align}
Using this notation, consider the Lagrangian
\begin{align}\label{ivanov_zupnik_lagrangian}
    \mathcal{L} = \frac{1}{2} \left( \varphi + \overbar{\varphi} \right) + \nu + \overbar{\nu} - 2 \left( V \cdot F + \overbar{V} \cdot \overbar{F} \right) + E \left( \nu \overbar{\nu} \right) \, ,
\end{align}
which depends on a function $E ( a )$ of one real variable $a = \nu \overbar{\nu}$ (not to be confused with the auxiliary PST scalar $a(x)$).

After integrating out the auxiliary fields $V$, $\overbar{V}$, any such Lagrangian gives rise to a duality-invariant theory of $4d$ electrodynamics. Conversely, given any Lagrangian for a duality-invariant theory which depends on $F_{\mu \nu}$ but not its derivatives, there exists some choice of function $E ( a )$ such that this given Lagrangian is equivalent to (\ref{ivanov_zupnik_lagrangian}).\footnote{Strictly speaking, this is only true for theories which reduce to Maxwell electrodynamics in some limit, and excludes exceptional cases like the Bialynicki-Birula theory.} Therefore, $4d$ duality-invariant theories of non-linear electrodynamics without higher derivative interactions are in one-to-one correspondence with interaction functions $E ( a )$ via the representation (\ref{ivanov_zupnik_lagrangian}). We refer to this as the ``$\nu$-frame representation.''

It is often convenient to write a theory in a different way by defining the complex Legendre transform of the interaction function $E$,
\begin{align}\label{nu_to_mu_1}
    H(\mu,\overbar{\mu}) = E(\nu,\overbar{\nu}) - \nu \mu -\overbar{\nu} \overbar{\mu} \, ,
\end{align}
where the variables $\mu$, $\overbar{\mu}$ are defined by
\begin{align}\label{nu_to_mu_2}
    \mu(\nu,\overbar{\nu}) = \partial_{\nu}E
    ~,\quad \overbar{\mu}(\nu,\overbar{\nu}) = \partial_{\overbar{\nu}} E \, .
\end{align}
In terms of these variables, one can rewrite the Lagrangian (\ref{ivanov_zupnik_lagrangian}) as
\begin{align}
    \mathcal{L} = \frac{\varphi(\mu-1)}{2(1+\mu)}+\frac{\overbar{\varphi}(\overbar{\mu}-1)}{2(1+\overbar{\mu})}+H ( \mu \overbar{\mu} ) \, ,
\end{align}
which we call the ``$\mu$-frame representation.''

There is a one-to-one correspondence between interaction functions $\mathcal{V} ( I )$ for $6d$ chiral tensor theories in the INZ formalism and interaction functions $E ( a )$ for $4d$ theories of duality-invariant electrodynamics in the $\nu$-frame representation.\footnote{Although note that $E(a)$ and $\mathcal{V} ( I )$ differ by a sign, since $E ( a )$ appears in the Lagrangian (\ref{ivanov_zupnik_lagrangian}) with a plus sign but $\mathcal{V} ( I )$ enters (\ref{INZNL}) with a minus sign.} For instance, we have already commented below equation (\ref{VMM}) that the $\nu$-frame presentation of the $4d$ ModMax theory, which is described by the interaction function
\begin{align}\label{IZ_4d_modmax}
    E ( a, \gamma ) = 2 \tanh \left( \frac{\gamma}{2} \right) \sqrt{ a } \, ,
\end{align}
is almost identical to the functional form (\ref{VMM}) of $\mathcal{V} ( I )$ for the $6d$ ModMax-like chiral tensor theory, although with the opposite sign and a different constant prefactor.

Likewise, one can also describe Born-Infeld electrodynamics using these auxiliary field formulations. It is simpler to write the explicit form of the BI interaction function in the $\mu$ frame, where it takes the form
\begin{align}\label{born_infeld_mu_frame}
    H ( b ) = \frac{1}{\lambda} \frac{2b}{b-1} \, , \qquad b = \mu \overbar{\mu} \, .
\end{align}
The $\nu$-frame description of the Born-Infeld theory is somewhat more unwieldy; one can write it in terms of the unique root $t ( a , \lambda )$ of the quartic equation
\begin{align}\label{t_quartic}
    t^4 + t^3 - \frac{\lambda^2 a}{4} = 0 \, ,
\end{align}
which has the property that $t ( a = 0 ) = -1$. In terms of this quantity, one has
\begin{align}\label{EBI_algebraic}
    E_{\text{BI}} ( a, \lambda ) = 2 \left( t^2 + 3 t + 1 \right) \, .
\end{align}
Note that, after making the identifications
\begin{align}
    b = \frac{2 I_h}{3 T^2} \, , \qquad \lambda = \frac{1}{T} \, ,
\end{align}
the $\mu$-frame interaction function (\ref{born_infeld_mu_frame}) can be written as
\begin{align}
    H ( I_h ) = 2 \cdot \left( \frac{2}{3 T} \frac{I_h}{\frac{2}{3 T^2} I_h - 1 } \right) = 2 \mathcal{W} ( I_h ) \, ,
\end{align}
which (up to a factor of $2$) is identical to the function $\mathcal{W} ( I_h )$ appearing on the right side of (\ref{ILIh}). Thus the expression for the $6d$ on-shell Born-Infeld-like Lagrangian density, written as a function of $I_h$, is proportional to the $\mu$-frame interaction function for the $4d$ Born-Infeld theory. Furthermore, the relations (\ref{nu_to_mu_1}) - (\ref{nu_to_mu_2}) which convert between the $\mu$ frame interaction function $H(b)$ and the $\nu$ frame interaction function $E(a)$ are also the same (up to constant factors) as those which relate $\mathcal{L}_{\text{BI}} \big\vert_{\text{on-shell}}$ to $\mathcal{V} ( I )$. As this relation expresses a Legendre transform in the variables $\mu$ and $\nu$, this is the sense in which we say that $\mathcal{W} ( I_h )$ in (\ref{ILIh}) is a ``Legendre transform'' of $\mathcal{V} ( I )$ from the variable $I$ to $I_h$.

We therefore expect that the INZ function $\mathcal{V} ( I )$ which describes the $6d$ BI-like theory should be equivalent, up to multiplicative factors and scaling of the argument, to the $\nu$-frame interaction function (\ref{EBI_algebraic}) for the $4d$ Born-Infeld theory. We will see that this is the case, and exhibit a closed form solution for this function, around equation (\ref{EBI_soln}).

\section{``Clone" formulation}\label{clone}
Let us now briefly consider yet another formulation of chiral $2n$-form theories in $4n+2$ spacetime dimensions \cite{Mkrtchyan:2019opf,Bansal:2021bis,Avetisyan:2022zza}. In $d=6$ the Lagrangian density of this formulation has the following form
\be\label{MkrtchyanL}
\mathcal L=-\frac {\sqrt{-g}}{4!}X_{\mu\nu\rho}X^{\mu\nu\rho}+\frac{a(x)}{12}\varepsilon^{\mu\nu\rho\sigma\lambda\kappa}F_{\mu\nu\rho}Q_{\sigma\lambda\kappa}-\sqrt{-g}\,\mathcal V(I(X^+)) \, ,
\ee
where
\be\label{Hmnr}
X_{\mu\nu\rho}=F_{\mu\nu\rho}+a(x)Q_{\mu\nu\rho},
\ee
$a(x)$ is a PST-like auxiliary scalar field and $Q_3=d\tilde A_2$ is the field strength of an additional two-form gauge field $\tilde A_2$ (whose presence explains the name ``clone" \cite{Evnin:2022kqn} of the formulation under consideration). The self-interaction term $\mathcal V(I)$ is an arbitrary function of the fourth order invariant of the self-dual part $X^+_3=\frac 12 (X_3+X_3^*)$ of $X_3$, namely
\be\label{IX+}
I(X^+)= X^+_{\mu\nu\rho}X^{+\nu\rho\lambda}X^+_{\lambda\kappa\sigma}X^{+\kappa\sigma\mu}.
\ee
In this formulation there are local symmetries (akin to \eqref{PST1} and \eqref{PST2}) which allow one to eliminate  $a(x)$ and $Q_3$ on the mass shell, thus leaving $F_3=dA_2$ the only independent physical field satisfying the non-linear self-duality condition equivalent to \eqref{NLsdM} (see \cite{Avetisyan:2022zza} for details).

The energy-momentum tensor derived by varying \eqref{MkrtchyanL} with respect to the metric has the following form
\be\label{EMMkrt}
T_{\mu\nu}=\frac 14\left(X_{\mu\rho\sigma}X_{\nu}{}^{\rho\sigma}-\frac 16 g_{\mu\nu} X^2\right)+\frac 12\left(X^{\lambda\rho}{}_{(\mu}\mathcal V_{\nu)\lambda\rho}-2g_{\mu\nu}\mathcal V\right)\,
\ee
where $\mathcal V_{\nu\lambda\rho}=6\frac{\partial\mathcal V}{\partial X^{+\nu\lambda\rho}}$ is anti-self-dual. It has the same form as given in \eqref{dV} but with $\Lambda$ replaced by $X^+$.

Upon splitting $X_{\mu\nu\rho}$ into its self-dual and anti-self dual parts, and using the identity
$$
X^{+\lambda\rho}{}_{(\mu}\mathcal V_{\nu)\lambda\rho}=4g_{\mu\nu}I\mathcal V_I,
$$
one can bring the energy-momentum tensor to the following form
\begin{align} \label{EMMkrt_flow2}
T_{\mu\nu} &= \left(\frac  14X^-_{\mu\rho\sigma}X^-_{\nu}{}^{\rho\sigma}+\frac 12X^{-\lambda\rho}{}_{(\mu}\mathcal V_{\nu)\lambda\rho}\right)+\left(\frac 14 X^+_{\mu\rho\sigma}X^+_{\nu}{}^{\rho\sigma}+ g_{\mu\nu}(2I\mathcal V_I-\mathcal V)\right)\,.
\end{align}
A consequence of the equations of motion and local symmetries in this formulation is that
\be\label{X-=VX+}
X^-_{\mu\rho\sigma}=-\mathcal V_{\mu\rho\sigma}(X^+)\,.
\ee
One can now perform a gauge fixing of local symmetries by setting $Q=0$ in \eqref{Hmnr}, so that
$X_{\mu\rho\sigma}=F_{\mu\rho\sigma}$, and then eq. \eqref{X-=VX+} reduces to
\be\label{msM}
\mathcal V_{\mu\rho\sigma}(F^+)=-F^-_{\mu\rho\sigma} \qquad \to \qquad F_{\mu\nu\rho}F^{\mu\nu\rho}=-48I\mathcal V_I(F^+)\,.
\ee
So, the non-linear self-duality condition and the on-shell energy-momentum tensor in this formulation coincide with those of the INZ-type formulation, eqs. \eqref{NLsdM} and  \eqref{EMTINZnl3}.
This, in  turn, relates the ``clone" formulation to the non-linear PST one, though in a less straightforward way.

\section{$\TT$-like flows of $6d$ chiral two-form theories}\label{TTbarf}

In this Section, we consider parameterized families of chiral two-form theories in six space-time dimensions whose derivatives with respect to these parameters can be expressed in terms of the energy-momentum tensor. That is, we study equations of the form
\begin{align}\label{lagrangian_flow_general}
    \frac{\partial \mathcal{L}^{(\lambda)}}{\partial \lambda} = f \left( T_{\mu \nu}^{(\lambda)} , \lambda \right) \, ,
\end{align}
in the Lagrangian formulation, or
\begin{align}\label{hamiltonian_flow_general}
    \frac{\partial \mathcal{H}^{(\lambda)}}{\partial \lambda} = g \left( T_{\mu \nu}^{(\lambda)} , \lambda \right) \, ,
\end{align}
in Hamiltonian language. We call any equation which takes either of these forms a ``stress tensor flow,'' or equivalently ``$\TT$-like flow,'' and we refer to the function $f$ or $g$ on the right side of these equations as the operator driving the flow. However, we should emphasize that our analysis will be entirely restricted to the classical level, so these operators are simply classical functions of the fields and not true quantum mechanical operators.

First let us make some general comments about such flows. It is straightforward to show quite generally that classical deformations of the Lagrangian are equivalent to those of the Hamiltonian, once the deforming operator is expressed in terms of the correct variables. To be precise, let $\mathcal{L} ( \phi^a , \partial_i \phi^a , \dot{\phi}^a )$ be any Lagrangian density describing a collection of fields $\phi^a$ in $d$ space-time dimensions. Here we use $a$ as an internal index which labels the different fields, while $i = 1 , \ldots, d-1$ is a spatial index, and $\dot{\phi}^a = \frac{\partial \phi^a}{\partial x^0}$. The fields $\phi^a$ may also carry arbitrary Lorentz indices, such as those of the $p$-form fields considered in this work, but we suppress such indices for simplicity. Now let
\begin{align}
    \mathcal{H} = \sum_{a} \pi^a \dot{\phi}^a - \mathcal{L}
\end{align}
be the corresponding Hamiltonian density. Suppose that the Lagrangian density depends on some parameter $\lambda$ and obeys a differential equation
\begin{align}
    \frac{\partial \mathcal{L}}{\partial \lambda} = \mathcal{O} ( \phi^a , \partial_i \phi^a , \dot{\phi}^a ) \, ,
\end{align}
where $\mathcal{O}$ is a function of the fields and their derivatives, and may also have explicit dependence on $\lambda$ which we will suppress in what follows. Then the Hamiltonian density satisfies the equation
\begin{align}\label{hamiltonian_flow}
    \frac{\partial \mathcal{H}}{\partial \lambda} = - \mathcal{O} \left( \phi^a , \partial_i \phi^a , \dot{\phi}^a \left( \phi^b , \partial_j \phi^b , \pi^b \right) \right) \, ,
\end{align}
where we write $\dot{\phi}^a \left( \phi^b , \partial_j \phi^b , \pi^b \right)$ to indicate the functional dependence of the time derivatives $\dot{\phi}^a$ on the canonical momenta $\pi^b$, fields $\phi^b$, and their spatial derivatives. For equation (\ref{hamiltonian_flow}) to hold, it is necessary to use the relationship between the time derivatives and canonical momenta within the theory $\mathcal{L} ( \lambda )$ or $\mathcal{H} ( \lambda )$, rather than the corresponding relation in the ``undeformed'' theory at $\lambda = 0$. A proof of this statement at leading order in $\lambda$ was given in Appendix A of \cite{Kruthoff:2020hsi}. The all-orders proof for $(0+1)$-dimensional theories (that is, particle mechanics) can be found in \cite{Ferko:2023iha}, although the proof for field theories in arbitrary spacetime dimension is identical.

The upshot of this simple observation is that we are free to consider stress tensor deformations of either the Lagrangian or the Hamiltonian, and the two are equivalent. The dictionary which translates between the Lagrangian flows (\ref{lagrangian_flow_general}) and Hamiltonian flows (\ref{hamiltonian_flow_general}) is simply taking $g = - f$. We will study flows in both of these formulations in this Section. For the class of models considered in this work, the relationship between Lagrangian and Hamiltonian flows is somewhat more transparent, since the interaction function $\mathcal{H} ( s, p )$ appearing in the Lagrangian PST formulation plays the role of the Hamiltonian, and thus it is clear that $\partial_\lambda \mathcal{L} = - \partial_\lambda \mathcal{H}$ for deformations of these models.

Our second general comment concerns the universality of flows constructed from the energy-momentum tensor. We have seen in the preceding sections that a $6d$ chiral two-form theory is characterized by a single interaction function of one real variable, which is called $\mathcal{V} ( I )$ in the INZ-type and clone formulations. (Alternatively, one can describe such a theory in terms of an interaction function $\mathcal{H} ( s, p )$ in the PST formulation, but this function must satisfy the partial differential equation \eqref{Li}, whose solutions are parameterized by a single function of one real variable.) Thus, one might ask why we consider flow equations of the form (\ref{lagrangian_flow_general}), rather than the seemingly more general form
\begin{align}\label{general_I_flow}
    \frac{\partial \mathcal{L}}{\partial \lambda} = f ( I , \lambda ) \, ,
\end{align}
where $I$ is the invariant used in the INZ-type and clone formalisms. Indeed, we have already seen a differential equation of this schematic form in equation (\ref{corrected_I}), which relates the $\gamma$ derivative of $\mathcal H_{\text{MM}}$ to $I$.

The reason is that any such flow equation (\ref{general_I_flow}) can be recast in the form of a stress tensor flow equation due to the functional dependence between the variable $I$ and the Lorentz scalars that can be constructed from the stress tensor. In fact, more is true: for any theory in this class, the two Lorentz scalars $\tensor{T}{^\mu_\mu}$ and $T^{\mu \nu} T_{\mu \nu}$ built from the stress tensor are themselves functionally dependent -- as we will show shortly -- so the variable $I$ can be taken to be dependent on one of these. We will choose the second invariant, $T^{\mu \nu} T_{\mu \nu}$, to write this relation, since it is non-zero even for conformal theories. Therefore, in any theory in the INZ-type or clone formulations, there exists a relation of the form
\begin{align}\label{stress_tensor_dependence}
    \Upsilon \left( I , T_{\mu \nu} T^{\mu \nu} \right) = 0 \, ,
\end{align}
for some function $\Upsilon$ of two variables. Therefore, one can locally invert this relation\footnote{More precisely, this is true with the possible exception of a discrete collection of exceptional points at which this inverse map becomes singular.} to express any function of $I$ as a function of the scalar combination $T_{\mu \nu} T^{\mu \nu}$.

The analogue of this statement for theories of duality-invariant nonlinear electrodynamics in four dimensions was discussed in \cite{Ferko:2023wyi}. It follows that any flow equation (\ref{general_I_flow}) can be brought to the form (\ref{lagrangian_flow_general}).

Given the equivalence between deformations driven by functions of $I$, and those driven by functions of the energy-momentum tensor, we prefer the latter. These stress tensor flow equations are expressed in terms of deforming operators which are defined directly using the data of a given theory (namely, its Hilbert stress tensor), rather than using an external function $f ( I )$ which does not make obvious reference to the theory under consideration.

\subsection{Hamiltonian flow equations}

Let us begin by discussing deformations of chiral tensor theories with actions of the form (\ref{PST}) described by a PST interaction function $\mathcal{H} ( s, p )$. As we have reviewed around equation (\ref{LH}), since $\mathcal{L}_{\text{PST}}$ becomes first-order in time derivatives upon fixing a particular choice of gauge, this interaction function can be viewed as the Hamiltonian of the model. In order to state our results most simply, in this subsection we will first work in the ``Hamiltonian PST'' formulation in which manifest Lorentz invariance is broken by singling out a preferred time coordinate; we also assume that we work in flat Minkowski space-time throughout this subsection. One can view this choice as working in the gauge
\begin{align}\label{vmu_gauge_fixed}
    v_\mu = \tensor{\delta}{^0_\mu} \, .
\end{align}
As in the previous sections, we will reserve Greek symbols $\mu$, $\nu$, etc., for $6d$ indices, including the time coordinate $x^0$, and use Latin indices such as $i$ and $j$ for spatial (five-dimensional) directions.

Owing to the relation $B_{\mu \nu} v^\nu = 0$, when (\ref{vmu_gauge_fixed}) is satisfied, we may restrict attention to only the spatial entries $B_{ij}$ since the other components $B_{0 \mu}$ are vanishing. A general Hamiltonian $\mathcal{H}$ is therefore a function of two variables, namely the $SO(5)$ scalar
\begin{align}\label{spatial_s}
    s = \frac{1}{4} B^{ij} B^{kl} \delta_{ik} \delta_{jl} \, ,
\end{align}
as well as the length of the spatial vector
\begin{align}\label{spatial_p}
    p_i = \frac{1}{8} \varepsilon_{i j k l m} B^{jk} B^{lm} \, .
\end{align}
Thus far, we have made no assumptions whatsoever about any constraints satisfied by the function $\mathcal{H} ( s, p )$. However, we saw in Section \ref{PSTf} that the interaction function in the Lagrangian PST formulation must satisfy the partial differential equation (\ref{Li}). In Hamiltonian language, this equation is required in order to guarantee the Lorentz invariance of the model.

The primary models of our interest in this work, which include the $6d$ chiral tensor versions of the Born-Infeld and ModMax theories, are Lorentz-invariant theories and therefore the corresponding functions $\mathcal{H} ( s, p )$ for these models satisfy (\ref{Li}). Nonetheless, we should point out that several past works have investigated $\TT$-like flows of \emph{non}-Lorentz-invariant field theories \cite{Cardy:2018jho,Cardy:2020olv,Ceschin:2020jto,Esper:2021hfq}. These explorations have been motivated, in part, by the fact that the argument of \cite{Zamolodchikov:2004ce} for the well-definedness and factorization of the two-dimensional $\TT$ operator does not require Lorentz invariance (or rotational symmetry, in Euclidean signature), but merely translation invariance. With an eye to potential future interest in results on $\TT$-like deformations of non-Lorentz-invariant $6d$ chiral tensor theories, we will present several general expressions without assuming that the condition (\ref{Li}) holds, only imposing it at the end to obtain simplified results in equation (\ref{LorentziOs}).

Given a Hamiltonian $\mathcal{H} ( s, p )$, one can compute \cite{Bandos:2020hgy} the components of the energy-momentum tensor which coincide with those of \eqref{EMPST} for $g_{\mu\nu}=\eta_{\mu\nu}$ and $v_\mu= \tensor{\delta}{_\mu^0}$,
\begin{align}
    T_{00} = \mathcal{H} \, , \quad T_{0 i} = - p_i \, , \quad T_{ij} = H_{ij} B^{jk} - \tensor{\delta}{^i_k} \left( \frac{1}{2} B_{j l} H^{l j} + \mathcal{H} \right) \, ,
\end{align}
where $H_{ij} = 2 \frac{\partial \mathcal{H}}{\partial B_{ij}}$ is defined in equation (\ref{Hmn}), here restricted to spatial indices.

The two combinations which we will use for constructing flows are then
\begin{align}\label{hamiltonian_stress_tensor_evaluated}
    \tensor{T}{^\mu_\mu} &=  6\left( s \mathcal{H}_s + p \mathcal{H}_p - \mathcal{H} \right)  \, ,  \\
    T^{\mu \nu} T_{\mu \nu} &= 6 \mathcal{H}^2 - 2 p^2  - 12 p \mathcal{H} \mathcal{H}_p + 8 p^2 \mathcal{H}_p^2 + 4 s \mathcal{H}_s \left( 4 p \mathcal{H}_p - 3 \mathcal{H} \right) - 4 \left( p^2 - 3 s^2 \right) \mathcal{H}_s^2 \, .\nonumber
\end{align}
A general stress tensor deformation of the Hamiltonian is then understood to mean a flow equation of the functional form
\begin{align}\label{hamiltonian_flow1}
    \frac{\partial \mathcal{H}}{\partial \lambda} = -\mathcal{O} \left( \tensor{T}{^\mu_\mu}, T^{\mu \nu} T_{\mu \nu} , \lambda \right)
    \,,
\end{align}
for some function $\mathcal{O}$, along with an initial condition $\mathcal{H} ( \lambda = 0 ) = \mathcal{H}_0$. In equation (\ref{hamiltonian_flow1}), the stress tensors in the arguments of $\mathcal{O}$ are evaluated from $\mathcal{H} ( \lambda )$ rather than from the seed Hamiltonian $\mathcal{H}_0$. Although we have indicated that a general deforming operator (\ref{hamiltonian_flow1}) may have explicit $\lambda$ dependence, in our primary cases of interest in this work, $\mathcal{O}$ will depend on $\lambda$ only through the combinations $\tensor{T}{^\mu_\mu}$ and $T^{\mu \nu} T_{\mu \nu}$.

Among the family of stress tensor deformations, we are especially interested in two special cases. The first is when the function $\mathcal{O}$ takes the form
\begin{align}\label{6d_TT_combination}
    \mathcal{O} \left( \tensor{T}{^\mu_\mu}, T^{\mu \nu} T_{\mu \nu} \right) &= \frac{1}{12} \left( T^{\mu \nu} T_{\mu \nu} - \frac{1}{3} \left( \tensor{T}{^\mu_\mu} \right)^2 \right) 
    \equiv \mathcal{O}_{T^2} \, ,
\end{align}
which is the six-dimensional version of the classical $\TT$-like deformation that is defined in $d$ space-time dimensions by the formula
\begin{align}
    \mathcal{O}_{T^2}^{(d)} = \frac{1}{2 d} \left( T^{\mu \nu} T_{\mu \nu} - \frac{2}{d} \left( \tensor{T}{^\mu_\mu} \right)^2 \right) \, .
\end{align}
Higher-dimensional classical flows of this form, driven by quadratic combinations of the stress tensor, have been considered by many authors \cite{Bonelli:2018kik,Conti:2022egv}; see \cite{Taylor:2018xcy} for a study of the obstruction to defining such operators at the quantum level. Here we make the particular choice of numerical coefficients which leads to flows that generate the Nambu-Goto action from free scalar in two space-time dimensions and the Born-Infeld action from the Maxwell theory in four space-time dimensions, following the conventions in \cite{Ferko:2023ruw}.

The second combination of interest is a version of the root-$\TT$ operator,
\begin{align}\label{root6d}
    \mathcal{R} = \sqrt{ \frac{1}{6} \left( T^{\mu \nu} T_{\mu \nu} - \frac{1}{6} \left( \tensor{T}{^\mu_\mu} \right)^2 \right) } \, ,
\end{align}
which likewise is the $d = 6$ case of the general definition
\begin{align}\label{general_root_TT}
    \mathcal{R}^{(d)} = \frac{1}{\sqrt{d}} \sqrt{ \widehat{T}^{\mu \nu} \widehat{T}_{\mu \nu} }  \, ,
\end{align}
where
\begin{align}
    \widehat{T}_{\mu \nu} = T_{\mu \nu} - \frac{1}{d} g_{\mu \nu} \tensor{T}{^\rho_\rho}
\end{align}
is the traceless part of the stress tensor. An operator proportional to (\ref{general_root_TT}) for general dimension $d$ was proposed in \cite{Ferko:2022cix}. As in the case of $\mathcal{O}_{T^2}^{(d)}$, the numerical prefactor is a choice of conventions, and here we follow the normalization used in \cite{Ferko:2023ruw}.

By using the expressions (\ref{hamiltonian_stress_tensor_evaluated}), we may express the two operators $\mathcal{O}_{T^2}$ and $\mathcal{R}$ in terms of the Hamiltonian and its derivatives:
\begin{align}\label{two_6d_operators_hamiltonian}
   \mathcal{O}_{T^2} &=  \frac{1}{2} \left((s^2-p^2) \mathcal H_s^2 - (\mathcal{H}- s \mathcal{H}_s - p \mathcal{H}_p )^2- {p^2\left( 1-\mathcal H_s^2-2sp^{-1}\mathcal H_s\mathcal H_p -\mathcal H_p^2\right )}\right) \, , \nonumber \\
    \mathcal{R}^2 &= \frac{1}{3} \left({ 3 \left( s^2 -  p^2 \right) \mathcal{H}_s^2 -p^2+ p^2 \left( \mathcal{H}_p^2  + 2 s p^{-1}\mathcal{H}_s \mathcal{H}_p + \mathcal H_s^2\right)} \right) \, .
\end{align}
We may now state one of our main results, which is that the six-dimensional interacting chiral tensor model describing the worldvolume gauge theory of an M5-brane satisfies a classical $\TT$-like flow. In fact, more is true: the entire two-parameter family of ModMax-Born-Infeld-like chiral tensor theories, defined by the Hamiltonian
\begin{align}\label{6d_modmax_born_infeld_hamiltonian}
    \mathcal{H} ( \gamma, \lambda ) = \frac{1}{\lambda} \left( \sqrt{ 1 + 2 \lambda \left( s \cosh ( \gamma ) - \sinh ( \gamma ) \sqrt{ s^2 - p^2 } \right) + \lambda^2 p^2 } - 1 \right) \, ,
\end{align}
satisfies a pair of commuting flow equations,
\begin{align}\label{two_commuting_flows}
    \frac{\partial \mathcal{H}}{\partial \lambda} = - \mathcal{O}_{T^2} \, , \qquad \frac{\partial \mathcal{H}}{\partial \gamma} = - \mathcal{R} \, .
\end{align}
The minus signs in (\ref{two_commuting_flows}) arise because we are deforming the Hamiltonian rather than the Lagrangian, which reverses the sign of the deforming operators as discussed around equation (\ref{hamiltonian_flow1}). This result is the six-dimensional analogue of the statement that the family of ModMax-Born-Infeld theories in four space-time dimensions satisfies two commuting $\TT$-like and root-$\TT$ like flow equations \cite{Babaei-Aghbolagh:2022uij,Ferko:2022iru,Ferko:2023ruw}.\footnote{A similar result holds in $2d$ for the so-called ``Modified-Nambu-Goto'' scalar theories, which reduce to the Nambu-Goto Lagrangian for $\lambda = 0$ and a ModMax-like scalar theory when $\lambda = 0$ \cite{Ferko:2022cix,Conti:2022egv,Babaei-Aghbolagh:2022leo}.} Note that the form of the root-$T\overbar{T}$ equation in \eqref{two_commuting_flows} implies that $\mathcal H_\gamma\leq 0$ in compliance with the comment in item (iv) of Section \eqref{PSTac}.

As we mentioned above, in this analysis we have not assumed that the Hamiltonian $\mathcal{H} ( s, p )$ satisfies the Lorentz invariance condition (\ref{Li}). However, the two-parameter family (\ref{6d_modmax_born_infeld_hamiltonian}) does satisfy this condition, and more generally one can show that any deformation of a Lorentz-invariant seed Hamiltonian $\mathcal{H}_0$ by a flow of the form (\ref{hamiltonian_flow1}) also preserves this condition. For Lorentz-invariant Hamiltonians, the two operators $\mathcal{O}_{T^2}$ and $\mathcal{R}$ of equation (\ref{two_6d_operators_hamiltonian}) simplify to
\begin{align}\label{LorentziOs}
    \mathcal{O}_{T^2} &=  \frac{1}{2} \left((s^2-p^2) \mathcal H_s^2- (\mathcal{H}- s \mathcal{H}_s - p \mathcal{H}_p )^2 \right) \, , \nonumber \\
    \mathcal{R}^2 &=  \left( s^2 -  p^2 \right) \mathcal{H}_s^2 \, .
\end{align}
In fact, when imposing the Lorentz-invariance condition, it turns out that, for a given $\mathcal H$, the two invariants $\tensor{T}{^\mu_\mu}$ and $T^{\mu \nu} T_{\mu \nu}$ become functionally dependent
\be\label{trTT=f(trT)}
T_{\mu\nu}T^{\mu\nu}=\frac 13(3\mathcal H+T_{\mu}{}^{\mu})^2+3(\mathcal H^2-2p^2),
\ee
and indeed that \emph{any} pair of Lorentz-invariant functions of $s$ and $p$ obey some functional relation. This will be discussed in further detail in Section \ref{sec:invariants}.

\subsection{Lagrangian flow equations} \label{Lflows}

We established, around equation (\ref{hamiltonian_flow}), that -- up to a sign -- deformations of the Lagrangian are equivalent to those of the Hamiltonian. It therefore follows that any Lagrangian formulation of the two-parameter family of ModMax-Born-Infeld-like chiral tensor theories will also admit flow equations similar to those in equation (\ref{two_commuting_flows}). Nonetheless, we find it instructive to verify this claim in the various formulations that have been treated in Sections \ref{PSTf}, \ref{sec:INZ}, and \ref{clone}. All three of these formalisms are equivalent and, by construction, preserve manifest Lorentz invariance. We now discuss flows in each formulation in turn.  One interesting aspect of our analysis is that, by investigating the Lagrangian analogue of Hamiltonian flows given above, we are able to obtain new explicit Lagrangian descriptions for theories with Born-Infeld-type interaction functions (see equations (\ref{hypergeometric_solution}) and (\ref{EBI_soln})) by solving sets of differential equations.

\subsubsection*{\ul{\it Lagrangian PST formulation}}
In this formulation (described in Section \ref{PSTf}), the generic Lagrangian flow equation is
\be\label{GeneralFlowPST}
\frac{\partial \mathcal{L}_{\text{PST}}}{\partial \lambda}=-\frac{\partial \mathcal{H}}{\partial \lambda} =   \mathcal{O} \left( \tensor{T}{^\mu_\mu}, T^{\mu \nu} T_{\mu \nu} , \lambda \right) \, .
\ee
Such a deformation modifies only the interaction function $\mathcal{H}(s,p)$ in the PST action \eqref{PST}, but not the term proportional to $E^{\mu \nu} B_{\mu \nu}$. The properties of these flows are essentially identical to those in the Hamiltonian PST formulation discussed above; deformations of the form (\ref{GeneralFlowPST}) reduce to these Hamiltonian flows in the gauge $v^\mu = \tensor{\delta}{^\mu_0}$, and conversely, writing the flows in Lagrangian language is merely restating these deformations in a manifestly Lorentz-invariant way by restoring the auxiliary field $v^\mu$.

In particular, the Lorentz scalars ${T}_\mu{}^{\mu}$ and $T^{\mu \nu} T_{\mu \nu}$ in the Lagrangian PST formulation have exactly the same form as in \eqref{hamiltonian_stress_tensor_evaluated}, but with $s$ and $p$ (defined in \eqref{sp}) now containing the auxiliary vector field $v^\mu$. Likewise, the expressions for the deforming operators $\mathcal{O}_{T^2}$ of
(\ref{6d_TT_combination}) and $\mathcal{R}$ of (\ref{root6d}) for these Lagrangian flows is exactly the same as in \eqref{LorentziOs} for Lorentz-invariant theories in the Hamiltonian formulation.

In the Lagrangian presentation, the interpretation of the condition (\ref{Li}) satisfied by the function $\mathcal{H} ( s, p )$ is that it guarantees PST gauge invariance of the Lagrangian (rather than Lorentz-invariance of the Hamiltonian). Just as before, it turns out that any deformation of the Lagrangian of the form (\ref{GeneralFlowPST}), which is driven by a function of the stress tensor, preserves this PST gauge-invariance condition. This is because, as we will discuss in detail in Section \ref{sec:invariants}, on the mass shell all functions of $T_{\mu \nu}$ are PST-invariant, $v^\mu$-independent and hence manifestly Lorentz invariant, due to the fact that the on-shell PST energy-momentum tensor does not depend on $v^\mu$ and is therefore fully-fledged Lorentz covariant expression in terms of the chiral field strength components, as was proved in Section \ref{PSTEMon}.

In summary, we conclude that all general statements made about energy-momentum flows in the Hamiltonian formulation are applicable to the Lagrangian PST formulation and vice versa.

\subsubsection*{\ul{\it INZ-type formulation}}

We will now consider flows in the INZ-type formalism, which includes an auxiliary self-dual three-form field $\Lambda_3$, as discussed in Section \ref{sec:INZ}.

To construct flows, we will begin with the off-shell stress tensor (\ref{EMTINZn2}) for a general model, which is described by an interaction function $\mathcal{V} ( I )$. This off-shell stress tensor, and invariants built from it such as $T^{\mu \nu} T_{\mu \nu}$ and $\tensor{T}{^\mu_\mu}$, depend upon both the field $B_{\mu \nu}$ and upon the auxiliary field $\Lambda_{\mu \nu \rho}$, where the latter is used to construct interactions via the invariant $I(\Lambda)$ defined in eq. \eqref{I}. To obtain closed flow equations for a deformed interaction function $\mathcal{V} ( I )$, we should replace $B_{\mu \nu}$ with $\Lambda_{\mu\nu\rho}$ using the auxiliary field equation of motion, which (after contraction with $v^\mu$) takes the form (\ref{L=B}). The logic is very similar to the study of flows  for $4d$ duality invariant non-linear electrodynamics described in \cite{Ferko:2023wyi} by using the auxiliary field formulation of \cite{Ivanov:2002ab,Ivanov:2003uj}.

After eliminating $B_{\mu \nu}$ in favor of $\Lambda_{\mu \nu \rho}$ in \eqref{EMTINZn2} in this way, one finds the following expressions for the two invariants constructed from the stress tensor as functions of $I$:
\begin{align}\label{trTTINZ}
    \tensor{T}{^\mu_\mu} &= 6 \left( 2 I \mathcal{V}_I - \mathcal V \right) \, , \nonumber \\
    T^{\mu \nu} T_{\mu \nu} &= 6 (2 I  \mathcal{V}_I-\mathcal{V})^2 + \frac{I}{16} \left( 1 - 96 I \mathcal{V}_I^2 \right)^2 \, .
\end{align}
Note that both of the  quantities (\ref{trTTINZ}) depend (via $I(\Lambda)$) only on the auxiliary self-dual field $\Lambda_3$, which is inert under the PST symmetry \eqref{PST2INZ}. Furthermore, by equation ({\ref{L=F*}), $\Lambda_3$ is equal to the self-dual part $F_3^+$ of the physical field strength  on the mass shell. Therefore, by construction, these quantities are independent of the auxiliary field $v^\mu$ on-shell. We will see later that any quantity which is independent of $v^\mu$ on-shell can be expressed in terms of stress tensor invariants such as those in (\ref{trTTINZ}).

Exactly as in the Hamiltonian formulation (and in the Lagrangian PST formalism), one finds that higher traces of the INZ energy-momentum tensor \eqref{EMTINZn2}, with $B_{\mu\nu}$ expressed as a function of $\Lambda_{\mu\nu}$ in \eqref{L=B}, are dependent upon these two invariants (\ref{trTTINZ}).

We can now construct the operators driving the $\TT$-like and root-$\TT$-like flows, eqs.\eqref{6d_TT_combination} and \eqref{root6d}, as functions of the invariant $I(\Lambda)$:
\begin{align}
    \mathcal{O}_{T^2}& =\frac 1{12}   \left(T^{\mu \nu} T_{\mu \nu} - \frac{1}{3} \left( \tensor{T}{^\mu_\mu} \right)^2 \right) \nonumber \\
    &= \frac 1{2} \left( \frac{I}{96} \left( 1 - 96 I \mathcal{V}_I^2 \right)^2- (2 I  \mathcal{V}_I-\mathcal{V})^2\right) \, , \nonumber \\
    \mathcal R^2 &=   \frac 16\left(T^{\mu \nu} T_{\mu \nu} - \frac{1}{6} \left( \tensor{T}{^\mu_\mu} \right)^2 \right) \nonumber \\
    &= \frac{I}{96} \left( 1 - 96 I \mathcal{V}_I^2 \right)^2 \, . \label{TTroot}
\end{align}
The root-$\TT$ flow equation will then be
\begin{align}
  -  \frac{\partial \mathcal{V}}{\partial \gamma}(\gamma,I) &= \mathcal R
    = \frac{1}{4 \sqrt{6}} \sqrt{I} |1 - 96 I \mathcal{V}_I^2 | \, ,
\end{align}
and the solution with initial condition $\mathcal{V} ( \gamma = 0 ) = 0$ is
\begin{align}\label{MMcase}
    \mathcal{V} ( \gamma, I ) = - \frac{1}{2 \sqrt{6}} \tanh \left( \frac{\gamma}{2} \right) \sqrt{ I } \, ,
\end{align}
which is exactly the $6d$ counterpart \eqref{VMM} of the ModMax interaction function in the $\nu$ frame of the four-dimensional Ivanov-Zupnik setting \cite{Kuzenko:2021cvx,Ferko:2023wyi} derived in Section \ref{examples} from the PST formulation.

Next let us consider the flow equation driven by the irrelevant operator $O_{T^2}$,
\be\label{IOT2}
    - \frac{\partial \mathcal{V} (\lambda,I) }{\partial \lambda} = \mathcal O_{T^2}^{(6d)} = \frac 1{2} \left( \frac{I}{96} \left( 1 - 96 I \mathcal{V}_I^2 \right)^2- (2 I  \mathcal{V}_I-\mathcal{V})^2\right) \, ,
\ee
again with the initial condition $\mathcal{V} ( \lambda = 0 ) = 0$. As in the ModMax case we have just considered, one might expect that the solution $\mathcal{V} ( I )$ to this $6d$ flow equation will correspond to a similar IZ-type interaction functions $E ( a )$ in the $\nu$ frame, as defined in equation (\ref{ivanov_zupnik_lagrangian}). As a check of this statement, one can verify that the flow equation (\ref{IOT2}) is identical, up to numerical factors, to the one satisfied by the $\nu$-frame Born-Infeld interaction function under a $T^2$ flow \cite{Ferko:2023wyi}, which we record here for comparison:
\begin{align}\label{EBI_flow_equation}
    \frac{\partial E_{\text{BI}} ( a , \lambda)}{\partial \lambda} = \mathcal{O}_{T^2}^{(4d)} = \frac{1}{2} \left( a\left(1-a(\mathcal{E}_{a})^2\right)^2
    - \left(2a\mathcal{E}_{a} - \mathcal{E} \right)^2 \right) \, .
\end{align}
It is straightforward to see that the differential equations (\ref{IOT2}) and (\ref{EBI_flow_equation}) are the same after reversing the sign of the function and identifying the variables as
\begin{align}\label{a_to_I}
    a = \frac{I}{96} \, .
\end{align}
We can study the solution to either of these equivalent flow equations perturbatively by first making the ansatz
\begin{align}
    \mathcal{V} ( \lambda, I ) = - \frac{1}{\lambda} f \left( \lambda^2 I \right) \, ,
\end{align}
and then defining $x = \frac{\lambda^2 I}{96}$. To the first few orders in $x$, the solution is
\begin{align}
    f(x) = \frac{x}{2} - \frac{x^2}{8} + \frac{3 x^3}{32} - \frac{13 x^4}{128} + \frac{17 x^5}{128} + \mathcal{O} ( x^6 ) \, .
\end{align}
As expected, this exactly matches the Taylor series expansion of the interaction function $E_{\text{BI}} (a , \lambda)$, which was defined implicitly in (\ref{EBI_algebraic}), for the Born-Infeld theory in the $4d$ Ivanov-Zupnik (IZ) formalism. This series expansion is recorded, for instance, in equation (6.30) of \cite{Ferko:2023wyi}. Thus we find that, up to the scaling of variables, the $\TT$-like and root-$\TT$-like flows in the $6d$ INZ formalism have precisely the same behavior as the corresponding flows in the $4d$ IZ formalism. Again, we note that the role of the interaction function $\mathcal{E}$ in $4d$ is played by the function $\mathcal{V} ( I )$ in $6d$.

It is also possible to present the solution to the flow equation (\ref{IOT2}) in another form, which has appeared in other studies of $\TT$-like flows. Remarkably, the differential equation (\ref{IOT2}) admits the following closed-form solution in terms of a  hypergeometric function:
\begin{align}\label{hypergeometric_solution}
    \mathcal{V} ( \lambda, I ) = \frac{3}{2 \lambda} \left( {}_{3} F_2 \left( - \frac{1}{2} , - \frac{1}{4} , \frac{1}{4} ; \frac{1}{3} , \frac{2}{3} ; - \frac{2}{81} \lambda^2 I \right) - 1 \right) \, .
\end{align}
The same hypergeometric function, with the same parameters, has appeared in the study of the $\TT$ deformation of Yang-Mills in two space-time dimensions \cite{Conti:2018jho,Brennan:2019azg}, and in deformations of one-dimensional theories by functions of the Hamiltonian \cite{Gross:2019ach,Gross:2019uxi,Ferko:2023ozb,Ferko:2023iha,Bhattacharyya:2023gvg}. In the present context, the reason for the appearance of this function can be understood from the fact that this hypergeometric function satisfies a certain fourth-order algebraic constraint equation.\footnote{The observation that, for special values of their parameters, some hypergeometric functions satisfy algebraic constraints of this form was first made by Schwarz \cite{Schwarz1873}.} Explicitly, if we define
\begin{align}
    F ( x ) = {}_{3} F_2 \left( - \frac{1}{2} , - \frac{1}{4} , \frac{1}{4} ; \frac{1}{3} , \frac{2}{3} ; - \frac{2}{81} x \right) - 1 \, ,
\end{align}
then one has the constraint
\begin{align}
    2 x + 3 \left( 1 - \sqrt{ 1 - 6 F } \right) \left( 3 + \sqrt{ 1 - 6 F } \right)^3 = 0 \, .
\end{align}
This function $F(x)$ can therefore be characterized as the root of an algebraic equation. Alternatively, one can write this quartic constraint in terms of the derivative
\begin{align}
    F' ( x ) = - \frac{1}{288} {}_3 F_2 \left( \frac{1}{2} , \frac{3}{4} , \frac{5}{4} ; \frac{4}{3} , \frac{5}{3} ; \frac{- 2}{81} x \right) \, ,
\end{align}
which satisfies
\begin{align}\label{quartic_no_roots}
    288 F' ( x ) + \left( 1 - 216 x \left( F'(x) \right)^2 \right)^2 = 0 \, .
\end{align}
The relationship between this hypergeometric function and the Born-Infeld theory, including the quartic identity (\ref{quartic_no_roots}) that the function satisfies, was discussed in \cite{Aschieri:2013nda}. The origin of this algebraic identity can be traced back to the quartic equation (\ref{t_quartic}) obeyed by the function
$t(a)$
in the $4d$ Ivanov-Zupnik formulation.

Since the structure of the $6d$ interaction function $\mathcal{V} ( I )$ studied here is almost identical to that of the $\nu$-frame interaction function describing the $4d$ Born-Infeld theory in the Ivanov-Zupnik formalism, a similar expression yields a closed-form result for the latter:
\begin{align}\label{EBI_soln}
    E_{\text{BI}} ( a ) = - \frac{3}{2 \lambda} \left( {}_{3} F_2 \left( - \frac{1}{2} , - \frac{1}{4} , \frac{1}{4} ; \frac{1}{3} , \frac{2}{3} ; - \frac{64}{27} \lambda^2 a \right) - 1 \right) \, .
\end{align}
This closed-form expression is exactly equivalent to the implicit definition (\ref{EBI_algebraic}) involving the root of a quartic. We see that the function  (\ref{EBI_soln}) is related to the $6d$ analogue (\ref{hypergeometric_solution}) by reversing the overall sign and replacing $a \to \frac{I}{96}$, as expected from (\ref{a_to_I}).

\subsubsection*{\ul{\it Clone formulation}}

Finally, let us now consider flows in the  ``clone'' formulation which was described in \hbox{Section \ref{clone}}.

Although this formalism is on-shell equivalent to the INZ one, stress tensor deformations will have slightly different features in this formulation due to the difference in field content. We saw above that the clone formulation has, in addition to the physical field $F_3 = d A_2$, another tensor field $Q_3 = d \widetilde{A}_2$ and an auxiliary scalar field $a(x)$, both of which appear in the stress tensor via the combination $X_3 = F_3 + a Q_3$. We will therefore need to use an implication of the equations of motion in order to express stress tensor flows as closed differential equations for the interaction function $\mathcal{V} ( I )$, where now $I(X^+)$ is a fourth order invariant  \eqref{IX+} built out of the self-dual part of $X_3$, much as we did when replacing $B_{\mu \nu}$ using the auxiliary field equation of motion for flows in the INZ formulation.

In particular, we recall that the equations of motion in the clone formulation imply the relation (\ref{X-=VX+}), which allows us to eliminate the anti-self-dual part of $X_3$. Let us stress that this equation is only \emph{one} implication of the equations of motion, and that imposing this condition is strictly weaker than assuming that all of the fields satisfy their equations of motion. Importantly, this assumption does \emph{not} imply the equation of motion for either the physical field $F_3$ or the tensor field $Q_3$. Therefore, when we use this condition to formulate flows, one might say that we are going ``partially on-shell'' in the sense that we assume one (but not all) of the implications of the equations of motion.

The energy-momentum tensor for a theory in the clone formalism with interaction function $\mathcal{V}(I(X^+))$ was given in \eqref{EMMkrt_flow2}, which we reproduce again for convenience:
\begin{align} \label{EMMkrt_flow3}
T_{\mu\nu} &= \left(\frac  14X^-_{\mu\rho\sigma}X^-_{\nu}{}^{\rho\sigma}+\frac 12X^{-\lambda\rho}{}_{(\mu}\mathcal V_{\nu)\lambda\rho}\right)+\left(\frac 14 X^+_{\mu\rho\sigma}X^+_{\nu}{}^{\rho\sigma}+ g_{\mu\nu}(2I\mathcal V_I-\mathcal V)\right)\,.
\end{align}
The explicit form of $\mathcal V_{\nu\lambda\rho}$ can be read from \eqref{dV} with $\Lambda_3$ replaced by $X^+_3$.

We can now use the ``partial on-shell'' relation \eqref{X-=VX+} and the identity
$$
\mathcal V^{\lambda\rho}{}_{(\mu}\mathcal V_{\nu)\lambda\rho}=96X^+_{\mu\rho\lambda}X_\nu^{+\,\rho\lambda}\, I(\mathcal V_I)^2\,
$$
to express $T_{\mu\nu}$ entirely as a function of $X^+_3$:
\begin{align} \label{EMMkrt_flow4}
T_{\mu\nu} &= \frac 14 X^+_{\mu\rho\sigma}X^+_{\nu}{}^{\rho\sigma}\left(1-24I(\mathcal V_I)^2\right) + g_{\mu\nu}(2I\mathcal V_I-\mathcal V)\,.
\end{align}
Then one can check that $T_\mu{}^\nu$ and $T_{\mu\nu}T^{\mu\nu}$ have exactly the same form as \eqref{trTTINZ} but now as functions of the forth order invariant $I(X^+)$. Therefore, all the general discussion regarding the stress energy flows in this formulation goes through exactly as in the INZ-type case. For instance, the solutions to the $\TT$-like and root-$\TT$-like flows in equations (\ref{MMcase}) and (\ref{hypergeometric_solution}) will be identical in the clone formulation, where all that has changed is that now $I$ depends on $X^+_3$ rather than $\Lambda_3$

An important lesson that we learned from the above consideration is that for constructing suitable operators defining stress-tensor flows one should use an appropriate implication of the equations of motion (algebraic in the INZ case, and dynamical in the ``clone" case) to bring the energy-momentum tensor to a form in which it only depends on the self-dual tensor from which the interaction function is built in these formulations.

\subsection{Generalities on deformations and  invariants}\label{sec:invariants}

There is a close formal analogy between stress tensor flows for six-dimensional chiral tensor theories of the form considered in this work, and  corresponding deformations of four-dimensional theories of duality-invariant electrodynamics. Some of the properties of stress tensor deformations in the latter context have been investigated in \cite{Ferko:2023wyi}; for instance, it was shown that any parameterized family of duality-invariant theories of electrodynamics in $4d$ can be described as a stress tensor flow. In this section we will briefly comment on the corresponding statements in the $6d$ chiral two-form context.

The properties of interest for this discussion can be stated in any of the three equivalent formulations described above.  Let us first focus on the PST formalism, for concreteness. Given a PST model with interaction function $\mathcal{H} ( s, p )$ satisfying the differential equation (\ref{Li}), consider an infinitesimal deformation of this interaction function,
\begin{align}
    \mathcal{H} ( s, p ) \to \mathcal{H} ( s, p ) + \lambda \mathcal{O} ( s, p ) \, .
\end{align}
In order to preserve the PST gauge invariance condition (\ref{Li}) to leading order in $\lambda$, this function $\mathcal{O} ( s, p )$ must satisfy
\begin{align}\label{PST_admissible_deformation}
    \mathcal{H}_s \mathcal{O}_s + \frac{s}{p} \left( \mathcal{H}_p \mathcal{O}_s + \mathcal{H}_s \mathcal{O}_p \right) + \mathcal{H}_p \mathcal{O}_p = 0 \, .
\end{align}
First, one can show by direct computation that if $\mathcal{O}$ depends on $s$ and $p$ only through the two quantities $\tensor{T}{^\mu_\mu}$ and $T^{\mu \nu} T_{\mu \nu}$ given in equation (\ref{hamiltonian_stress_tensor_evaluated}), then $\mathcal{O}$ identically satisfies the differential equation (\ref{PST_admissible_deformation}), assuming that the undeformed interaction $\mathcal{H}$ itself obeys (\ref{Li}). This means that any Lorentz scalar constructed from the energy-momentum tensor gives rise to a ``consistent deformation'' of the PST interaction function, in the sense that such a deformation preserves PST gauge invariance to leading order. One can also argue for this conclusion by noting that the stress tensor is independent of $v^\mu$ on-shell, as we demonstrated around equation (\ref{EMPSTOS}).

Indeed, the condition (\ref{PST_admissible_deformation}) is actually equivalent to the statement that the operator $\mathcal{O}$ is independent of the auxiliary field $v_\mu$ on the mass shell. That is, the constraint
\begin{align}
    \delta_v \mathcal{O} \big\vert = 0 \, ,
\end{align}
where $\big\vert$ indicates that the fields satisfy their equations of motion, implies that
\begin{align}\label{epsilon_H_O_B}
    \varepsilon^{\mu \nu \rho \sigma \kappa \delta} v_\nu H_{\rho \sigma} \frac{\partial \mathcal{O}}{\partial B^{\kappa \delta}} = 0 \, ,
\end{align}
which, after some algebraic manipulations, yields the condition
\begin{align}\label{later_equivalent_condition}
    0 &= \left[ \mathcal{O}_s \left( \mathcal{H}_s + \frac{s}{p} \mathcal{H}_p \right) + p^{-1} \mathcal{O}_p \left( p \mathcal{H}_p + s \mathcal{H}_s \right) \right] p^\mu \delta v_\mu \, .
\end{align}
Equation (\ref{later_equivalent_condition}) is satisfied for arbitrary $\delta v_\rho$ if and only if (\ref{PST_admissible_deformation}) holds. Furthermore, note that (\ref{epsilon_H_O_B}) can also be obtained by expanding the PST gauge invariance condition (\ref{HH=BB}) under a deformation by the operator $\mathcal{O}$, which means that this condition also expresses the assumption that $\mathcal{O}$ be invariant under such gauge transformations. Therefore, we may characterize a ``consistent deformation'' $\mathcal{O}$ in any of the following equivalent ways:
\begin{enumerate}
    \item The interaction function $\mathcal{H} ( s, p ) + \lambda \mathcal{O} ( s, p ) $ satisfies the differential equation (\ref{Li}) to leading order in $\lambda$.

    \item The operator $\mathcal{O} ( s, p )$ obeys the ``consistent deformation'' condition (\ref{PST_admissible_deformation}).

    \item The function $\mathcal{O} ( s, p )$ is independent of the auxiliary field $v_\mu$ on the mass shell.

    \item The operator $\mathcal{O} ( s, p )$ is invariant under the PST gauge transformations (\ref{PST2}).
\end{enumerate}
In Section \ref{invobs}, we will generalize these notions of consistent deformations to theories of chiral $(2n)$-forms in $d = 4 n + 2$ dimensions.

The statement that an invariant observable $\mathcal{O}$ generates a consistent deformation of a chiral tensor theory, to leading order in the deformation parameter, extends to all orders in a natural way. Consider a one-parameter family of PST interaction functions $\mathcal{H} ( s, p ; \lambda )$ which obey a differential equation
\begin{align}\label{consistent_flow_equation}
    \frac{\partial \mathcal{H} ( s, p; \lambda ) }{\partial \lambda} = \mathcal{O} ( s, p ; \lambda ) \, ,
\end{align}
along with an initial condition $\mathcal{H} ( s, p ; 0 )$ which satisfies (\ref{Li}). Then all of the functions $\mathcal{H} ( s, p ; \lambda )$ will satisfy the gauge invariance condition (\ref{Li}) if, at each value of $\lambda$, the function $\mathcal{O} ( s, p ; \lambda )$ obeys the differential equation (\ref{PST_admissible_deformation}) with respect to the interaction function $\mathcal{H} ( s, p ; \lambda )$ at the same value of $\lambda$. This condition guarantees that the solution to any stress tensor flow equation, with an appropriately chosen initial condition, gives a consistent model in the PST formulation. We have already seen one example in the solution (\ref{6d_modmax_born_infeld_hamiltonian}), the ModMax-Born-Infeld interaction function, which is generated by stress tensor deformations of a free model and which indeed obeys (\ref{Li}).

Furthermore, given any two invariant observables $\mathcal{O}_1$ and $\mathcal{O}_2$, we claim that there exists a functional relation of the form
\begin{align}
    \Upsilon \left( \mathcal{O}_1 , \mathcal{O}_2 \right) = 0 \, .
\end{align}
One can prove this statement in several ways; one approach, following the proof of Theorem 2 in \cite{Ferko:2023wyi},
begins by defining two vector fields
\begin{align}
    \vec{f}_1 = \frac{\partial \mathcal{O}_1}{\partial s} \partial_s + \frac{\partial \mathcal{O}_1}{\partial p} \partial_p \, , \qquad \vec{f}_2 = \frac{\partial \mathcal{O}_2}{\partial s} \partial_s + \frac{\partial \mathcal{O}_2}{\partial p} \partial_p \, ,
\end{align}
in the $(s, p)$ plane. One can check that both of these vector fields are non-vanishing, assuming that the interaction function $\mathcal{H} ( s, p )$ is not a constant. Since the operators $\mathcal{O}_1$, $\mathcal{O}_2$ both satisfy the condition (\ref{PST_admissible_deformation}), they are each orthogonal to the vector field
\begin{align}
    \vec{v} = \left( s \mathcal{H}_p + p \mathcal{H}_s \right) \partial_s + \left( s \mathcal{H}_s + p \mathcal{H}_p \right) \partial_p \, .
\end{align}
But if two vector fields in two dimensions are both orthogonal to a third vector field, then the two must be parallel, which implies that $\vec{f}_1 = \alpha \vec{f}_2$ for some constant $\alpha$. It follows that the two operators $\mathcal{O}_1$, $\mathcal{O}_2$ are functionally dependent, as claimed.

A different way to prove this functional dependence is to use the INZ description of a chiral tensor theory. In this formulation, any two consistent deformations $\mathcal{O}_1$, $\mathcal{O}_2$ must be functions only of the quartic invariant $I$. But by the inverse function theorem, on an open interval around a fixed value $I = I_0$, each of the functions $\mathcal{O}_1 ( I )$, $\mathcal{O}_2 ( I )$ can be inverted to write $I ( \mathcal{O}_1 )$, $I ( \mathcal{O}_2 )$ (so long as the derivatives of these functions are non-vanishing). We may then express either of these operators in terms of the other, for instance by writing
\begin{align}
    \mathcal{O}_2 ( I ) = \mathcal{O}_2 \left( I \left( \mathcal{O}_1 \right) \right) \, .
\end{align}
Thus one can, away from exceptional points, locally invert to express one of the operators in terms of the other, which again means that they are functionally dependent.

Since any Lorentz scalar constructed from the stress tensor is an invariant observable, the relation (\ref{stress_tensor_dependence}) which we stated earlier follows as a special case of this argument. In particular, this means that \emph{any} invariant observable $\mathcal{O}$ can be expressed as a function of the stress tensor, which means that a general consistent deformation (\ref{consistent_flow_equation}) can always -- at least locally and away from singular points -- be written in a form
\begin{align}\label{deformations_are_TT_flows}
    \frac{\partial \mathcal{H} ( s, p ; \lambda )}{\partial \lambda} = \mathcal{O} \left( s , p ; \lambda \right) \overset{!}{=} f \left( T_{\mu \nu}^{(\lambda)} , \lambda \right) \, .
\end{align}
We conclude that any parameterized family $\mathcal{H} ( s, p ; \lambda )$ of chiral tensor theories can be viewed as a generalized $\TT$-like flow, since taking a derivative with respect to $\lambda$ must necessarily yield a family of invariant observables $\mathcal{O}^{(\lambda)}$ which can then be expressed in terms of the stress tensor as in (\ref{deformations_are_TT_flows}).

One could also make an entirely analogous argument within the equivalent clone formulation. Here the only difference is that the invariant observables are functions of $I ( X^+ )$, which is a quartic invariant constructed from the self-dual part of $X_3$, rather than functions of the corresponding fourth-order invariant built from $\Lambda_3$.

All of the preceding remarks apply to manifestly Lorentz-invariant Lagrangian formulations of $6d$ chiral tensor theories, such as the PST, INZ, and clone formalisms. To conclude this section, let us briefly point out that entirely analogous statements hold within the Hamiltonian formulation, where the role of an ``invariant observable'' is now played by a \emph{Lorentz}-invariant observable, and the notion of ``consistent deformation'' now refers to a deformation which preserves Lorentz invariance. For instance, a function $\mathcal{O} ( B_{ij} )$ of the spatial $2$-form $B_{ij}$ is Lorentz-invariant if and only if it satisfies the condition
\begin{align}
    \varepsilon^{l m i j k} H_{j k} \frac{\partial \mathcal{O}}{\partial B^{lm} } = 0 \, ,
\end{align}
which is equivalent to the relation
\begin{align}\label{lorentz_invariant_function}
    0 = \mathcal{H}_s \mathcal{O}_s + \frac{s}{p} \left( \mathcal{H}_p \mathcal{O}_s + \mathcal{H}_s \mathcal{O}_p \right) + \mathcal{H}_p \mathcal{O}_p \, ,
\end{align}
where in this equation the symbols $s$ and $p$ now refer to the $SO(5)$ scalars (\ref{spatial_s}) and (\ref{spatial_p}). Likewise, every Lorentz-invariant function (\ref{lorentz_invariant_function}) gives rise to a deformation of the Hamiltonian as $\mathcal{H} ( s, p ) \to \mathcal{H} ( s, p ) + \lambda \mathcal{O} ( s, p )$ which preserves Lorentz invariance to leading order in $\lambda$, and any two such Lorentz-invariant observables obey a functional relation, which means that any such $\mathcal{O}$ can be expressed in terms of the stress tensor.

\section{Self-interacting chiral $2n$-forms in $d=4n+2$ dimensions}
\label{hdINZ}

This section is devoted to the study of general models for self-interacting chiral $2n$-forms in $d= 4n + 2\equiv 2p +2$ dimensions, with $n \in \mathbb N$. First, we will review the work
\cite{Buratti:2019guq}, in which Buratti, Lechner and Melotti (BLM) extended the PST formalism beyond six dimensions. Then, we will describe consistent deformations
of models for self-interacting chiral gauge $p$-forms. Finally, we will present a new (INZ-type) formulation for such theories.

Let $A_{\mu ( p ) } = A_{\m_1 \dots \m_p}$ be a gauge $p$-form potential\footnote{Throughout this section, we often make use of the condensed notation  $ T_{\mu ( k ) } $ and $T^{\m(k)}$ for rank-$k$ antisymmetric tensors $T_{\mu_1 \ldots \mu_k} = T_{[\mu_1 \ldots \mu_k]} $  and $T^{\mu_1 \ldots \mu_k} = T^{[\mu_1 \ldots \mu_k]} $, respectively.} on a time orientable space-time $\cM^d$ with metric $g_{\m\n}$,
and let
\begin{align}
    F_{\mu ( p + 1 ) } = ( p + 1 ) \partial_{[\mu_1} A_{\mu_2 \ldots \mu_{p+1} ] }
\end{align}
be the corresponding gauge-invariant field strength.
As in the $d=6$ case, we introduce
a normalized timelike vector field\footnote{The existence of such a vector field is guaranteed on any paracompact time orientable space-time $(\cM^d, g_{\m\n}$), see e.g. \cite{Hawking:1973uf, Wald:1984rg}.}
$v^\mu$,
\begin{align}
    v^\mu v_\mu = - 1 \, .
\end{align}
Making use of $v^\m$ allows us to associate with $ F_{\mu ( p + 1 ) } $
the electric field
\begin{align} \label{electric}
    E_{\mu ( p ) } = F_{\mu_1 \ldots \mu_p \nu } v^\nu \, ,  \qquad
    E_{\mu_1 \ldots \mu_{p-1} \s } v^\s = 0 \, ,
\end{align}
and the magnetic field
\begin{align}\label{B_constraint}
    B_{\mu ( p ) } = {F}^*_{\mu_1 \ldots \mu_p \nu} v^\nu\, , \qquad
    B_{\mu_1 \ldots \mu_{p-1} \s } v^\s = 0 \, .
\end{align}
Here we have introduced the dual field strength
\begin{align}
    {F}^{*\,\mu ( p + 1 ) } := \frac{1}{(p+1)!} {\bm \varepsilon}^{\mu_1 \ldots \mu_{p+1} \nu_1 \ldots \nu_{p+1}} F_{\nu_1 \ldots \nu_{p+1}} \, ,
    \qquad
    (F^*)^* = F \, ,
\end{align}
where ${\bm \ve}^{\m(d)} $ denotes the Levi-Civita tensor
\bea
{\bm \ve}^{\m_1 \dots \m_d} := \frac{1}{ \sqrt{-g}} \ve^{\m_1 \dots \m_d} \quad \implies \quad
{\bm \ve}_{\m_1 \dots \m_d} =  \sqrt{-g} \ve_{\m_1 \dots \m_d} \, ,
\label{Levi-Civita}
\eea
with $\ve^{\m(d)} $ and $\ve_{\m(d)} $ being the Levi-Civita tensors in Minkowski space.

Choosing $v^\mu$ to be the first element of a basis for the tangent space $T_q \mathcal{M}^d$ at a point $q \in \cM^d$ gives
\begin{subequations}
\begin{align}\label{F_expansion}
    F_{\mu ( p + 1 ) } &= - ( p + 1 ) E_{[ \mu_1 \ldots \mu_{p}} v_{\mu_{p+1} ]} - \frac{1}{p!} {\bm \varepsilon}_{\mu_1 \ldots \mu_{p+1} \nu_1 \ldots \nu_{p+1}} B^{\nu_1 \ldots \nu_p} v^{\nu_{p+1}} \, ,\\
\label{F_tilde_identity}
    {F}^*_{\mu ( p+1 ) } &= - ( p + 1 ) B_{[ \mu_1 \ldots \mu_p} v_{\mu_{p+1} ] } - \frac{1}{p!} {\bm \varepsilon}_{\mu_1 \ldots \mu_{p+1} \nu_1 \ldots \nu_{p+1}} E^{\nu_1 \ldots \nu_p} v^{\nu_{p+1}} \, .
\end{align}
\end{subequations}
These identities lead to
\begin{align}
    F \cdot F &:= F^{\mu_1 \ldots \mu_{p+1}} F_{\mu_1 \ldots \mu_{p+1}}
    = ( p + 1 ) \left( B \cdot B - E \cdot E \right) \, .
\end{align}


\subsection{Extending the PST formalism for non-linear chiral $p$-forms beyond six dimensions}
\label{BLMreview}

To describe the dynamics of a self-interacting chiral gauge $p$-form,
the authors of \cite{Buratti:2019guq} postulated the action
\begin{align}\label{BLM_action}
    S [ A, a ] =
       \int \rd^d x \sqrt{-g}
     \left[ \frac{1}{2 p!} E \cdot B - \cH ( B_{\m(p)}, g_{\m\n} ) \right] \, , \qquad
     v_\mu = \frac{\partial_\mu a}{\sqrt{ - \partial a \cdot \partial a } } \, .
\end{align}
The existence of a scalar field $a(x)$, such that the vector field $v^\m $ defined by \eqref{BLM_action} is past directed and timelike, implies non-trivial conditions on the global causal structure of space-time.
A time-oriented space-time $(\cM^d, g_{\m\n})$ is called stably causal\footnote{Following \cite{Hawking:1973uf, Wald:1984rg}, a space-time $(\cM^d, g_{\m\n})$ is said to be stably causal if there exists a timelike vector field $t^\m $ such that the metric $\tilde{g}_{\m\n} = g_{\m\n} - t_\m t_\n$ possesses no closed timelike curves.} if and only if there exists a smooth function $\cT $ on $\cM^d$ such that $\nabla^\m \cT$
is a past directed timelike vector field \cite{Hawking:1973uf, Wald:1984rg}. Therefore, due to the explicit structure of the timelike vector field $v^\m$, eq. \eqref{BLM_action}, the space-time on which the theory is defined is stably causal. It is also known that (i) every stably causal space-time is strongly causal and  (ii) every globally hyperbolic space-time is strongly causal, see
\cite{Hawking:1973uf, Wald:1984rg, Chrusciel:2020fql} for the technical details.\footnote{However, not every stably causal space-time is globally hyperbolic, see
\cite{Beem:1996xpa}.}
For the globally hyperbolic space-times, the following theorem\footnote{Here we use a reformulation of the
Bernal-Sanchez theorem given in the 2004 lecture notes ``Lorentzian Geometry'' by  Christian B\"ar \cite{bar_notes}.}
holds \cite{Bernal:2004gm}:
\begin{theorem}
Every globally hyperbolic space-time $(\cM, g)$ is isometric to
\bea
\Big( {\mathbb R} \times {\mathfrak S} \, ,  -\b \rd \cT^2 + g_\cT \Big)~,
\eea
where
$\cT: {\mathbb R} \times {\mathfrak S} \to \mathbb R$ is smooth and positive, and
$g_\cT$ is a smooth family of Riemannian metrics on $\mathfrak S$. Furthermore,
$\{ \cT_0 \} \times {\mathfrak S} $ is a smooth and spacelike Cauchy hypersurface
for all $\cT_0$.
\end{theorem}
In order to introduce the Hamiltonian formulation of General Relativity (the ADM formalism), space-time has to be globally hyperbolic, and thus a time function $\cT$ exists. In other words, the existence of the vector $v^\mu$ with the above properties ensures the existence of the Hamiltonian for the system under consideration.

The scalar function $\cH ( B_{\m(d)},  g_{\m\n} )$, associated with the Hamiltonian density of the theory for $\partial_\mu a = \tensor{\delta}{_\mu^0}$,
depends on all possible contractions of $B_{\m(p)} $ with itself, such as
\begin{align}
    \tr \big( \widehat{B}^k \big) \, ,
    \qquad    \widehat{B} = \left( \tensor{B}{_{\mu_1}_{\ldots} _{\mu_{n}}^{\nu_1}^{\ldots}^{\nu_{n}}} \right) \, ,
    \quad k > 1 \, .
\end{align}
This means that $ H_{\mu (p) } = p!\frac{\partial \cH ( B, g ) }{\partial B^{\mu (p)} }$,
defined by
\begin{align} \label{H-der}
    \delta_B \cH ( B , g ) = \frac{1}{p!} \delta B^{\mu_1 \ldots \mu_p} H_{\mu_1 \ldots \mu_p} \, ,
\end{align}
is orthogonal to the vector field $v^\nu$, that is
\begin{align}\label{H_dot_v_constraint}
    H_{\mu_1 \ldots \mu_{p-1} \nu} v^\nu = 0 \, .
\end{align}
 $\cH(B , g)$ should also satisfy a condition which is dynamical in the sense that it guarantees invariance under (properly generalised) PST gauge transformations. To derive this condition, we first compute the variation of \eqref{BLM_action} under arbitrary displacements
\begin{align}
    A_{\mu ( p ) } \to A_{\mu ( p ) } + \delta A_{\mu ( p ) } \, , \qquad
    a \to a + \delta a \, .
    \label{113}
\end{align}
Varying $\cH(B, g) $ gives
\begin{align}\label{delta_H_expression}
    \delta \cH ( B , g) = \frac{1}{(p!)^2} \Big(
    \frac{1}{p+1} {\bm \varepsilon}^{\mu_1 \ldots \mu_p \rho \nu_1 \ldots \nu_{p+1}} \delta F_{\nu_1 \ldots \nu_{p+1}} v_\rho
       + {\bm \varepsilon}^{\mu_1 \ldots \mu_p \sigma \rho \nu_1 \ldots \nu_p}
    E_{\nu_1 \ldots \nu_p} v_\sigma \delta v_\rho \Big) H_{\mu_1 \ldots \mu_p}
     \, .
\end{align}
The kinetic term in the action (\ref{BLM_action}) can be varied using the steps described in detail in \cite{Buratti:2019guq}.
With the notation $\int = \int \rd^d x \sqrt{-g}$,
the outcome is
\begin{align}
    \delta \int \frac{1}{2} E \cdot B =& \int \delta {F}^{*\,\mu_1 \ldots \mu_{p+1}} E_{\mu_1 \ldots \mu_p} v_{\mu_{p+1}} \non \\
&    + \frac{1}{2p!} \int {\bm \varepsilon}^{\mu_1 \ldots \mu_p \sigma_1 \ldots \sigma_p \rho \nu} \left( B_{\mu_1 \ldots \mu_p} B_{\sigma_1 \ldots \sigma_p} + E_{\mu_1 \ldots \mu_p} E_{\sigma_1 \ldots \sigma_p} \right) v_\rho \delta v_\nu \, .
\end{align}
Combining this with the expression (\ref{delta_H_expression}) for $\delta \cH ( B , g)$, we obtain
\begin{align}
    \delta S &= \frac{1}{p!} \int \delta {F}^{*\,\mu_1 \ldots \mu_{p+1}} \left( E_{\mu_1 \ldots \mu_p} - H_{\mu_1 \ldots \mu_p} \right) v_{\mu_{p+1}}
        \\
    &\quad + \frac{1}{\left( p! \right)^2 } \int \varepsilon^{\mu_1 \ldots \mu_p \nu_1 \ldots \nu_p \sigma \rho} \left( \frac{1}{2} E_{\mu_1 \ldots \mu_p} E_{\nu_1 \ldots \nu_p} + \frac{1}{2} B_{\mu_1 \ldots \mu_p} B_{\nu_1 \ldots \nu_p} - H_{\mu_1 \ldots \mu_p} E_{\nu_1 \ldots \nu_p} \right) v_\sigma \delta v_\rho \, . \non
\end{align}
It is useful to introduce
\begin{align}
    \mathbb{E}_{\mu ( p ) } = E_{\mu ( p ) } - H_{\mu ( p ) } \, .
\end{align}
Then, the variation of the action can be rewritten as
\begin{align}
    \delta S &= \frac{1}{p!} \int \delta {F}^{*\,\mu_1 \ldots \mu_{p+1}}
    \mathbb{E}_{\mu_1 \ldots \mu_p} v_{\mu_{p+1}}
    \\
    &\quad + \frac{1}{\left( p! \right)^2 } \int \varepsilon^{\mu_1 \ldots \mu_p \nu_1 \ldots \nu_p \sigma \rho} \left( \frac{1}{2} \mathbb{E}_{\mu_1 \ldots \mu_p}
    \mathbb{E}_{\nu_1 \ldots \nu_p} - \frac{1}{2} H_{\mu_1 \ldots \mu_p}
    H_{\nu_1 \ldots \nu_p} + \frac{1}{2} B_{\mu_1 \ldots \mu_p} B_{\nu_1 \ldots \nu_p} \right) v_\sigma \delta v_\rho \, .
    \non
\end{align}
Integrating by parts,
the first term can be rewritten as follows:
\begin{align}
    \frac{1}{p!} \int \delta {F}^{*\,\mu_1 \ldots \mu_{p+1}}
    \mathbb{E}_{\mu_1 \ldots \mu_p} v_{\mu_{p+1}}
    = - \frac{1}{\left( p ! \right)^2} \int \varepsilon^{\mu_1 \ldots \mu_p \nu_1 \ldots \nu_p \sigma \rho} \delta A_{\nu_1 \ldots \nu_p} \partial_\rho
    \left( \mathbb{E}_{\mu_1 \ldots \mu_p} v_\sigma \right) \, .
\end{align}
As a simple application of the above consideration, we read off the equation of motion for the gauge $p$-form $A_{\m(p)} $
\bea
\partial_{[\m_1} \left( \mathbb{E}_{\mu_2 \ldots \mu_{p+1}} v_{\m_{p+2} ]} \right) =0\,.
\label{AEoM}
\eea

The two types of the PST gauge transformations extend from six to to higher dimensions as follows. The first one is
\begin{align} \label{PSTGT1}
         \delta A_{\mu ( p ) } = p v_{[\mu_1} \psi_{\mu_2 \ldots \mu_p]} \, , \qquad
    \delta a = 0 \,.
\end{align}
The  gauge freedom associated with this symmetry may be fixed in such a way that  \eqref{AEoM}  turns into
 \bea
  \mathbb{E}_{\mu ( p ) } = E_{\mu ( p ) } - H_{\mu ( p ) } =0
\eea
which will be assumed below.

The second PST transformation is
\begin{align} \label{PSTGT2}
         \delta A_{\mu ( p ) } = - \frac{\varphi}{\sqrt{- \partial a \partial a}} \mathbb{E}_{\mu ( p ) } \, , \qquad
    \delta a = \varphi \, .
\end{align}
The $\varphi$-variation \eqref{PSTGT2} of the action proves to be \cite{Buratti:2019guq}
\begin{align}
    \delta_{\varphi} S = \frac{1}{2 ( p! )^2} \int {\bm \varepsilon}^{\mu_1 \ldots \mu_p \nu_1 \ldots \nu_p \sigma \rho} \left( B_{\mu_1 \ldots \mu_p} B_{\nu_1 \ldots \nu_p}
    - H_{\mu_1 \ldots \mu_p} H_{\nu_1 \ldots \nu_p} \right) v_\sigma \delta v_\rho \, .
\end{align}
This vanishes if
\begin{align}
   {\bm  \varepsilon}^{\sigma \rho \mu_1 \ldots \mu_p \nu_1 \ldots \nu_p}
    \left( B_{\mu_1 \ldots \mu_p} B_{\nu_1 \ldots \nu_p} - H_{\mu_1 \ldots \mu_p} H_{\nu_1 \ldots \nu_p} \right) v_\rho = 0 \, .
\end{align}
Since $B_{\mu ( p )}$ and $H_{\mu ( p ) }$ are orthogonal to $v^\mu$, in accordance with (\ref{B_constraint}) and (\ref{H_dot_v_constraint}), the above condition is equivalent to the equation
\begin{align} \label{MasterEquation}
    B_{[\mu_1 \ldots \mu_p} B_{\mu_{p+1} \ldots \mu_{2p} ] }
    = H_{[ \mu_1 \ldots \mu_p } H_{\mu_{p+1} \ldots \mu_{2p} ] } \, ,
\end{align}
which is the dynamical condition on  $\cH ( B_{\m(d)},   g_{\m\n} )$.
Under this condition, the equation of motion for $a$ is identically satisfied
if the equation of motion for $A_{\m(p)}$, eq. \eqref{AEoM}, holds.

Every solution of the equation  \eqref{MasterEquation}
generates a consistent model for a chiral gauge $p$-form.
The case of a free chiral $p$-form corresponds to
\bea
\cH^{\rm free}_{\rm PST}
\left(B_{\m(p)}, g_{\m\n}\right) = \frac{1}{2 p!} B_{\mu (p)} B^{\mu (p)} \, .
\eea
Representing
\bea
\cH = \frac{1}{2 p!} B_{\mu (p)} B^{\mu (p)}  + \D \cH\, ,
\eea
\eqref{MasterEquation} takes the form
\bea
  B_{[\mu_1 \ldots \mu_p} \D H_{\mu_{p+1} \ldots \mu_{2p} ] }
    +\hf  \D H_{[ \mu_1 \ldots \mu_p } \D H_{\mu_{p+1} \ldots \mu_{2p} ] } =0 \, ,
\eea
which is useful for setting up a perturbative scheme to compute $\D \cH$.

In accordance with \eqref{PSTGT2}, the scalar field $a (x)$ is a purely gauge degree of freedom. Its significance, in particular, is that it defines a $(d-1) + 1$ splitting, which foliates curved space-time $\cM^d$ into spacelike hypersurfaces  defined by the level sets of  $a(x)$.


\subsection{Invariant observables and consistent deformations}\label{invobs}

Given a scalar function $\cO(B_{\m(p)} , g_{\m\n}) $, its partial derivative
$ \cO_{\mu (p) } = \frac{\partial \cO ( B, g ) }{\partial B^{\mu (p)} }$ is
defined by
\begin{subequations}
\begin{align}
    \delta_B \cO ( B, g ) = \frac{1}{p!} \delta B^{\mu_1 \ldots \mu_p} \cO_{\mu_1 \ldots \mu_p} \,  .
\end{align}
By construction, it is orthogonal to the vector field $v$, that is
\begin{align}\label{O_dot_v_constraint}
    \cO_{\mu_1 \ldots \mu_{p-1} \nu} v^\nu = 0 \, .
\end{align}
\end{subequations}
The function $\cO(B)$ is said to be an `invariant observable' if it obeys the first-order differential equation
\begin{align}
   \cO_{[ \mu_1 \ldots \mu_p } H_{\mu_{p+1} \ldots \mu_{2p} ] } =0 \, .
   \label{PhysicalObservable}
\end{align}
The reason for the name is that on the mass shell such quantities are $v^\mu$-field independent and hence Lorentz (or general coordinate) invariant.

Indeed, let us give a small displacement to the timelike vector field, $v_\m \to v_\m +\d v_\m$, with $v^\m \d v_\m =0$. Then $B^{\m(p)}$ varies as
\bea
\d_v B^{\m(p)} &=&   {F}^{*\,\mu_1 \ldots \mu_p \nu} \d v_\nu
=- ( p + 1 ) B^{[ \mu_1 \ldots \mu_p} v^{\n ] } \d v_\n
- \frac{1}{p!} {\bm \varepsilon}^{\mu_1 \ldots \mu_{p} \r \nu_1 \ldots \nu_{p+1}} E_{\nu_1 \ldots \nu_p} v_{\nu_{p+1}} \d v_\r
\non \\
&=& -  B^{ \n [ \m_1 \dots \m_{p-1} } v^{\m_p ]} \d v_\n
- \frac{1}{p!} {\bm \varepsilon}^{\r \mu_1 \ldots \mu_{p}  \nu_1 \ldots \nu_{p+1}} E_{\nu_1 \ldots \nu_p} v_{\nu_{p+1}} \d v_\r \, .
\eea
Due to \eqref{O_dot_v_constraint}, the first term in this variation does not contribute to
$\d_v \cO$,
\bea
\d_v \cO = \frac{1}{p!} \delta_v B^{\mu_1 \ldots \mu_p} \cO_{\mu_1 \ldots \mu_p}
= - \frac{1}{(p!)^2} {\bm \varepsilon}^{\r \mu_1 \ldots \mu_{p}  \nu_1 \ldots \nu_{p+1}} \cO_{\mu_1 \ldots \mu_p} E_{\nu_1 \ldots \nu_p} v_{\nu_{p+1}}
\d v_\r \,.
\eea
On the mass shell, $E_{\m(p)} = H_{\m(p)}$, and therefore the variation takes the form
\bea
\d_v \cO \big|= - \frac{1}{(p!)^2} {\bm \varepsilon}^{\r \s \mu_1 \ldots \mu_{p}  \nu_1 \ldots \nu_{p}} \cO_{\mu_1 \ldots \mu_p} H_{\nu_1 \ldots \nu_p} v_{\s}
\d v_\r \, ,
\eea
where $\cO |$ means that the fields obey their equation of motion.
Requiring this variation to vanish for arbitrary $\d v_\r$ leads to
the condition
\bea
{\bm \varepsilon}^{\r \s \mu_1 \ldots \mu_{p}  \nu_1 \ldots \nu_{p}} \cO_{\mu_1 \ldots \mu_p} H_{\nu_1 \ldots \nu_p} v_{\s} =0\, ,
\label{v-ind}
\eea
which is equivalent to \eqref{PhysicalObservable}. Vice versa, if $\cO$ obeys eq.  \eqref{PhysicalObservable}, then the variation $\d_v \cO \big|$ vanishes and $\cO$ is $v$-independent on the mass shell.

There is another equivalent interpretation of \eqref{PhysicalObservable} as the condition of Lorentz (or general coordinate) invariance of $\cO$. This property is naturally visualised either by choosing $\cM^d$ to be Minkowski space, ${\mathbb M}^d$,  or by making use of a Lorentz-invariant formulation for gravity-matter systems in which: (i) the gravitational field is described by a vielbein $e_\m{}^a (x)$ such that the space-time metric $g_{\m\n}(x)$ is a composite field, $g_{\m\n}= e_\m{}^a e_\n{}^b \eta_{ab}$, with $\eta_{ab} $ the Minkowski metric; and (ii) each matter field is a scalar with respect to general coordinate transformations and a tensor with respect to the local Lorentz group.
For simplicity, we restrict our analysis to ${\mathbb M}^d$.
Given a scalar function $\cO(B_{\m(p)} , \eta_{\m\n}) $, it is Lorentz invariant if both
$F_{\m(p+1)} $ and $v_\m$ transform according to their tensorial structure.
We can choose a gauge $v_\m = \d_\m{}^0$, and then $B^{\m(p)}$ defined by \eqref{B_constraint} has only space components,
\bea
B^{\m(p)} \to B^{i(p)} = \frac{1}{(p+1)!}  \ve^{i_1 \dots i_p 0 j_1 \dots j_{p+1} } F_{j_1 \dots j_{p+1} }
= -   \frac{1}{(p+1)!}  \ve^{i_1 \dots i_p  j_1 \dots j_{p+1} } F_{j_1 \dots j_{p+1} } \, .
\eea
In this gauge,  $\cO(B_{\m(p)} , \eta_{\m\n}) $ turns into $\cO(B_{i(p)} , \d_{i j})  \equiv \cO(B^{i(p)} ) $.

Consider an infinitesimal Lorentz boost generated by parameters
$\o_{0i} = - \o_{i0} \equiv \o_i$. It acts on $F_{ i(p+1)} $ as follows
\bea
\d_\o F_{ i(p+1)} = -(p+1) \o_{[i_1} E_{i_2 \dots i_{p+1}]} \quad
\implies \quad
\d_\o B^{ i(p)} = \frac{1}{p!} \ve^{i_1 \dots i_p j_1 \dots j_{p+1} }
\o_{j_1} E_{j_2 \dots j_{p+1}} \, ,
\eea
with $E^{i(p)} = F^{i(p)0}$. On the mass shell, $\d_\o B^{ i(p)} $ can be rewritten
as
\bea
\d_\o B^{ i(p)} \big|= \frac{1}{p!} \ve^{i_1 \dots i_p j_1 \dots j_{p+1} }
\o_{j_1} H_{j_2 \dots j_{p+1}} \, .
\eea
Now, the condition of Lorentz invariance of $\cO(B^{i(p)} ) |$ is
\bea
 \ve^{i_1 \dots i_p j_1 \dots j_{p} k }
\frac{\pa \cO}{\pa B^{i_1 \dots i_p} } H_{j_1 \dots j_{p}} =0\, ,
\eea
which coincides with \eqref{v-ind}.

Invariant observables naturally arise as follows. The interaction term in \eqref{BLM_action}
may depend on a parameter $\g$,
\bea
    S [ A, a; \g ] =
        \int \rd^d x \sqrt{-g}
     \left[ \frac{1}{2 p!} E \cdot B - \cH ( B_{\m(p)}, g_{\m\n}; \g ) \right]  \, ,
\eea
such that $\cH ( B_{\m(p)}, g_{\m\n} ; \g ) $ is a solution of the equation \eqref{MasterEquation} for every value of the parameter. Differentiating \eqref{MasterEquation} with respect to $\g$ leads to \eqref{PhysicalObservable} in which
\bea
\cO  = \frac{\pa}{\pa \g} \cH(B, g;\g)\, .
\eea
And vice versa, invariant observables generate consistent flows in the space of field theories describing the dynamics of self-interacting chiral $p$-forms in the following sense. Let $\cH^{(\g)}  ( B_{\m(p)}, g_{\m\n} ) $ and $\cO^{ (\g) } ( B_{\m(p)}, g_{\m\n})$
be two scalar functions that depend on a real parameter $\g$
and satisfy the following conditions:
\begin{itemize}
\item $\cH^{(\g)}$ and $\cO^{(\g)}$  obey the equations
\bea
 \frac{\pa}{\pa \g} \cH^{(\g)} = \cO^{(\g)} \, ,\qquad
   \cO^{(\g)}_{[ \mu_1 \ldots \mu_p } H^{(\g)}_{\mu_{p+1} \ldots \mu_{2p} ] } =0 \, ;
 \eea
\item  $\cH^{(0)}  ( B_{\m(p)}, g_{\m\n} ) $ is a solution of \eqref{MasterEquation}.
\end{itemize}
Then $\cH^{(\g)}  ( B_{\m(p)}, g_{\m\n} ) $ is a solution of \eqref{MasterEquation} at every value of the parameter $\g$.

Guided by the $d=6$ results obtained in this paper, we may conjecture their extension to $d= 4n +2 >6 $ dimensions, specifically: (i) every physical observable may be realised as a function of the energy-momentum tensor $T_{\m\n}$; and (ii) the number of functionally independent invariant observables constructed from $T_{\m\n}$, is equal to the number of functionally independent Lorentz invariants constructed from a self-dual $(p+1)$-form $\L_{\m(p+1)}$  that plays an important role in the next subsection. The proof of this conjecture is beyond the scope of this paper.


\subsection{Extending the INZ-type approach beyond six dimensions} \label{INZ-d>6}

To develop an INZ-type formulation, we introduce a general self-dual $(p+1)$-form field
$\Lambda_{\mu ( p+ 1 ) }$, ${\L}^* = \L$,
which plays the role of an auxiliary field. Integrating out the auxiliary field
should lead to the model \eqref{BLM_action} which we have studied earlier.

To start with, we discuss a free model.
The proposed free action is
\begin{align}
\label{FreeAction}
    S_{\rm free} [ A , a , \Lambda ] &= \frac{1}{p!} \int \left[ \frac{1}{2} E \cdot B - \frac{1}{2} B \cdot B + ( B + \lambda )\cdot  ( B + \lambda ) ] \right] \nonumber \\
    &= S_{\text{free}} [ A , a ] + \frac{1}{p!} \int ( B + \lambda ) \cdot ( B + \lambda ) \, ,
\end{align}
where we have defined
\begin{align}
    \lambda_{\mu ( p ) } = \Lambda_{\mu_1 \ldots \mu_p \nu} v^\nu \, .
\end{align}
The equation of motion for $\L_{\m(p+1)}$ is
\begin{align}
    \left( B + \lambda \right)^{[\mu_1 \ldots \mu_p} v^{\mu_{p+1} ] } ={\bm \varepsilon}^{\mu_1 \ldots \mu_{p+1} \nu_1 \ldots \nu_{p+1}} \left( B + \lambda \right)_{\nu_1 \ldots \nu_p} v_{\nu_{p+1}} \, .
\end{align}
Contracting with $v_{\mu_{p+1}}$ and lowering the indices gives
\begin{align}
    B_{\mu ( p ) } + \lambda_{\mu ( p ) } =0\, ,
\end{align}
and the action \eqref{FreeAction} turns into $S_{\text{free}} [ A , a ] $.
Since ${\Lambda}^* = \Lambda$,  the Lagrange multiplier $\L_{\m(p+1)}$
becomes a function of $B_{\m(p)}$,
\begin{align}
    \Lambda_{\mu ( p + 1 ) } = ( p + 1 ) B_{[ \mu_1 \ldots \mu_p} v_{\mu_{p+1} ] }
    + \frac{1}{p!} {\bm \varepsilon}_{\mu_1 \ldots \mu_{p+1} \nu_1 \ldots \nu_{p+1}} B^{\nu_1 \ldots \nu_p} v^{\nu_{p+1}} \, .
\end{align}
We have shown that the model with action \eqref{FreeAction} is equivalent to that described by $S_{\text{free}} [ A , a ] $.

Our next task is to show that the action  \eqref{FreeAction} is invariant under
a simple generalization of the gauge transformation \eqref{PSTGT2} given by
\begin{align} \label{PSTGT3}
    \delta A_{\mu ( p ) } = \frac{\varphi}{\sqrt{ - \partial a \partial a}} \left[ \mathfrak{E} + 2 \left( B + \lambda \right) \right]_{\mu(p)} \, , \qquad \delta a = \varphi \, , \qquad
    \d \L_{\m(p+1)} =0\, ,
\end{align}
where we have denoted
\begin{align}
    \mathfrak{E}_{\mu ( p ) } = E_{\mu ( p ) } - B_{\mu ( p ) } \, .
\end{align}
First of all, let us give the fields $A_{\m(p)} $ and $a$ arbitrary small disturbances, as in \eqref{113}, while keeping the Lagrange multiplier $\L_{\m(p+1)}$ fixed, and compute the corresponding variation of the action  \eqref{FreeAction}.
Routine calculations give
\begin{align}
    &\delta_{A, a} \int \left( B + \lambda \right) \cdot \left( B + \lambda \right) \nonumber \\
    &\quad =\frac{2}{p!} \int {\bm \varepsilon}^{\mu ( p ) \nu ( p ) \rho \sigma} \left( \mathfrak{E} + B + \lambda \right)_{\nu ( p ) } \left( B + \lambda \right)_{\mu ( p ) } v_\rho \delta v_\sigma \nonumber \\
    &\qquad - \frac{2}{p!} \int {\bm \varepsilon}^{\mu_1 \ldots \mu_{p+1} \nu_1 \ldots \nu_{p+1} } \delta A_{\nu_1 \ldots \nu_p} \partial_{\nu_{p+1}} \left[ \left( B + \lambda \right)_{\mu_1 \ldots \mu_p} v_{\mu_{p+1}} \right] \, .
\end{align}
Combining this with the variation of $S_{\text{free}} [ A, a ] $ gives
\begin{align}
    & \delta_{A,a} \left[ S_{\text{free}} [ A, a ] + \frac{1}{p!}
    \int \left( B + \lambda \right) \cdot \left( B + \lambda \right) \right] \nonumber \\
    &\quad = - \frac{1}{(p!)^2} \int {\bm \varepsilon}^{\mu_1 \ldots \mu_{p+1} \nu_1 \ldots \nu_{p+1}} \delta A_{\nu_1 \ldots \nu_p} \partial_{\nu_{p+1}} \left[
    \mathfrak{E} + 2 ( B + \lambda )
    \right]_{\mu_1 \ldots \mu_p} v_{\mu_{p+1}} \nonumber \\
    &\qquad
    + \frac{1}{2 (p!)^2} \int {\bm \varepsilon}^{\mu_1 \ldots \mu_p \nu_1 \ldots \nu_p \rho \sigma} \left[ \mathfrak{E} + 2 \left( B + \lambda \right) \right]_{\mu_1  \dots \m_p  }
\left[ \mathfrak{E} + 2 ( B + \lambda ) \right]_{\nu_1 \dots \n_p }
    v_\rho \delta v_\sigma \, .
\end{align}
With this result, the proof of invariance of the action under \eqref{PSTGT3}
 is identical to that discussed in subsection \ref{BLMreview}.
 One may also see that the action \eqref{FreeAction}
  is invariant under the following generalisation  of
  \eqref{PSTGT1}
  \begin{align} \label{PSTGT4}
       \delta A_{\mu ( p ) } = p v_{[\mu_1} \psi_{\mu_2 \ldots \mu_p]} \, , \qquad
    \delta a = 0 \, , \qquad \d \L_{\m(p+1)}=0\, .
\end{align}
Since the auxiliary field is inert under the gauge transformations \eqref{PSTGT3}
and \eqref{PSTGT4}, we can immediately turn on self-interactions of the form
\begin{align}
\label{Interaction}
    S [ A , a , \Lambda ] &= \frac{1}{p!} \int \left[ \frac{1}{2} E \cdot B - \frac{1}{2} B \cdot B + ( B + \lambda )\cdot  ( B + \lambda ) ] \right]
    - \int   \mathcal{V} \left(  \L_{\m(p+1)} , g_{\m\n}\right) \, .
\end{align}


\subsection{Weyl transformations}

In conclusion, we discuss the Weyl transformation laws of different fields under consideration. A Weyl transformation acts on the metric and its inverse
by the rule
\begin{align}
    g_{\mu \nu} \to \re^{- 2 \sigma  } g_{\mu \nu} \, , \qquad g^{\mu \nu} \to \re^{2 \sigma  } g^{\mu \nu} \, ,
\end{align}
with the local scale parameter $\s = \s(x)$ being arbitrary.
It follows that  the Levi-Civita tensors \eqref{Levi-Civita} transform as
\begin{align}
    \bm{\varepsilon}^{\mu_1 \dots \m_d} \to \re^{d \sigma} \bm{\varepsilon}^{\mu_1 \ldots \mu_d} \, , \qquad
     \bm{\varepsilon}_{\mu_1 \dots \m_d} \to \re^{-d \sigma} \bm{\varepsilon}_{\mu_1 \ldots \mu_d} \, .
\end{align}
The gauge $p$-form $A_{\m(p)}$ is Weyl inert, and thus the Weyl transformation laws of  the field strength $F_{\m (d+1)}$ and its contravariant counterpart $F^{\m(d+1) }$
are
\begin{align}
    F_{\mu (p+1)} \to F_{\mu (p+1)} \, , \qquad  F^{\mu (p+1)} \to \re^{d \s} F^{\mu (p+1)} \, .
\end{align}
For the dual field strength $ {F}^{*\,\mu (p+1)} $ and its covariant counterpart
$F_{\mu (p+1)}^{\ast} $ we get
\begin{align}
    {F}^{*\,\mu (p+1)} \to \re^{d \sigma} {F}^{*\,\mu (p+1)} \, , \qquad
    {F}^*_{\mu (p+1)}  \to {F}^*_{\mu (p+1)} \, .
\end{align}
The scalar field $a(x)$ in \eqref{BLM_action}
must be Weyl neutral in order for $v_\m$ to be primary
and thus
\begin{align}
    v_\mu \to \re^{- \sigma} v_\mu \, , \qquad
    v^\mu \to \re^{\sigma} v^\mu
\, . \end{align}
These transformation laws imply that the electric field \eqref{electric}
and its contravariant counterpart transform as
\begin{align}
    E_{\mu (p)} \to \re^{\sigma} E_{\mu (p)} \, , \qquad
    E^{\mu (p)} \to \re^{(d-1) \sigma} E^{\mu (p)}
 \, .
 \end{align}
Similar transformation laws hold for the magnetic field \eqref{B_constraint}
and its contravariant counterpart:
\begin{align}
   B_{\mu(p)} \to \re^{\sigma} B_{\mu (p)} \, ,\qquad
    B^{\mu (p)} \to e^{(d-1) \sigma} B^{\mu (p)} \, .
   \end{align}
The above relations imply that the free PST Lagrangian density,
\begin{align}
    \sqrt{-g} \mathcal{L}^{\text{free}}_{\text{PST}} = \sqrt{-g}  \frac{1}{2 p!}
    \Big(  B_{\mu (p)} E^{\mu (p)} -B_{\mu (p)} B^{\mu (p)} \Big) \, ,
\end{align}
is Weyl invariant.

Now, consider the Lagrangian density in \eqref{BLM_action},
\begin{align}
    \sqrt{-g} \mathcal{L}_{\text{PST}} = \sqrt{-g} \Big( \frac{1}{2p!} B_{\mu (p)} F^{\mu (p)} - \mathcal{H} ( B_{\mu (p)} , g_{\m\n} )\Big) \, .
\end{align}
In order for $\sqrt{-g} \mathcal{L}_{\text{PST}}$ to be Weyl invariant, $\mathcal{H} $ must be a homogeneous function,
\begin{align}
    \mathcal{H} \left( \re^{\sigma} B_{\mu(p)}, \re^{- 2 \sigma  } g_{\mu \nu}  \right) = \re^{d \sigma} \mathcal{H} \left( B_{\mu (p)} , g_{\m\n} \right) \, .
\end{align}
Then $H_{\m (p)}$ defined by \eqref{H-der}
is a primary field,
\begin{align}
    H_{\mu (p)} \to \re^{\sigma} H_{\mu(p)} \, , \qquad
    H^{\mu (p)} \to \re^{(d-1) \sigma} H^{\mu (p)} \
\, . \end{align}

Let us turn to the INZ formulation,
\begin{align}
    \mathcal{L}_{\text{INZ}} = \mathcal{L}_{\text{PST}}^{\text{free}} + \frac{1}{p!} \left( B_{\mu (p)} + \lambda_{\mu (p)} \right)\left( B^{\mu (p)} + \lambda^{\mu (p)} \right)
     - \mathcal{V} \left(  \L_{\m(p+1)} , g_{\m\n}\right) \, ,
\end{align}
with $\lambda_{\mu (p)} = \Lambda_{\mu (p)\nu } v^\nu$.
In order for $B_{\mu (p)}$ and $\lambda_{\mu (p)}$ to have the same Weyl transformation law, $\Lambda_{\mu (p+1)}$ must be Weyl neutral,
\begin{align}
    \Lambda_{\mu (p+1)} \to \Lambda_{\mu (p+1)} \, , \qquad
    \Lambda^{\mu (p+1)} \to \re^{d \sigma} \Lambda^{\mu (d+1)}
\, . \end{align}
In order for $\sqrt{-g} \mathcal{L}_{\text{INZ}}$ to be Weyl-invariant, $\mathcal{V} $ must be a homogeneous function,
\begin{align}\label{confd}
    \mathcal{V} \left( \L_{\m(p+1)},  \re^{- 2 \sigma  } g_{\mu \nu} \right) = \re^{d \sigma} \mathcal{V} ( \L_{\m(p+1)} , g_{\m\n} ) \, .
\end{align}
In the $d=6$ case, there is a unique solution to this condition given in \eqref{VMM}.
It corresponds to the ModMax theory, but not to the Bialynicki-Birula theory, as was explained in Section \ref{examples}. In $d=10$  and higher space-time dimensions there are more solutions of \eqref{confd} (see e.g. \cite{Avetisyan:2022zza}), which would be of interest to study.

\section{Conclusion}
\label{Conclusion}

In this paper, for the aim of developing a general approach to the construction of $\TT$-like deformations of interacting chiral form theories in six and higher space-time dimensions, we have reviewed the PST formulation of these theories, in particular the structure of the PST energy-momentum tensor, and related it to two other Lagrangian formulations of chiral $p$-forms, the `clone' formulation \cite{Avetisyan:2022zza,Evnin:2022kqn} and a novel formulation which we obtained by generalizing the duality-symmetric construction of non-linear $4d$ electrodynamics by Ivanov, Nurmagambetov and Zupnik \cite{Ivanov:2014nya} to $d\geq 6$ .

We have shown that in all these formulations (which are on-shell equivalent, except for a specific case of the Bialynicki-Birula theory, as explained in Section \ref{examples}), the chiral $p$-form equations of motion  (i.e.~non-linear self-duality conditions), the on-shell values of their Lagrangians, and the energy-momentum tensors are manifestly Lorentz (or diffeomorphism) invariant and independent of the auxiliary fields. Using these results we have introduced the general notion of `invariant observables', the quantities which satisfy a certain invariance condition that makes them gauge- and Lorentz-invariant and independent of the auxiliary fields on the mass shell. The invariant observables were shown to be functions of the components of the energy-momentum tensor and as such are natural operators for constructing different deformations of chiral form theories satisfying corresponding $\TT$-like flow equations. We have studied such stress tensor flows in all these formulations and, as an explicit example, established that the ModMax-Born-Infeld chiral tensor theory satisfies two commuting flow equations driven by $6d$ analogues of the $\TT$ and root-$\TT$ operators, see equations \eqref{TTbar_flow} and \eqref{root_TTbar_flow}, respectively. We have also argued that, extending $4d$ results of \cite{Ferko:2023wyi}, such stress tensor deformations are, in some sense, generic, since \emph{any} parameterized family of Lorentz-invariant $6d$ chiral tensor theories satisfy some generalized $\TT$-like flow equation. Finally, in Appendices \ref{AppendixA} - \ref{AppendixB}, we then presented some extensions of our analyses by discussing $\TT$-like flows for self-interacting gauge $(2n-1)$-forms in $d=4n$ dimensions by using three approaches, and by discussing  deformations of $\sU(1)$ duality-invariant supersymmetric theories in $d=4$.

Further directions in which our results would be of interest to generalize include (i) a detailed study of $\TT$-like deformations of duality-symmetric and/or chiral $p$-form theories in $d\geq 8$, with $d=10$ being the most interesting case in the context of string theory; (ii) the study of the possibility of finding supersymmetric generalizations of the non-linear $p$-form theories; and (iii) investigating possible connections between stress tensor deformations of $6d$ chiral tensor theories and theories of modified gravity. In what follows, we briefly discuss each of these directions.

\subsubsection*{\ul{\it $\TT$-like deformations in $d \geq 8$}}

To date, there has been very little work on classical $\TT$-like flows in spacetime dimensions $d \geq 8$ (although see \cite{Babaei-Aghbolagh:2020kjg} for some results concerning deformations of $4$-form field strengths in eight dimensions up to $\mathcal{O} ( \lambda^2 )$). In Appendix \ref{AppendixA} of the present work, we have made some initial remarks concerning such deformations in $d = 4n$ dimensions, but this is only scratching the surface. It would be very interesting to see whether more of the results concerning $T^2$ flows in lower spacetime dimensions have natural analogues in $d \geq 8$. For instance, one future direction is to verify the conjecture that every duality-invariant observable $\mathcal{O}$ can be expressed as a function of the stress tensor in $d \geq 8$, as we mention below equation (\ref{higher_d_conditions}). As another example, in $2d$, it is possible to couple two theories in a ``universal'' way by performing sequential $\TT$ flows \cite{Ferko:2022dpg}; one might ask whether such a procedure also gives rise to natural interrelations between theories in higher dimensions.

As we have emphasized, $\TT$-like deformations in $d = 2, 4$, and $6$ seem to give rise to theories related to strings and branes, including the Nambu-Goto action, the Born-Infeld theory, and the interacting chiral tensor theory on the M5-brane worldvolume. It would be very interesting if this pattern continues in $d = 10$. For instance, one might ask whether a generalized stress tensor deformations of some terms in the action for type IIA or type IIB supergravity generates a physically relevant series of higher-derivative corrections, such as $\alpha'$ corrections that are required by string theory. If so, stress tensor flows might provide an organizing principle for generating and understanding such stringy corrections.

\subsubsection*{\ul{\it Supersymmetry}}

By now, the supersymmetric extensions of generic non-linear electrodynamics theories were constructed only in $d=4$ with the use of the $\cN=1$ and $\cN=2$ superfield formalism
\cite{Kuzenko:2000tg, Kuzenko:2000uh, Kuzenko:2002vk, Kuzenko:2012ht, Kuzenko:2013gr, Ivanov:2013ppa}
(see Appendix \ref{AppendixB} for some details).\footnote{The formalism of $\sU(1)$ duality rotations has recently been extended
to higher-spin conformal gauge fields on conformally flat backgrounds and some of their $\cN$-extended  superconformal cousins \cite{Kuzenko:2021qcx, Kuzenko:2023ebe}.}
In $d=6$ the only available superfield Lagrangian is for a free chiral two-form multiplet constructed in an $N=(1,0)$, $d=6$ harmonic superspace in \cite{Kozyrev:2022dri}. Component actions for $d=6$ supergravity coupled to chiral tensor multiplets were constructed in \cite{Dall'Agata:1997db,Riccioni:1998pj,DePol:2000re}, while the M5-brane \cite{Howe:1996yn,Howe:1997fb,Bandos:1997ui,Aganagic:1997zq,Cederwall:1997gg,Ko:2013dka,Vanichchapongjaroen:2020wza} and its NS5-brane counterpart \cite{Bandos:2000az} remain the only examples of supersymmetric self-interacting chiral two-form theories in $d=6$. In $d=10$ the full supersymmetric duality-invariant actions are available for IIB \cite{Dall'Agata:1998va} and IIA \cite{Bandos:2003et} supergravities. It would be of interest to study whether techniques developed in this paper might be useful for the construction of duality invariant higher order (supersymmetric) corrections to type II, $d=10$ supergravitites. In particular, it would be intriguing to see whether one can use $\TT$-like deformations to build supersymmetric interacting chiral form theories as deformations of free theories.

Another direction, for which supersymmetry might play an important role, is to find a version of the six-dimensional $\TT$-like deformation considered in this work which is defined at the quantum level. It is well-known \cite{Taylor:2018xcy} that no bilinear in stress tensor operators can give rise to a well-defined local operator by point-splitting in $d > 2$; one way to see this is to note that, even near a CFT seed, the leading term in our classical deformation has
\begin{align}
    \left\langle T_{\mu \nu} ( x ) T^{\mu \nu} ( 0 ) \right\rangle = \frac{d-2}{x^{2d}} + \text{regular} \, ,
\end{align}
and thus there is no hope of applying an argument like that of \cite{Zamolodchikov:2004ce} to find a combination of stress tensors which is independent of $x$ except in $d = 2$. However, in a supersymmetric theory, one might hope to identify some combination of bilinears of fields in the stress tensor multiplet (and their derivatives) with the property that the appropriate point-split bilocal combination is actually independent of the distance between the insertion points, and therefore attempt to define a local supercurrent-squared operator by point-splitting.

\subsubsection*{\ul{\it Connections to Gravity}}

Several authors have pointed out that the two-dimensional $\TT$ deformation can be interpreted as a coupling to some kind of gravity \cite{Dubovsky:2017cnj,Dubovsky:2018bmo} or random geometry \cite{Cardy:2018sdv}, or relatedly, that it may be viewed as a certain field-dependent diffeomorphism \cite{Conti:2018tca,Conti:2022egv}.  Generalizations of this statement in higher dimensions \cite{Morone:2024ffm} include connections to modified gravity theories like Eddington-Inspired-Born-Infeld.\footnote{See also \cite{Floss:2023nod} for another interesting observation on the relation between massive gravity and non-linear electrodynamics as its `effective theory'.} In the latter work, the eigenvalue structure of the stress tensor for $4d$ Abelian gauge theories played an important role: in particular, the stress tensor for such a theory always has two pairs of degenerate eigenvalues. We have seen that the stress tensor of a chiral tensor theory in $6d$ has a similar degeneracy, as encoded for instance in the relation
\begin{align}
    \tr ( T^3 ) = \frac{1}{2} \tr ( T^2 ) \tr ( T ) - \frac{1}{18} \left( \tr ( T ) \right)^3 \, ,
\end{align}
which holds for the energy-momentum tensor of any $6d$ chiral tensor theory. The degeneracy of the eigenvalues also implies
\begin{align}
    \det ( T ) = - \frac{1}{6^3} \left( \tr ( T^2 ) - \frac{1}{3} \left( \tr ( T ) \right)^2 \right)^3 \, ,
\end{align}
which is reminiscent of the corresponding relation for gauge theories pointed out in \cite{Conti:2018jho}. In particular, this means that the $6d$ flow equation driven by $\mathcal{O}_{T^2}$ which we considered in this work can also be presented in the form
\begin{align}
    \frac{\partial \mathcal{L}}{\partial \lambda} = - \frac{1}{2} \left( \det ( T ) \right)^{1/3} \, ,
\end{align}
which is driven by a power of the determinant of the energy-momentum tensor. It would be very interesting to see whether interacting $6d$ chiral tensor theories, such as the one with Born-Infeld-type interaction function which describes the worldvolume gauge theory of the M5-brane, has a connection to modified gravity theories as in the $4d$ case of \cite{Morone:2024ffm}.

\section*{Acknowledgements}
The authors are grateful to Igor Bandos, Jessica Hutomo, Yi Pang, Savdeep Sethi, Alessandro Sfondrini, and Roberto Tateo for useful discussions, and Oleg Evnin and Karapet Mkrtchyan for email correspondence. C. F., S. M. K., D. S. and G. T.-M. acknowledge the kind hospitality and financial support extended to them at the MATRIX Program ``New Deformations of Quantum Field and Gravity Theories,'' between 22 Jan and 2 Feb 2024, during the final stage of this project. The work of S. M. K. and D. S. was partially supported by
the Australian Research Council, project No. DP230101629. D. S. thanks the Department of Physics, UWA for kind hospitality during his visit on Jan 8-21, 2024.
The work of D. S. was also partially supported by Spanish AEI MCIN and FEDER (ERDF EU) under grant PID2021-125700NB-C21 and by the Basque Government Grant IT1628-22.
G. T.-M. has been supported by the Australian Research Council (ARC) Future Fellowship FT180100353, ARC Discovery Project DP240101409, and the Capacity Building Package of the University of Queensland.

\appendix
\section{$\TT$-like flows for self-interacting gauge $(2n-1)$-forms in $d=4n$ dimensions}
\label{AppendixA}

This Appendix is devoted to a general discussion of $\TT$-like flows in $\sU(1)$ duality-invariant theories of self-interacting gauge $(2n-1)$-forms in $d=4n$ dimensions.

\subsection{Gaillard-Zumino-Gibbons-Rasheed-type formalism}
\label{AppendixA1}

As is known, the Gaillard-Zumino-Gibbons-Rasheed  (GZGR) formalism
\cite{Gaillard:1981rj, Gibbons:1995cv, Gibbons:1995ap, Gaillard:1997zr, Gaillard:1997rt}
to describe duality-invariant
models for nonlinear electrodynamics in four dimensions
was also extended to higher dimensions in
\cite{Gibbons:1995cv, Tanii:1998px, Araki:1998nn, Aschieri:1999jr} (for reviews see \cite{Kuzenko:2000uh, Tanii:2014gaa}).
This formalism is useful in the framework of $\TT$-like flows of these theories.

In a curved space $\cM^d$  of even dimension $d=4n$, with $n$ a positive integer,
we  consider
a self-interacting theory of a  gauge $p$-form $A_{\m_1 \dots \m_{p}}$ (for $p=2n-1$)
such that its Lagrangian, $L=L (F)$, is a function of the field strength
$F_{\m_1 \dots \m_{p+1}} =(p+1)  \pa_{ [\m_1} A_{\m_2 \dots \m_{p+1} ] }$.
In order for this theory to possess $\sU(1)$ duality invariance,
the Lagrangian must satisfy the self-duality equation \cite{Araki:1998nn}
\bea
 G^{\m_1 \dots \m_{p+1}} G^*_{\m_1 \dots \m_{p+1}}
+  F^{\m_1 \dots \m_{p+1}} F^*_{\m_1 \dots \m_{p+1}} =0~,
\label{self-duality}
\eea
where we have introduced\footnote{A partial derivative of $L(F)$ is defined by
$ \d L(F) = \d F_{\m_1 \dots \m_{p+1}} \frac{\pa L (F)}{\pa F_{\m_1 \dots \m_{p+1}}}$.}
\bea
{G}^{*\,\m_1 \dots \m_{p+1}} (F)= (p+1)!
\frac{\pa L (F)}{\pa F_{\m_1 \dots \m_{p+1}}}~.
\eea
As usual, the notation  $ F^*$ is used  for the Hodge dual of $F$,
\bea
{F}^{*\,\m_1 \dots \m_{p+1}} =
\frac{1}{(p+1)!} \, {\bm \ve}^{\m_1 \dots \m_{p+1} \n_1 \dots \n_{p+1}} \,
F_{\n_1 \dots \n_{p+1} }~.
\eea
Every solution of \eqref{self-duality} defines a  $\sU(1)$ duality-invariant theory.
An infinitesimal U(1) duality transformation is given by
\bea
\d   \left( \begin{array}{c}  G \\ F \end{array} \right)
=  \left( \begin{array}{cr} 0 &
- c \\  c  &  0 \end{array} \right)
\left( \begin{array}{c}  G \\ F  \end{array} \right) ~,
\label{duality-tran}
\eea
with $c \in \mathbb R$ a constant  parameter.

A function $\cO (F)$ is said to be a duality-invariant observable if it obeys the first-order differential equation
\bea
\frac{\pa \cO (F)}{\pa F_{\m_1 \dots \m_{p+1}}} G_{\m_1 \dots \m_{p+1}} =0~.
\eea
In accordance with \eqref{duality-tran}, $\cO(F)$ is inert under the duality transformations. An example of a duality-invariant observable is the energy-momentum tensor $T_{\m\n}$. This follows from a simple generalization of the arguments given in
\cite{Gibbons:1995cv,Gaillard:1997zr,Gaillard:1997rt}. Specifically, let $L(F; \g)$ be the Lagrangian of a $\sU(1)$ duality-invariant theory, where $\g$ is a duality-invariant parameter; $L(F; \g)$  is a solution of \eqref{self-duality} for every value of $\g$.
Then, the function $\pa L(F; \g) / \pa \g$ is  duality invariant.

Duality-invariant observables generate consistent flows in the space of field theories describing the dynamics of self-interacting gauge $p$-forms in the following sense. Let $L^{(\g)}  (F ) $ and $\cO^{ (\g) } ( F)$
be two scalar functions that depend on a real parameter $\g \in (-\e, \e)\subset \mathbb R$ and satisfy the following conditions:
\begin{itemize}
\item $L^{(\g)}$ and $\cO^{(\g)}$  obey the equations
\bea\label{higher_d_conditions}
 \frac{\pa}{\pa \g} L^{(\g)} = \cO^{(\g)} \, ,\qquad
\frac{\pa \cO^{(\g)} (F)}{\pa F_{\m_1 \dots \m_{p+1}}}G^{(\g)}_{\m_1 \dots \m_{p+1}} =0~.
 \eea
\item  $L^{(0)}  ( F ) $ is a solution of \eqref{self-duality}.
\end{itemize}
Then $L^{(\g)}  ( F ) $ is a solution of \eqref{self-duality} at every value of the parameter $\g$.

In the four-dimensional case, $p =1$, it was demonstrated in \cite{Ferko:2023wyi} that every scalar duality-invariant observable $\cO$ is a function of the energy-momentum tensor, $\cO = \cO(T_{\m\n})$. It is an interesting problem to determine whether this property also holds beyond four dimensions.

\subsection{Ivanov-Zupnik-type formalism}
\label{AppendixA2}

Ref. \cite{Kuzenko:2019nlm} described a reformulation of the above theory, which for $p=1$ coincides with the Ivanov-Zupnik formulation \cite{Ivanov:2001ec, Ivanov:2002ab, Ivanov:2003uj}.
In this reformulation, along with the field strength $F_{\m_1\dots \m_{p+1}}$, the Lagrangian $L(F, \L)$  depends on an auxiliary unconstrained rank-$(p+1)$
antisymmetric tensor $\L_{\m_1 \dots \m_{p+1}}$
and has the form
\bea
L(F,\L) = \frac{1}{(p+1)!} \Big\{ \hf F \cdot F +  \L \cdot \L - 2 \L \cdot F\Big\}
+ L_{\rm int} (\L) ~,
\label{first-order}
\eea
where we have denoted
\bea
 \L\cdot F:= \L^{\m_1 \dots \m_{p+1}} F_{\m_1 \dots \m_{p+1}}~.
 \eea
 The last term in \eqref{first-order},  $L_{\rm int} (\L) $, is at least quartic
 in $\L_{\m_1 \dots \m_{p+1}}$.
It is assumed that the equation of motion for $V$,
\bea
\frac{\pa}{\pa \L^{\m_1 \dots \m_{p+1}} } L(F,\L) =0~,
\eea
allows one to integrate out the auxiliary field $\L$, resulting in a Lagrangian $L(F)$.

The self-duality equation \eqref{self-duality} proves
to be equivalent to the following condition on the self-interaction in \eqref{first-order}
\bea
 \L^*_{\m_1 \dots \m_{p+1}} \frac{\pa}{\pa \L_{\m_1 \dots \m_{p+1}} } L_{\rm int} (\L) =0~.
\eea
Introducing (anti) self-dual components of $\L$,
\bea
\L_\pm^{\m_1\dots \m_{p+1}} = \hf \Big( \L^{\m_1\dots \m_{p+1}}
\pm \ri  \L^{* \,\m_1\dots\m_{p+1}} \Big) ~, \qquad
 \L^*_\pm = \mp\ri \L_\pm ~,\qquad \L = \L_+ +\L_-~,
\eea
the above condition turns into
\bea
\Big(
 \L_+^{\m_1 \dots \m_{p+1}} \frac{\pa}{\pa \L_+^{\m_1 \dots \m_{p+1}} }
 - \L_-^{\m_1 \dots \m_{p+1}} \frac{\pa}{\pa \L_-^{\m_1 \dots \m_{p+1}} } \Big)L_{\rm int} (\L_+, \L_-)
=0~.
 \eea
This means that $ L_{\rm int} (\L_+, \L_-) $ is invariant under $\sU(1)$ phase transformations,
\bea
L_{\rm int} (\re^{\ri \vf}  \L_+, \re^{-\ri \vf} \L_-)  = L_{\rm int} (\L_+, \L_-) ~, \qquad
\vf \in {\mathbb R}~.
\label{manifestU(1)}
\eea
It should be pointed out that the duality transformation \eqref{duality-tran}  acts on $\L$ as
\bea
\d \L = c \L^*~.
\eea
In four dimensions, the most general solution to the condition \eqref{manifestU(1)}
 is given by
\bea
 L_{\rm int} (\L_+, \L_-) = f (|\L_+ \cdot \L_+ |)
 ~,
 \eea
 with $f(x)$ a real function
of one variable.
Similar solutions exist in higher dimensions. However more general
self-interactions are possible beyond four dimensions.

The description of consistent deformations of $\sU(1)$ duality-invariant self-interacting gauge $p$-forms
is simple in the above approach based on
the first-order formulation  \eqref{first-order}. One considers a Lagrangian of the form
\bea
L^{(\g)}(F,\L) = \frac{1}{p!} \Big\{ \hf F \cdot F +  \L \cdot \L - 2 \L \cdot F\Big\}
+ L^{(\g)}_{\rm int} (\L) ~,
\label{first-order-deformed}
\eea
where $L^{(\g)}_{\rm int} (\L) $ is required to obey the condition \eqref{manifestU(1)}.


\subsection{PST-type formalism}
\label{AppendixA3}

Another way to construct duality invariant theories for $(2n-1)$-gauge potentials in $d=4n$ dimensions is provided by the PST formalism \cite{Buratti:2019cbm}. It relies on the introduction of a pair of $p$-form potentials, $A_{\m_1 \dots \m_{p}}^a$, where $p=2n-1$ and $a=1,2$. In this case the $U(1)$ symmetry is manifest in that it is realized as an $SO(2)$-duality rotation of the index $a$. The corresponding pair of field strengths is given by
\begin{equation}
  F_{\mu_1 \ldots \mu_{p+1}}^a   = ( p + 1 ) \partial_{[\mu_1} A_{\mu_2 \ldots \mu_{p+1} ]}^a.
\end{equation}
In addition we introduce the electric and magnetic pairs of $p$-forms
\begin{align}
    E_{\mu_1 \ldots \mu_p}^a = F_{\mu_1 \ldots \mu_p \nu }^a v^\nu \,  ,\qquad B_{\mu_1 \ldots \mu_p}^a =  F^{*a}_{\mu_1 \ldots \mu_p \nu } v^\nu \,.  \qquad
\end{align}
In the PST approach the action, involving also the scalar auxiliary field $a$, then assumes the form (for simplicity we work in a flat background)
\begin{equation}
\label{action1}
    S [ A, a ] =
       \int \rd^d x
     \left[ \frac{1}{2 p!}\varepsilon^{ab} E^a \cdot B^b - \cH ( B )\right] \, , \qquad
     v_\mu = \frac{\partial_\mu a}{\sqrt{ - \partial a \cdot \partial a } } \, ,
\end{equation}
where the Hamiltonian $\cH(B)$ is an $SO(2)$-invariant function of $B^a$. In this case the condition for Lorentz-invariance of the action \eqref{action1} reads
\begin{equation}\label{condlor}
    \varepsilon^{ab} \left(B_{[\mu_1 \ldots \mu_p}^a B_{\mu_{p+1} \ldots \mu_{2p} ] }^b
    - H_{[ \mu_1 \ldots \mu_p }^a H_{\mu_{p+1} \ldots \mu_{2p} ] }^b\right)=0 \,,
\end{equation}
since it ensures that the auxiliary field $a$ becomes a pure gauge degree of freedom \cite{Buratti:2019cbm}. Above we have defined the partial derivatives $H_{\mu_1 \ldots \mu_p}^a=\partial\cH/\partial B^{a\mu_1 \ldots \mu_p}$
of the Hamiltonian through the relation
\begin{equation}
    \delta \cH (B) = \frac{1}{p!} \delta B^{a\mu_1 \ldots \mu_p} H_{\mu_1 \ldots \mu_p}^a\, .
\end{equation}
The (PST gauge-fixed) equation of motion of the $p$-form potential $A^a$ derived from the action \eqref{action1} is
\begin{equation}\label{eom1}
E_{\mu_1 \ldots \mu_p}^a = \varepsilon^{ab}H_{\mu_1 \ldots \mu_p}^b.
\end{equation}

A Lorentz- and duality-invariant observable $\cO(B)$ is an $SO(2)$-invariant function of the magnetic fields $B^a$, such that
\begin{equation}\label{condlor1}
\varepsilon^{ab}\cO_{[ \mu_1 \ldots \mu_p }^a H_{\mu_{p+1} \ldots \mu_{2p} ] }^b =0 \, ,
\end{equation}
where
\begin{equation}
\cO_{ \mu_1 \ldots \mu_p }^a=\frac{\partial\cO}{\partial B^{a\mu_1 \ldots \mu_p}}.
\end{equation}
In particular, with the same steps performed in Section \ref{invobs}, one can show that on-shell, i.e. when \eqref{eom1} holds, thanks to \eqref{condlor1} the operator
$\cO(B)$ is invariant under infinitesimal variations of the four-vector $v^\mu$. This signals, once again, that the field $a(x)$ is non-propagating and drops out from the dynamics.

Let us now consider a deformed $SO(2)$-invariant Hamiltonian $\cH ( B,\lambda)$ which depends on a continuous parameter and tends to $\cH ( B )$ as $\lambda\rightarrow0$. Then the  condition \eqref{condlor} for the decoupling of $a$ is replaced by
\begin{equation}\label{condlor2}
    \varepsilon^{ab} \left(B_{[\mu_1 \ldots \mu_p}^a B_{\mu_{p+1} \ldots \mu_{2p} ] }^b
    - H_{[ \mu_1 \ldots \mu_p }^{a(\lambda)} H_{\mu_{p+1} \ldots \mu_{2p} ] }^{b(\lambda)} \right)=0 \,.
\end{equation}
This deformation can be interpreted as a consistent flow in the space of field theories if there exists an $SO(2)$-invariant operator $\cO(B,\lambda)$ such that the latter and $\cH(B,\g)$ satisfy the conditions:
\begin{itemize}
\item $\cH{(B,\g)}$ and $\cO{(B,\g)}$  obey the equations
\bea
 \frac{\pa}{\pa \g} \cH(B,\g) = \cO(B,\g) \, ,\qquad
  \varepsilon^{ab}\,\cO^{a(\g)}_{[ \mu_1 \ldots \mu_p } H^{b(\g)}_{\mu_{p+1} \ldots \mu_{2p} ] } =0 \, ;\label{coch1}
 \eea
\item  $\cH( B,0)\equiv \cH( B) $ is a solution of \eqref{condlor}.
\end{itemize}
In fact, the derivative w.r.t. $\lambda$ of \eqref{condlor2} corresponds to the second relation in \eqref{coch1}, and the second condition above for $\cH(B,0)$ enforces the initial condition for \eqref{condlor2}.
According to this scheme, in a $d=8$ dimensional space-time the $SO(2)$-invariant quartic interactions, $\cO(B,\lambda) \sim \lambda B^4$, found in \cite{Buratti:2019cbm} can be interpreted as first order flows (in $\lambda$) of the free theory of duality invariant three-form potentials. In fact, at this order, the quartic interactions found in \cite{Buratti:2019cbm} satisfy, actually by definition, the Lorentz-invariance condition \eqref{condlor1}

The analysis of this section can in principle be reproduced by a dimensional reduction from $d=4n+2$ to $d=4n$ of the corresponding analysis of Section  \ref{invobs}, in which the only non-vanishing components of the $(4n+2)$-dimensional chiral gauge potential $A_{\mu_1 \cdots \mu_{2n}}$ are the gauge fields $A^{a}{}_ {M_1\cdots M_{2n-1}}$ with $a=(4n,4n+1)$ and $M_i=0,\cdots,4n-1$.

\section{Deformations of $\sU(1)$ duality-invariant supersymmetric  theories in $d=4$} \label{AppendixB}

In this Appendix we briefly describe how consistent deformations can be defined in the family of $\sU(1)$ duality-invariant models for non-linear $\cN=1$, $d=4$ supersymmetric electrodynamics \cite{Kuzenko:2000tg, Kuzenko:2000uh}. The formalism of  supersymmetric duality rotations in Minkowski superspace \cite{Kuzenko:2000tg, Kuzenko:2000uh} was extended to supergravity in \cite{Kuzenko:2002vk} and the reader is referred to that paper for more technical details. We only point out that the supergravity covariant derivatives are denoted
$\cD_A = (\cD_a, \cD_\a , \bar \cD^\ad)$, and the covariantly chiral analogue of the scalar curvature is denoted $R$, $\bar \cD_\ad R =0$.

We consider a theory for an Abelian
vector multiplet in curved superspace and denote by $S[W , {\bar W}]$
the corresponding action functional. The action depends
on the covariantly chiral spinor field strength $W_\a$
 and its conjugate ${\bar W}_\ad$,
\bea
W_\a = -\frac{1}{4}\,  (\bar \cD^2 -4R)
 \cD_\a  V~, \qquad \bar \cD_\bd W_\a=0~,
\eea
which are constructed  in terms of a real unconstrained gauge prepotential $V$.
The prepotential is defined modulo gauge transformations
\bea
\d_\l V = \l + \bar \l ~, \qquad \bar \cD_\ad \l =0~,
\eea
such that $ W_\a $ and ${\bar W}_\ad$ are gauge invariant.
The gauge-invariant  field strengths  obey
the Bianchi identity
\bea
\cD^\a W_\a = \bar \cD_\ad {\bar W}^\ad~,
\label{eq:bianchi}
\eea
and thus $W_\a$ is a reduced chiral superfield.
The action $S[W , {\bar W}]$ is assumed to have no dependence on  $\cD^\a W_\a $, 
and therefore it
can unambiguously be defined
as a functional of a {\it general}
 chiral superfield $W_\a$ and its conjugate ${\bar W}_\ad$.
Then, defining
\bea
{\rm i}\,M_\a := 2\, \frac{\d }{\d W^\a}\,S[W , {\bar W}]~,
\eea
the equation of motion for $V $ is
\bea
\cD^\a M_\a = \cDB_\ad {\bar M}^\ad~.
\label{eq:eom}
\eea
Here the variational derivative $\d S/\d W^\a $ is defined by
\bea
\d S =  \int \rd^4 x \,{\rm d}^2 \q \,\cE\, \d W^\a \frac{\d S}{\d W^\a}+{\rm c.c.}~,
\eea
where $\cE$ denotes the chiral integration measure, and $W^\a$ is viewed to be an unrestricted covariantly chiral spinor.

Since the Bianchi identity (\ref{eq:bianchi}) and the equation of
motion (\ref{eq:eom}) have the same functional form, one may
consider $\sU(1)$ duality rotations
\bea
\d W_\a = c M_\a ~,
\qquad \d M_\a = - c W_\a~,
\label{DualRot}
\eea
with $c \in {\mathbb R}$ a constant parameter. The theory possesses this $\sU(1)$ duality invariance if and only if the action obeys the self-duality equation
\bea
{\rm Im} \int \rd^4 x \rd^2 \q  \,\cE \Big\{ W^\a W_\a  +M^\a M_\a \Big\} =0~.
\label{SSDE}
\eea

A functional $\cO[W, \bar W]$ is said to be duality invariant if it is inert under \eqref{DualRot},
\bea
 \int \rd^4 x \,{\rm d}^2 \q \,\cE\, \d M^\a \frac{\d \cO}{\d W^\a}+{\rm c.c.}=0~.
 \eea
 Let us recall examples of duality-invariant observables given in
  \cite{Kuzenko:2000tg, Kuzenko:2000uh, Kuzenko:2002vk}.
 Let $S^{(\l)} := S[W , {\bar W}; \l ]$ be a $\sU(1)$ duality-invariant theory,
 with $\l$ a duality-inert parameter. Then $\pa S^{(\l)} /\pa \l$ is duality invariant.
 This property implies that the supercurrent multiplet, i.e. the supersymmetric generalization of the energy-momentum tensor, is duality invariant.
We recall that this multiplet is described
by the supercurrent $J_a = {\bar J}_a$ and the
covaraintly chiral supertrace $T$, $\cDB_\ad T=0$, which  are defined in terms of
covariantized variational derivatives with respect to the supergravity
prepotentials (see, e.g.,  \cite{Buchbinder:1998qv} for the technical details),
\bea
J_a = \frac{\D S}{\D H^a} ~,
\qquad \quad
T = \frac{\D S}{\D \vf} ~,
\eea
and satisfy the conservation equation
\bea
\cDB^{\ad}J_{\a \ad} = -\frac{2}{3}\,\cD_\a T~,
\label{conservation}
\eea
when the matter superfields are put on the mass shell.
The supercurrent multiplet was computed in  \cite{Kuzenko:2002vk}
for the general family of non-linear vector multiplet theories of the form \begin{subequations}\label{VMMM}
\bea
S[W,{\bar W}] &=&
\frac{1}{4} \int  \rd^4 x \rd^2 \q  \,\cE \, W^2 +{\rm c.c.}
+ \frac14  \int \rd^4 x \rd^2 \q \rd^2\bar \q \,E \,
{W^2\,{\bar W}^2}\,
\L\left({u},{\bar u}\right)~,
\eea
where $W^2 =W^\a W_\a$,  $u  := \frac{1}{8} (\cD^2 - 4  \bar R)  W^2$. The restriction of the model to be self-dual requires
the interaction function $\L(u,\bar u)$ to satisfy the differential equation
\bea
{\rm Im} \,\Big\{ \frac{\pa (u \, \L) }{\pa u}
- \bar{u}\, \left(\frac{\pa (u \, \L) }{\pa u}\right)^2
\Big\} = 0~. \qquad
\label{differential}
\eea
\end{subequations}

Duality-invariant observables generate consistent flows in the space of $\sU(1)$ duality-invariant non-linear vector multiplet theories. Let  $S^{(\l)}[W,{\bar W}] $
and $\cO^{(\l)}[W,{\bar W}] $ be two scalar functionals that depend on a real parameter
$\l$ and satisfy the following conditions:
\begin{itemize}
\item $S^{(\l)}[W,{\bar W}] $ and $\cO^{(\l)}[W,{\bar W}] $ obey the equations
\bea
\frac{\pa}{\pa \l} S^{(\l)}[W,{\bar W}] = \cO^{(\l)}[W,{\bar W}] ~,
\quad
  \int \rd^4 x \,{\rm d}^2 \q \,\cE\, \d M^\a \frac{\d  }{\d W^\a}  \cO^{(\l)}
  +{\rm c.c.}=0~.
\eea
\item $S^{(0)}[W,{\bar W}] $ is a solution of the self-duality equation \eqref{SSDE}.
\end{itemize}
Then $S^{(\l)}[W,{\bar W}] $ is a solution of the self-duality equation \eqref{SSDE}
for any value of $\l$.
Within the family of duality-invariant models \eqref{VMMM},
it is natural to conjecture that consistent deformations are generated by duality-invariant observables of the form $\cO [W, \bar W] = {\mathfrak O} [ J_a, T, \bar T]$, since these theories are in a one-to-one correspondence \cite{Kuzenko:2000tg, Kuzenko:2000uh}
with the models for $\sU(1)$ duality-invariant electrodynamics
\bea
L(F_{\m\n})  = -\hf \, \Big( \o + \bar{\o} \Big) +
\o \, \bar{\o} \; \L (\o, \bar{\o} )~,
\label{B.14}
\eea
where $\o$ and $\bar \o$ denote two invariants of the electromagnetic field
\bea
\o= \a +\ri \b~, \qquad
\a = \frac{1}{4} F^{\m\n} F_{\m\n}~, \qquad
\b = \frac{1}{4} F^{\m\n} {F}^*_{\m\n} ~.
\eea
The interaction function $\L (\o, \bar{\o} )$ in \eqref{B.14}
obeys the same equation \eqref{SSDE}.
Examples of consistent $\TT$ flows in duality-invariant models for non-linear supersymmetric electrodynamics were constructed in \cite{Ferko:2019oyv,Ferko:2022iru,Ferko:2023ruw}.


\providecommand{\href}[2]{#2}\begingroup\raggedright\endgroup

\end{document}